\documentclass[aapm,mph,reprint,graphicx,onecolumn]{revtex4-1}
\draft 
\usepackage{amsfonts}
\usepackage{amsmath}
\usepackage{bm}
\usepackage{graphicx}
\usepackage{epstopdf}
\usepackage{subfigure}
\usepackage{algorithmic}
\usepackage{algorithm}
\usepackage{multirow}
\usepackage{url}
\usepackage{float}
\usepackage[mathlines]{lineno}

\usepackage[colorlinks=true]{hyperref}
\hypersetup{urlcolor=black, citecolor=blue,linkcolor=red,}
\begin{document}


\title{Image-domain multi-material decomposition for dual-energy CT based on
correlation and sparsity of material images } 



\author{Qiaoqiao Ding}
\affiliation{ School of Mathematical Sciences, Shanghai Jiao Tong University,\\ 800, Dongchuan Road, Shanghai, 200240, China}
\author{Tianye Niu}
\affiliation{Sir Run Run Shaw Hospital and Institute of Translational Medicine, Key Laboratory of Biomedical Engineering of Ministry of Education, Zhejiang University, Hangzhou, Zhejiang, 310009, China}
\author{Xiaoqun Zhang}
\affiliation{ School of Mathematical Sciences and Institute of Natural Sciences, Shanghai Jiao Tong University, 800, Dongchuan Road, Shanghai, 200240, China}
\author{Yong Long}
\email[]{yong.long@sjtu.edu.cn}
\affiliation{University of Michigan-Shanghai Jiao Tong University Joint Institute, Shanghai Jiao Tong University, 800 Dongchuan Road, Shanghai,  200240, China}
\date{\today}
\begin{abstract}
\noindent \textbf{Purpose:}
Dual energy CT (DECT) enhances tissue characterization because it can produce
images of basis materials such as soft-tissue and bone. DECT is of great
interest in applications to medical imaging, security inspection and
nondestructive testing. Theoretically, two materials with different linear
attenuation coefficients can be accurately reconstructed using DECT technique.
However, the ability to reconstruct three or more basis
materials is clinically and industrially important.
Under the assumption that there are at most three materials in each pixel,
there are a few methods that estimate multiple material images from DECT measurements  by enforcing sum-to-one
and a box constraint ([0 1]) derived from both the volume and mass conservation
assumption.
The recently proposed image-domain multi-material decomposition (MMD) method
introduces edge-preserving regularization for each material image which neglects the relations among material images, and enforced the assumption that there are at most three materials in each pixel using a time-consuming loop over all possible material-triplet in each iteration of optimizing its cost function.
We propose a new image-domain MMD method for DECT that considers
the prior information that different material images have common
edges and encourages sparsity of material composition in each pixel using regularization.
\\

\noindent \textbf{Method:}
The proposed PWLS-TNV-$\ell_0$ method uses penalized weighted least-square
(PWLS) reconstruction with three regularization terms. The first term is a
total nuclear norm (TNV) that accounts for the image property that basis
material images share common or complementary boundaries and each material
image is piecewise constant. The second term is a $\ell_0$ norm that encourages
each pixel containing a small subset of material types out of several possible
materials. The third term is a characteristic function based on sum-to-one and
box constraint derived from the volume and mass conservation assumption. We
apply the Alternating Direction Method of Multipliers (ADMM) to optimize the
cost function of the PWLS-TNV-$\ell_0$ method.
\\

\noindent \textbf{Result:}
We evaluated the proposed method on a simulated digital phantom,
Catphan\copyright600 phantom and patient's pelvis data.
%
We implemented two existing image-domain MMD methods for DECT,
the Direct Inversion \cite{mendonca2014a}
and the PWLS-EP-LOOP method \cite{xue2017statistical}.
We initialized the PWLS-TNV-$\ell_0$ method
and the PWLS-EP-LOOP method
with the results of
the Direct Inversion method
and compared performance of the proposed method
with that of the PWLS-EP-LOOP method.
The proposed method lowered bias of decomposed material fractions
 by $84.47\%$ in the digital phantom study,
by $99.50\%$ in the Catphan\copyright600 phantom study,
and by $99.64\%$ in the pelvis patient study,
respectively, compared to the PWLS-EP-LOOP method.
The proposed method reduced noise standard deviation (STD)
by $52.21\%$ in the Catphan\copyright600 phantom study,
and by $16.74\%$ in the patient's pelvis study,
compared to the PWLS-EP-LOOP method.
The proposed method increased volume fraction accuracy by
$6.04\%, 20.55\%$ and $13.46\%$
for the digital phantom, the Catphan\copyright600 phantom
and the patient's pelvis study, respectively,
compared to the PWLS-EP-LOOP method.
%
Compared with the PWLS-EP-LOOP method,
the root mean square percentage error (RMSE($\%$))
of electron densities in the
Catphan\copyright600 phantom was decreased about $7.39\%$.
 \\
 \\
\noindent \textbf{Conclusions:}
We proposed an image-domain MMD method, PWLS-TNV-$\ell_0$, for DECT. PWLS-TNV-$\ell_0$ method takes low rank property of material image gradients, sparsity of material composition and mass and volume conservation into consideration.  The proposed method suppresses noise, reduces crosstalk, and improves accuracy in the decomposed material images, compared to the PWLS-EP-LOOP method.
\\
\\
\noindent \textbf{ keywords:} Dual energy CT (DECT), Spectral CT,
Multi-material decomposition (MMD),  Total nuclear norm (TNV), Penalized
weighted least-square (PWLS)
\end{abstract}


\maketitle 
\section{Introduction}
Dual energy CT (DECT)
enhances tissue characterization which is of great
interest in applications of medical imaging, security inspection and
nondestructive testing.
In principle, with DECT measurements acquired at
low and high energies
only two basis materials can be accurately reconstructed
\cite{niu2014iterative,alvarez1976energy,alvarez1977x,macovski1976energy,
marshall1981initial,stonestrom1981framework}.
In reality a scanned object often contains multiple basis materials and many
clinical and industrial applications desire multi-material
images \cite{laidevant2010compositional,liu2009quantitative}.
A natural thought is to
utilize spectral CT that acquires multi-energy measurements
to achieve multiple basis material images.
However, spectral CT requires either multiple scans
which results in high radiation to patients and needs
complex processing (\emph{e.g.}, registration)
of CT images at different energies \cite{liu2016ticmr},
or specialized scanners which are expensive and not available clinically yet,
such as energy-sensitive photon-counting detectors
\cite{bornefalk2010photon,gao2011multi,shikhaliev2011photon,ding2014high,li2015spectral}.
In this work, we focus on multi-material decomposition (MMD) using
DECT measurements obtained from commercial available conventional DECT
scanners.

Multi-material decomposition from DECT measurements is
an ill-posed problem since multiple sets of images
are estimated from two sets of measurements
associated with low and high energies.
Several methods have been proposed
to reconstruct multi-material images from DECT measurements
\cite{mendonca2014a,lamb2015stratification,long2014multi-material,xue2017statistical}.
Mendonca \emph{et al.} \cite{mendonca2014a} proposed an image-domain
MMD method that decomposes FBP images at low- and high-energy reconstructed
from a DECT scan into multiple images of basis materials.
This method uses a material triplet library
(\emph{e.g.}, blood-air-fat, fat-blood-contrast agent),
finds the optimal material triplet for each pixel,
and then decompose each pixel into
the basis materials that correspond to the best material triplet.
It uses mass and volume conservation assumption, and a constraint
that each pixel contains at most three materials out of several possible
materials to help solve the ill-posed problem of estimating multiple images
from DECT measurements.
The decomposed multiple material images by this method
have been successfully applied to applications of virtual
non-contrast-enhanced (VNC) images, fatty liver disease, and liver fibrosis
\cite{mendonca2014a,lamb2015stratification}.
However, this method estimates volume fractions of basis materials from
linear attenuation coefficient (LAC) pairs at high and low energies pixel by
pixel without considering the noise statistics of DECT measurements and
prior information of material images,
such as piecewise constant property of material images
and similarity between different material images.
%
Using similar constraints that help
estimating multiple material images from DECT scans,
Long and Fessler \cite{long2014multi-material}
proposed a penalized-likelihood (PL) method with
edge-preserving  for each material
to directly reconstruct multiple basis material images
from DECT measurements.
This PL method significantly reduced
noise, streak and cross-talk artifacts in
the reconstructed basis material images.
However, this PL method is computationally
expensive mainly due to the forward and back-projection
between multiple material images and DECT sinograms at low and high energies.
Xue \emph{et al.} \cite{xue2017statistical}
proposed a statistical image-domain MMD method that
uses penalized weighted least-square (PWLS) estimation
with edge-preserving
(EP) regularizers for each material.
We call this method the PWLS-EP-LOOP method hereafter.
Compared to the image-domain direct inversion method in \cite{mendonca2014a},
the PWLS-EP-LOOP method reduces noise and
improves the accuracy of decomposed volume fractions.
Because it is an
image-domain method without forward and back-projection,
it is computationally
more practical than the PL method.
To enforce sum-to-one and a box constraint $([0 ,1])$ derived from both
volume and mass conservation assumption
\cite{mendonca2014a, long2014multi-material},
the aforementioned three methods loop over material triples
in a material triplet library
formed from several basis materials of interest,
and uses a criterion to determine the optimal
material triplet for each pixel.
Without considering the prior information
that different material images have common edges,
the edge-preserving regularization of
the PL and PWLS-EP-LOOP method is imposed
on each material image.

In this paper, we propose a PWLS-TNV-$\ell_0$ method whose
cost function consists of a weighted least square data term and three
regularization terms.
The first term is total nuclear norm (TNV) regularization derived from
image property that basis material images share common or complementary
boundaries.
The second term is a $\ell_0$ norm that encourages each pixel containing a
small subset of material types out of several possible materials and each
material image is piecewise constant.
The third term is a characteristic function based on sum-to-one and a box
constraint accounting for the volume and mass conservation assumption.
We apply  the  Alternating Direction Method of Multipliers method
(ADMM, also known as split Bregman method \cite{goldstein2009the})
to solve the optimization problem of the PWLS-TNV-$\ell_0$ method.
We solve the subproblems of ADMM for the PWLS-TNV-$\ell_0$ method
using Conjugate Gradient(CG),
Singular Value Thresholding (SVT) \cite{cai2010singular},
Hard Thresholding (HT) \cite{blumensath2008iterative,Trzasko2007Sparse} and
projection onto convex sets.
We evaluate the  proposed PWLS-TNV-$\ell_0$ method  on
simulated digital phantom, Catphan\copyright600 phantom
and patient data, and results demonstrate that the proposed method
suppresses noise, decreases crosstalk and improves accuracy in decomposed material images,
compared to the PWLS-EP-LOOP method.

This paper is organized as follows. Section \ref{method} describes the
PWLS-TNV-$\ell_0$ method and the ADMM algorithm that minimizes its cost
function. Section \ref{result} presents experiments and results.
Section \ref{discussion} discusses the propose models and future work.
Finally, we draw our conclusions in Section \ref{conclusion}.

%
%

%

\section{Method}
\label{method}
\subsection{DECT model}
 For dual energy CT,  we can obtain a two-channel image
 $\bm{y}=(\bm{y}^T_H,\bm{y}^T_L)^T\in \mathbb{R}^{2 N_p} $,   where
 $\bm{y}_H,\bm{y}_L \in \mathbb{R}^{N_p}$ are
 attenuation images at high- and low-energy
 respectively and $N_p$ is the number of pixels.
 With mass and volume conservation assumption  \cite{mendonca2014a},
the spatially- and energy-dependent attenuation image
$\bm{y}$ satisfy
\begin{eqnarray}
 \left(
 \begin{array}{ccc}
 \bm{y}_H \\
 \bm{y}_L
  \end{array}
  \right)
 =
 \left(
 \begin{array}{ccc}
 \sum_{l=1}^{L_0}\mu_{lH}\bm{x}_l \\
 \sum_{l=1}^{L_0}\mu_{lL}\bm{x}_l
  \end{array}
  \right),\label{MVC}
 \end{eqnarray}
where $\mu_{lH}$ and $\mu_{lL}$ denote the linear attenuation coefficient of
the $l$-th material at the high- and low-energy respectively,
$\bm{x}_l=(x_{l1},x_{l2},...,x_{ln},...,x_{lN_p})\in \mathbb{R}^{N_p}$
denotes the volume fraction of the $l$-th  material and
$L_0$ is the number of materials.
According to volume conservation,
the volume fraction
$
\bm{x} = (\bm{x}^T_1, \bm{x}^T_2,\ldots,\bm{x}^T_{L_0})^T
\in \mathbb{R}^{L_0  N_P }
$
satisfies sum-to-one and box constraints,
\begin{eqnarray}
 \left\{
 \begin{array}{ccc}
  \sum\limits_{l=1}^{L_0} x_{lj}=1, \quad \forall j \\
  0 \leq x_{lj} \leq 1, \quad \forall l, j.
  \end{array}
  \right.\label{MVCC}
 \end{eqnarray}
We rewrite (\ref{MVC}) in the matrix form as
\begin{eqnarray}
\bm{y}
  =\bm{A}\bm{x},
\label{eq:yax}
\end{eqnarray}
where $\bm{A}\in\mathbb{R}^{2N_p\times L_0N_p}$ is
\begin{eqnarray}
\bm{A}=\bm{A}_0 \otimes \bm{I}_{N_p}.
\end{eqnarray}
 Here, $\otimes$ denotes the Kronecker producter. $\bm{A}_0$ is the  material composition matrix
 \begin{eqnarray}
\bm{A}_0=
 \left(
 \begin{array}{ccc}
 \mu_{1H}&\mu_{2H} \cdots \mu_{L_0H}  \\
 \mu_{1L}&\mu_{2L} \cdots \mu_{L_0L}
  \end{array}
  \right),
 \end{eqnarray}
and  $\bm{I}_{N_p}$ is the $N_p\times N_p$ identity matrix. In this paper, we obtain $\mu_{lH}$, $\mu_{lL}$ by the same method in \cite{szczykutowicz2010dual,granton2008implementation,niu2014iterative}.
   Firstly, we manually select two uniform regions of interest (ROIs) in the CT images that contain the $l$-th basis material.
  Then, we compute the average CT values in the two ROIs as $\mu_{lH}$ and $\mu_{lL}$  of the decomposition matrix $\bm{A}_0$ .
\subsection{Variational model}
\label{method,noise}
In practice the acquired attenuation image $\bm{y}$
is corrupted with noise, \emph{i.e.},
\begin{equation}
\bm{y}=\bm{A}\bm{x}+\bm{\varepsilon},
\label{eq:fp}
\end{equation}
where $\bm{\varepsilon} \in \mathbb{R}^{2 N_p}$
is assumed to be additive white noise, \emph{i.e.},
\begin{eqnarray}
\bm{\varepsilon}\sim
 N(\bm{0}, \Sigma)
\end{eqnarray}
where $\bm{0}$ is the zero vector in $\mathbb{R}^{2N_p}$ and
$\Sigma$ is the covariance matrix of $\bm{\varepsilon}$.

We propose to use a penalized weighted least-square (PWLS) method to
estimate multi-material images $\bm{x}$ from DECT images $\bm{y}$.
The probability density function (pdf) of $\bm{y}$ is
\begin{align}
&p(\bm{y}| \bm{x})\nonumber\\
=&\frac{1}{(2\pi)^{N_p}|\Sigma|^{\frac{1}{2}}}\exp{(-\frac{(\bm{y}-\bm{A}\bm{x})^T
 \Sigma^{-1} (\bm{y}-\bm{A}\bm{x})}{2})}.
\end{align}
According to maximum-likelihood (ML) estimate,
the negative log-likelihood is,
\begin{align}
\bar{L}(\bm{x})
&=\frac{1}{2}\|\bm{y}-\bm{A}\bm{x}\|_{\Sigma^{-1}}^2.
\end{align}
We assume the noise in each pixel is uncorrelated and
every pixel in the high- or low-energy CT image
has the same noise variance
as in our pervious work
\cite{niu2014iterative,xue2017statistical}, \emph{i.e.},
\begin{equation}
\Sigma =
\mathrm{diag}(\sigma_H^2 \bm{I}_{N_p},\sigma_L^2 \bm{I}_{N_p}),
\end{equation}
where $\sigma_H^2$ and $\sigma_L^2$ are the noise variance for
the high-energy CT image $\bm{y}_H$ and low-energy image $\bm{y}_L$
respectively.
To estimate $\sigma_H^2$ and $\sigma_L^2$ we select a homogeneous region
with a single material in the high- and low-energy image and
calculate their numerical variances respectively.

The PWLS problem that estimates fraction images $\bm{x}$
from noisy DECT images $\bm{y}$ takes the following form
\begin{align}
\hat{\bm{x}}=\arg\min_{\bm{x}} \Psi(\bm{x}),
\quad \Psi({\bm{x}}) \ \triangleq \ \bar{L}(\bm{x})+R(\bm{x}).
\label{cost}
\end{align}
We propose to use the following regularization term $R(\bm{x})$
\begin{equation}
R(\bm{x})=\beta_1 R_1(\bm{x})+\beta_2 R_2(\bm{x})+ R_3(\bm{x}),
\end{equation}
where the parameters $\beta_1$ and $\beta_2$ control the noise and
resolution tradeoff,
$R_1(\bm{x})$ is a total nuclear norm (TNV),
$R_2(\bm{x})$ is a $\ell_0$ norm and
$R_3(\bm{x})$ is a characteristic function based on sum-to-one
and box constraints in (\ref{MVCC}).
The three regularization terms will be
explained in Section \ref{LRReg}, \ref{SPReg} and \ref{VMCReg}
respectively.

\subsubsection{Low rankness of image gradients}
\label{LRReg}
The first regularization term  $R_1(\bm{x})$ is designed to describe the correlation of material images.
In practice, each region of an object typically contains several materials,
and the material images share similar or complementary boundary structures. When a region contains more than one  material,
the fraction images of these materials share similar structure information. Structure information of an image can be captured by the image gradient. Thus, we can use the correlation of image gradient among different material images. This is realized by imposing low rankness of the generalized gradient matrix at each pixel location, for which we use \textit{total~nuclear~variation} (TNV) as a regularization. This regularization form was previously proposed  in \cite{rigie2014ageneralized,rigie2015joint} and the sum of the nuclear norm  of Jacobian matrix of multi-channel image were penalized to reconstruct  color images. Here, the same idea is employed to take into account of the structure correlation of fraction images of material.

 More specifically, the generalized gradient matrix $(\bm{Dx})_j\in \mathbb{R}^{L_0\times N_d}$
at the $j$th-pixel is defined as
 \begin{eqnarray}
 (\bm{Dx})_j=
   \left(
  \begin{array}{cccc}
   (\bm{J_1x_1})_{j}       &(\bm{J_2x_1})_j        & \cdots
   &(\bm{J_{N_d}x_1})_j\\
   (\bm{J_1x_2})_{j}       &(\bm{J_2 x_2})_{j}     & \cdots
   &(\bm{J_{N_d}x_2})_{j}\\
   \vdots               &\vdots                   & \ddots     &\vdots
   \\
   (\bm{J_1x_{L_0}})_j   &(\bm{J_2 x_{L_0}})_j    & \cdots
   &(\bm{J_{N_d}x_{L_0}})_j\\
  \end{array}
   \right),
 \end{eqnarray}
where $\bm{J_dx_l}$ denotes the finite difference in the $d$-th direction on
the fraction image of the $l$-th  material $\bm{x}_l$, and $N_d$ is the number of directions. The regularization term is written as
\begin{equation}
 R_1(\bm{x})=\sum\limits_{j=1}^{N_P} \|(\bm{Dx})_j\|_{\ast}\triangleq\|\bm{Dx}\|_{\ast}\triangleq R_{TNV}(\bm{x}),
\end{equation}
where $\|\cdot\|_{\ast}$ denotes the nuclear norm of the matrix. The matrix $\bm{Dx}$ can be also viewed as a 3D matrix
of size  $L_0\times N_d\times  N_p$ and the nuclear norm is computed at each pixel.
\subsubsection{Sparsity}
\label{SPReg}
The second regularization considers the number of materials at each pixel is small as locally human organs often consist of few kinds of materials and the fraction is piecewise constant. Let $\bm{x}_j \triangleq (x_{1j},x_{2j}, \cdots ,x_{L_0j})^T $ be  the material fraction image vector at the $j$-th pixel.  We  use $\ell_0 $ norm of the gradient of $\bm{x}$
as regularization,
\emph{i.e.},
\begin{equation}
R_2(\bm{x})=\sum_{j=1}^{N_P}\|(\nabla\bm{x})_{j}\|_0=\|\nabla\bm{x}\|_0.
\end{equation}
Here, $\nabla\bm{x}=(\nabla\bm{x}^T_l,\nabla\bm{x}^T_2,\cdots
\nabla\bm{x}^T_{L_0})^T$.
If   the discrete gradient is computed in two directions, then $\nabla\bm{x}\in
\mathbb{R}^{ L_0N_p\times2} $ and $(\nabla\bm{x})_{j} \in \mathbb{R}^{
L_0\times2}$.

\subsubsection{Volume and mass conservation}
\label{VMCReg}
In addition, one can assume that volume and mass of the material fraction is conserved, \emph{i.e.} $x_l$ satisfies sum-to-one and the box constraint
given in \eqref{MVCC}. The regularization term $R_3$ is used to account for these constraints,
\emph{i.e.},
\begin{align}
R_3(\bm{x})=\chi_{S}(\bm{x})=\begin{cases}0,\quad \bm{x} \in S \\ \infty, \quad \mathrm{else} ,\end{cases}
\end{align}
where
$S=\{\bm{x} :\sum\limits_{l=1}^{L_0} x_{lj}=1,0 \leq x_{lj} \leq 1, j=1,\cdots,
N_p\}$ and
$\chi_{S}(\cdot)$ is the characteristic function.

In summary, the so-called PWLS-TNV-$\ell_0$  variational model is written as
\begin{align}
\arg\min_{\bm{x}}\frac{1}{2}\|\bm{y}-\bm{A}\bm{x}\|_{\Sigma^{-1}}^2+\beta_1 \|\bm{Dx}\|_{\ast}
+\beta_2 \|\nabla\bm{x}\|_0+ \chi_{S}(\bm{x}). \label{model}
\end{align}

\subsection{Optimization Method}
\label{method,Optimization}
The proposed PWLS-TNV-$\ell_0$ model
\eqref{model}
is a complex problem to solve directly
due to its non-convexity, non-smoothness and multiple regularization terms.
We apply the Alternating Direction Method of Multipliers (ADMM) (also known as
split Bregman \cite{goldstein2009the}) algorithm to solve it.
By introducing auxiliary variables $\bm{u}\in \mathbb{R}^{L_{0}\times
N_{d}\times N_{p}},\bm{v}\in \mathbb{R}^{L_{0} N_{p}\times 2}$
and $\bm{w}\in \mathbb{R}^{L_{0} N_{p}}$, we acquire the
following equivalent constrained problem:
\begin{align}
\arg\min_{\bm{x},\bm{u},\bm{v},\bm{w}}&\frac{1}{2}\|\bm{y}-\bm{A}\bm{x}\|_{\Sigma^{-1}}^2+\beta_1\|\bm{u}\|_\ast+\beta_2\|\bm{v}\|_0+\chi_{S}(\bm{w})\nonumber\\
s.t.~~~& \bm{u}=\bm{Dx},~~~\bm{v}=\nabla\bm{x},~~~\bm{w}=\bm{x}.\label{model1}
\end{align}

To simplify, problem \eqref{model1} can be formulated as the following general form
 \begin{align}
\arg \min_{\bm{x},\bm{z}}\bar{L}(\bm{x})+R(\bm{z})~~s.t.~~ \bm{z}=\bm{Kx}\label{model2}
\end{align}
where
$\bm{z}\triangleq (\bm{u},\bm{v},
\bm{w})^T,\bm{K}\triangleq(\bm{D},\bm{\nabla},\bm{I})^T .$
Here, the variables are understood   in vector form and the transformation are considered as operators.
The ADMM scheme for solving
\eqref{model2}
alternates between optimizing $\bm{x}$ and $\bm{z}$
and updating the dual variable $\bm{p}$:
\begin{align}
\bm{x}^{n+1}&=\arg\min_{\bm{x}} \bar{L}(\bm{x})+\langle \bm{p}^{n},\bm{K}\bm{x}
-\bm{z}^n\rangle\nonumber\\
&+\frac{\bm{\gamma}}{2}\|\bm{K}\bm{x} -\bm{z}^n\|^2_{2},\label{sub1}\\
\bm{z}^{n+1}&=\arg\min_{\bm{z}}R(\bm{z})+\langle \bm{p}^n,\bm{K}\bm{x}^{n+1} -\bm{z}\rangle\nonumber\\
&+\frac{\bm{\gamma}}{2}\|\bm{K}\bm{x}^{n+1} -\bm{z}\|^2_{2},\label{sub2}\\
\bm{p}^{n+1}&=\bm{p}^n+\bm{\gamma}(\bm{K}\bm{x}^{n+1}-\bm{z}^{n+1}),\label{sub3}
\end{align}
where $\bm{p}=(\bm{p}_1,\bm{p}_2,\bm{p}_3)^T$ with
$\bm{p}_1\in \mathbb{R}^{L_{0}\times N_{d}\times N_{p}}$,
$\bm{p}_2\in \mathbb{R}^{L_0 N_{p}\times2}$ and
$\bm{p}_3\in \mathbb{R}^{L_0 N_{p}}$ have the same size as
$ \bm{D}\bm{x}$, $\bm{\nabla}\bm{x}$ and $\bm{x}$ respectively,
$ \langle \cdot, \cdot \rangle$ denotes inner product,
and
$\bm{\gamma}=(\gamma_1, \gamma_2, \gamma_3)>0$ is the penalty parameters vector in \eqref{sub2}.

In the following, we present solutions for the subproblems \eqref{sub1}, \eqref{sub2} and \eqref{sub3}.
Since \eqref{sub1} is quadratic and differentiable on $\bm{x}$, it is equal to solve a linear system  to obtain $\bm{x}^{n+1}$, \emph{i.e.},
 \begin{align}
G  \bm{x}&=\bm{A}^T\Sigma^{-1} \bm{y} +\bm{D}^T(\gamma_1\bm{u}^n-\bm{p}^n_1 )\nonumber\\
&+\nabla^T(\gamma_2\bm{v}^n-\bm{p}^n_2)+\gamma_3\bm{w}^n-\bm{p}^n_3,\label{solx}
\end{align}
where $G\bm{x}=\bm{A}^T\Sigma^{-1}\bm{A}\bm{x}+\gamma_1 \bm{D}^T\bm{D}\bm{x}+\gamma_2\nabla^T\nabla\bm{x}+\gamma_3\bm{x}$.
It is easy to see that this is a linear system  that can be solved by conjugate gradient method efficiently.

Due to the structure of $R(\bm{z})$ and $\bm{K}$,
the optimization problem \eqref{sub2} is  separable in terms of
$\bm{u},\bm{v}$ and $\bm{w}$.
The subproblems of  $\bm{u},\bm{v}$ and $\bm{w}$  are as follows:
 \begin{align}
 \bm{u}^{n+1}&=\arg\min_{\bm{u}}\beta_1\|\bm{u}\|_\ast+\frac{\gamma_1}{2}\|\bm{u}-\bm{D}\bm{x}^{n+1}-\frac{\bm{p}^n_1}{\gamma_1}\|_2^2,\label{a}\\
\bm{v}^{n+1}&= \arg\min_{\bm{v}}   \beta_2\|\bm{v}\|_0   +\frac{\gamma_2}{2}\|\bm{v}-\nabla\bm{x}^{n+1}-\frac{\bm{p}^n_2}{\gamma_2}\|_2^2 , \label{bb}\\
 \bm{w}^{n+1}&  = \arg\min_{\bm{w}} \chi_{S}(\bm{w})
 +\frac{\gamma_3}{2}\|\bm{w}-\bm{x}^{n+1}-\frac{\bm{p}^n_3}{\gamma_3}\|_2^2.
 \label{d}
\end{align}
The subproblem \eqref{a} can be solved by Singular Value Thresholding (SVT)
\cite{cai2010singular}, \eqref{bb} can be solved by Hard Thresholding (HT)
\cite{blumensath2008iterative,Trzasko2007Sparse} (HT)  and \eqref{d} can be
solved using projection on to a simplex
\cite{kyrillidis2012sparse,chen2011projection}.
Let $\mathcal{D}$, $\mathcal{H}$ and $\mathcal{P}$ denote the SVT operator,
HT operator and projection operator respectively, and then
we can obtain
 \begin{align}
&\bm{u}^{n+1}(:,:,j)=\mathcal{D}_{\frac{\beta_1}{\gamma_1}}([\bm{D}\bm{x}^{n+1}+\frac{\bm{p}^n_1}{\gamma_1}](:,:,j)),
\quad \forall j
\label{Subsola}\\
&\bm{v}^{n+1}=\mathcal{H}_{\frac{\beta_2}{\gamma_2}}(\nabla\bm{x}^{n+1}+\frac{\bm{p}^n_2}{\gamma_2}),\label{Subsolb}\\
&(\bm{w}^{n+1})_j=\mathcal{P}_{1^+}((\bm{x}^{n+1}+\frac{\bm{p}^n_3}{\gamma_3})_j
 ), \quad \forall j
.\label{Subsolc}
\end{align}
The details of the three operators are shown in Appendix.

Algorithm \ref{algADMM} summarizes the optimization algorithm of PWLS-TNV-$\ell_0$.
 \begin{algorithm}[H]
\caption{PWLS-TNV-$\ell_0$  }
\label{algADMM}
\begin{algorithmic}
\STATE \textbf{Input.}
$\bm{y}_H, \bm{y}_L, \beta_1, \beta_2,  A, \bm{\gamma}_1, \bm{\gamma}_2,
\bm{\gamma}_3$
\STATE \textbf{Initial}
$\bm{p}^{0}=(\bm{p}_1^{0},\bm{p}_2^{0},\bm{p}_3^{0},)^T$,
$\bm{u}^0=\bm{D}\bm{x}^{0},\bm{v}^0=\bm{\nabla}\bm{x}^{0},
\bm{w}^0=\bm{x}^{0}$, $\rm{Maxiter}, \rm{tol}$, $n=1$
\WHILE{$error>\rm{tol}, n<\rm{Maxiter}$}
\STATE Solve linear system \eqref{solx} by CG.
\STATE
update $\bm{u}^{n+1}$ using \eqref{Subsola}.
\STATE update $\bm{v}^{n+1}$ using \eqref{Subsolb}.
\STATE update $\bm{w}^{n+1}$ using \eqref{Subsolc}.
\STATE $\bm{p}_1^{n+1}=\bm{p}_1^n+\gamma_1(\bm{D}\bm{x}^{n+1}-\bm{u}^{n+1})$
\STATE $\bm{p}_2^{n+1}=\bm{p}_2^n+\gamma_2(\nabla\bm{x}^{n+1}-\bm{v}^{n+1})$
\STATE $\bm{p}_3^{n+1}=\bm{p}_3^n+\gamma_3(\bm{x}^{n+1}-\bm{w}^{n+1})$
\STATE $n=n+1$ and compute $error$
\ENDWHILE
\end{algorithmic}
\end{algorithm}

\section{Results}
\label{result}

We evaluated the proposed method, PWLS-TNV-$\ell_0$,
with simulated digital phantom, Catphan\copyright600 phantom and
patient's pelvis data, and compared its performance with
those of  direct inversion method \cite{mendoncca2010multi,mendonca2014flexible}
and the PWLS-EP-LOOP method \cite{xue2017statistical}.

\subsection{Evaluation Metrics}

To quantify the quality of decomposed material images,
we calculate the mean and standard deviation (STD) of
pixels within a uniform region of interest (ROI)
in material images, and the volume fraction(VF) accuracy
of all material images.
The mean $\bar{x}_l$ and  $\mathrm{STD}_l$ of the $l$-th material image are
defined as
  \begin{align}
\bar{x}_l \ \triangleq \ \frac{\sum_{j=1}^{M}x_{lj} }{M},
\end{align}
and
\begin{align}
\mathrm{STD}_l \  \triangleq \
\sqrt{\frac{1}{M}\sum_{j=1}^{M}(x_{lj}-\bar{x}_l)^2},
\end{align}
where
$x_{lj}$ is the fraction value of the $j$-th pixel in the ROI of the $l$-th
material image and $M$ is the total number of pixels in the selected ROI.
The VF accuracy of all materials in  ROIs is defined as
\begin{align}
\mathrm{VF} \ \triangleq \ (1-\frac{1}{L_0}\sum_{l=1}^{L_0}
\frac{|\bar{x}_l^{\mathrm{truth}}-\bar{x}_l|}{\bar{x}_l^{\mathrm{truth}}})\times
 100\%
,
\end{align}
where $\bar{x}_l^{\mathrm{truth}}$
is the mean of the $l$-th true fraction image in a ROI.

In the Catphan\copyright600 phantom study,
we also use the electron density to evaluate the decomposition accuracy.
We define the electron density $\bm{\rho}_e$ of an object as
\begin{align}
\bm{\rho}_e \ \triangleq \ \sum_{l=1}^{L_0}\rho_l \bm{x}_l,
\end{align}
where $\bm{x}_l$ is the $l$-th material image
and $\rho_l$ is the electron density of the $l$-th material.
In each rod, the average percentage error of
electron density is calculated as
\begin{align}
E(\%)=\frac{|\bar{\rho}_e-\rho_e^{\mathrm{truth}}|}{\rho_e^{\mathrm{truth}}}\times 100\%
,\end{align}
where $\bar{\rho}_e$ is the average electron density of decomposed material
images in a rod and $\rho_e^{\mathrm{truth}}$ is the true electron
density in a rod with a single material.
We calculate the Root Mean Square percentage Errors (RMSE(\%))
of electron density in all rods to
qualify the decomposition accuracy.
The RMSE is defined as
 \begin{align}
\mathrm{RMSE} \ \triangleq \
\sqrt{\frac{1}{N}\sum_{n=1}^{N}(\frac{|(\bar{\rho}_e)_n-(\rho_e^{\mathrm{truth}})_n|}{(\rho_e^{\mathrm{truth}})_n})^2
 } ,
 \end{align}
  where $N$ denotes the number of rods,
   $(\bar{\rho}_e)_n$ is the average electron density of
   the decomposed results in the $n$-th rod and
   $(\rho_e^{\mathrm{truth}})_n$
   is the true electron density in the $n$-th rod.
\subsection{Digital phantom study}
\begin{figure*}[htbp!]
\centering
\subfigure{\includegraphics[width=0.3\linewidth, height=0.3\linewidth ]{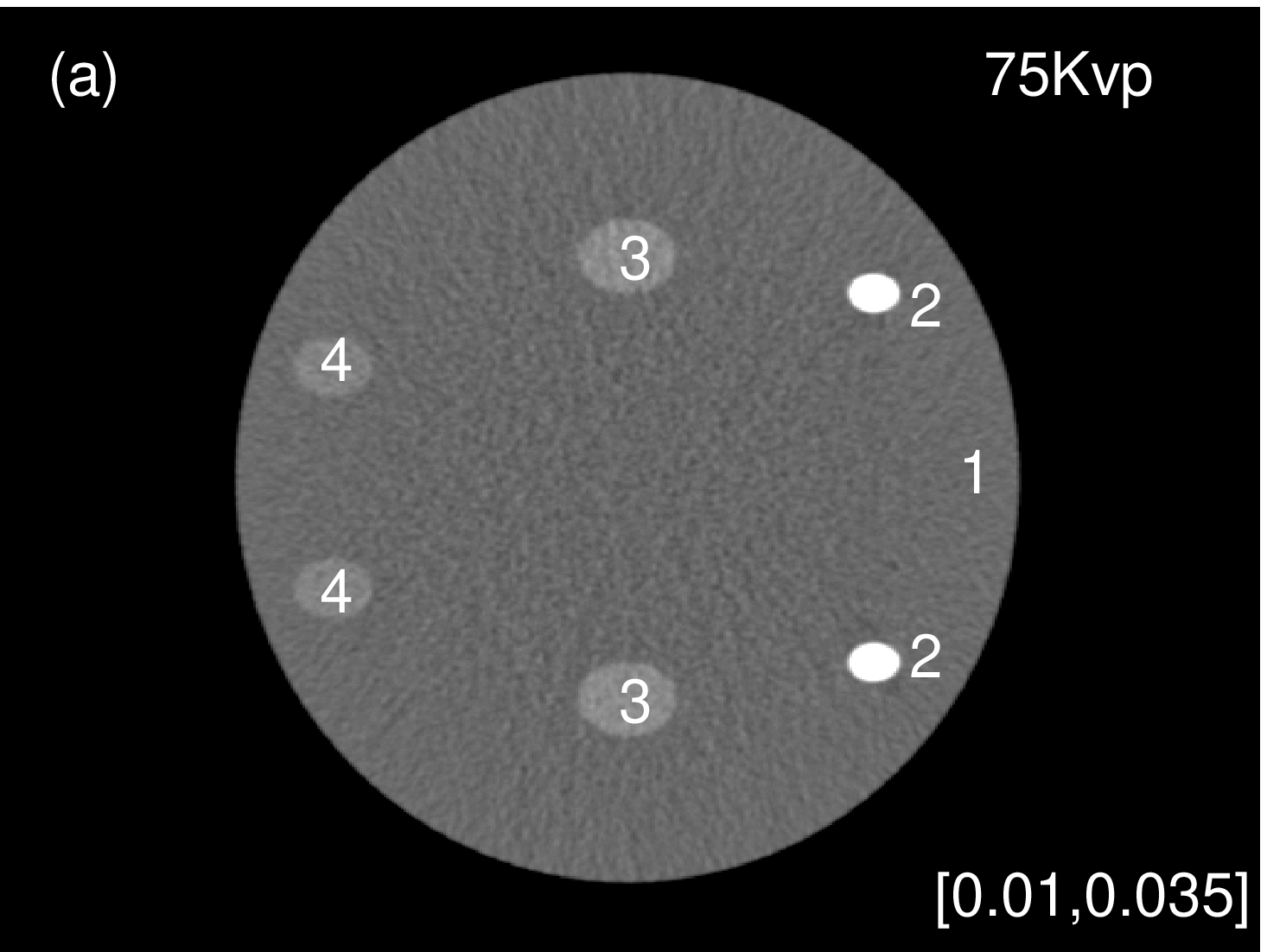}}
\subfigure{\includegraphics[width=0.3\linewidth, height=0.3\linewidth ]{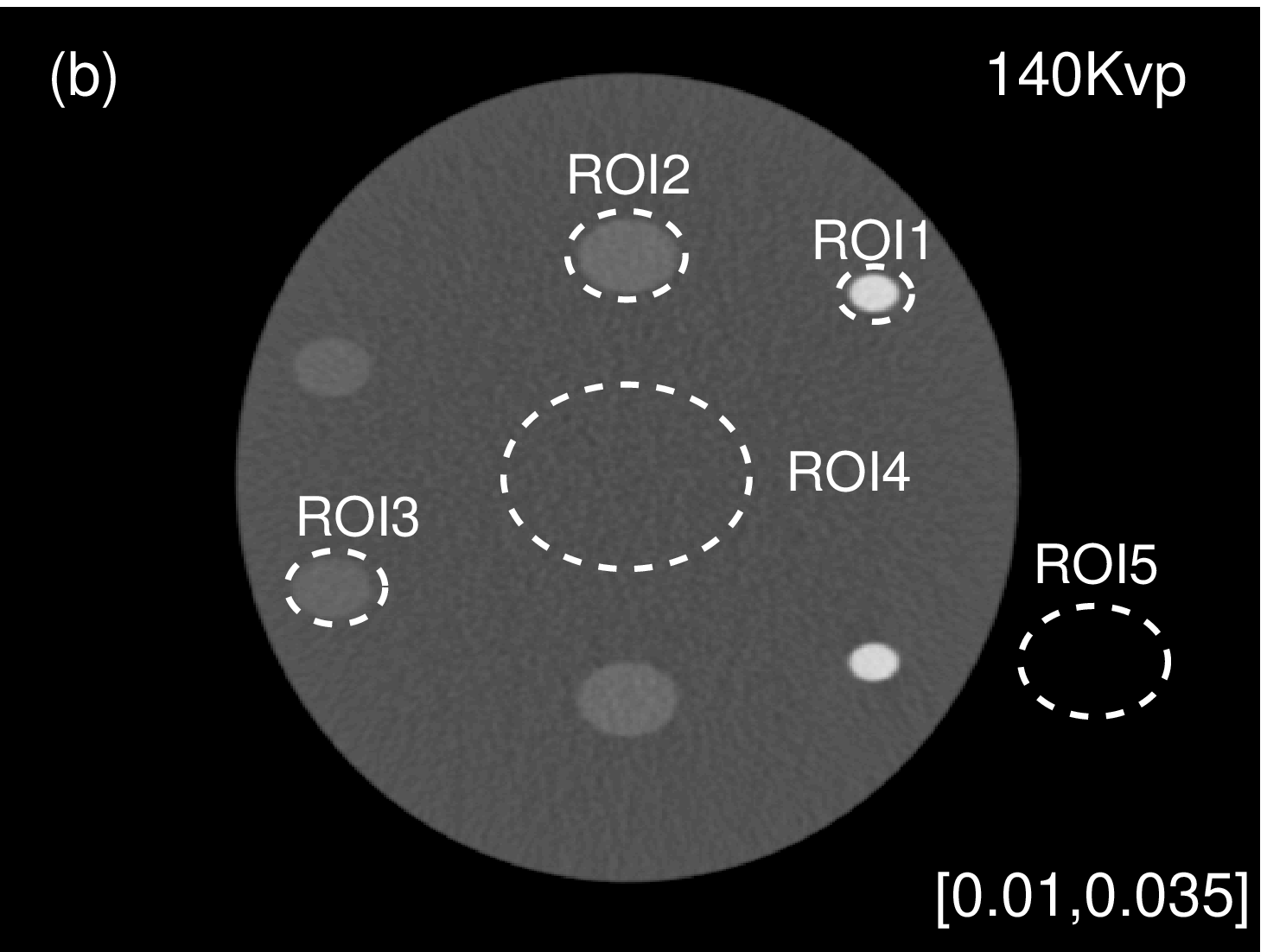}}
\caption{CT images of the digital phantom: (a) The low-energy: 75 kVp and (b) The high-energy: 140 kVp. Display window is $[0.01, 0.035] \mathrm{mm}^{-1}$. The components
of ROIs are bone (ROI1), muscle (ROI2), mixture (ROI3), fat (ROI4) and air (ROI5).}
\label{DigitalPhantom}
\end{figure*}

Fig. \ref{DigitalPhantom}(a) shows
the generated digital phantom that consists of four types of materials:
fat, bone, muscle and air.
Fat was selected as the background which is labeled as $\# 1$.
Bone was labeled as
$\# 2$ and muscle was labeled as $\# 3$.
Area $\# 4$ contains both fat and muscle with
a proportion of fat to muscle being $3:7$.
Mixed materials within one area would better
evaluate the decomposition accuracy of the MMD methods.

We obtained linear attenuation coefficients (LAC)
of the four basis materials from the National Institute of Standards and
Technology (NIST) database
\footnote{NIST,X-Ray Mass Attenuation Coefficients.(\url{https://www.nist.gov/pml/x-ray-mass-attenuation-coefficients})}.
We simulated a fan-beam CT geometry with
source to detector distance of $1500$~mm,
source to rotation center distance of $1000$~mm,
a detector size of $1024\times 768$ with
$0.388 \times 0.388~ \mathrm{mm}^2$ per detector pixel
and $676$ projection views over $[0^\circ, 360^\circ)$.
We generated DECT measurements
at $75$~kVp and $140$~kVp spectra with $12$~mm Al filter, respectively.
We simulated the high- and low-energy spectra of incident X-ray photons
using Siemens simulator
\footnote{Siemens.(\url{https://bps-healthcare.siemens.com/cv_oem/radIn.asp})}.
The projection data was corrupted with Poisson noise and the standard
filtered back projection (FBP) method
\cite{kak2001principles,natterer2001mathematics} was applied
to reconstruct high- and low-energy attenuation CT images of size $512 \times
512$, where the physical pixel size is $0.5\times0.5~\mathrm{mm}^2$.

\begin{figure*}[htbp!]
\begin{center}
\begin{tabular}{c@{\hspace{0pt}}c@{\hspace{0pt}}c@{\hspace{0pt}}c@{\hspace{0pt}}c@{\hspace{0pt}}c@{\hspace{0pt}}c@{\hspace{0pt}}c}
\includegraphics[width=.2\linewidth,height=.2\linewidth]{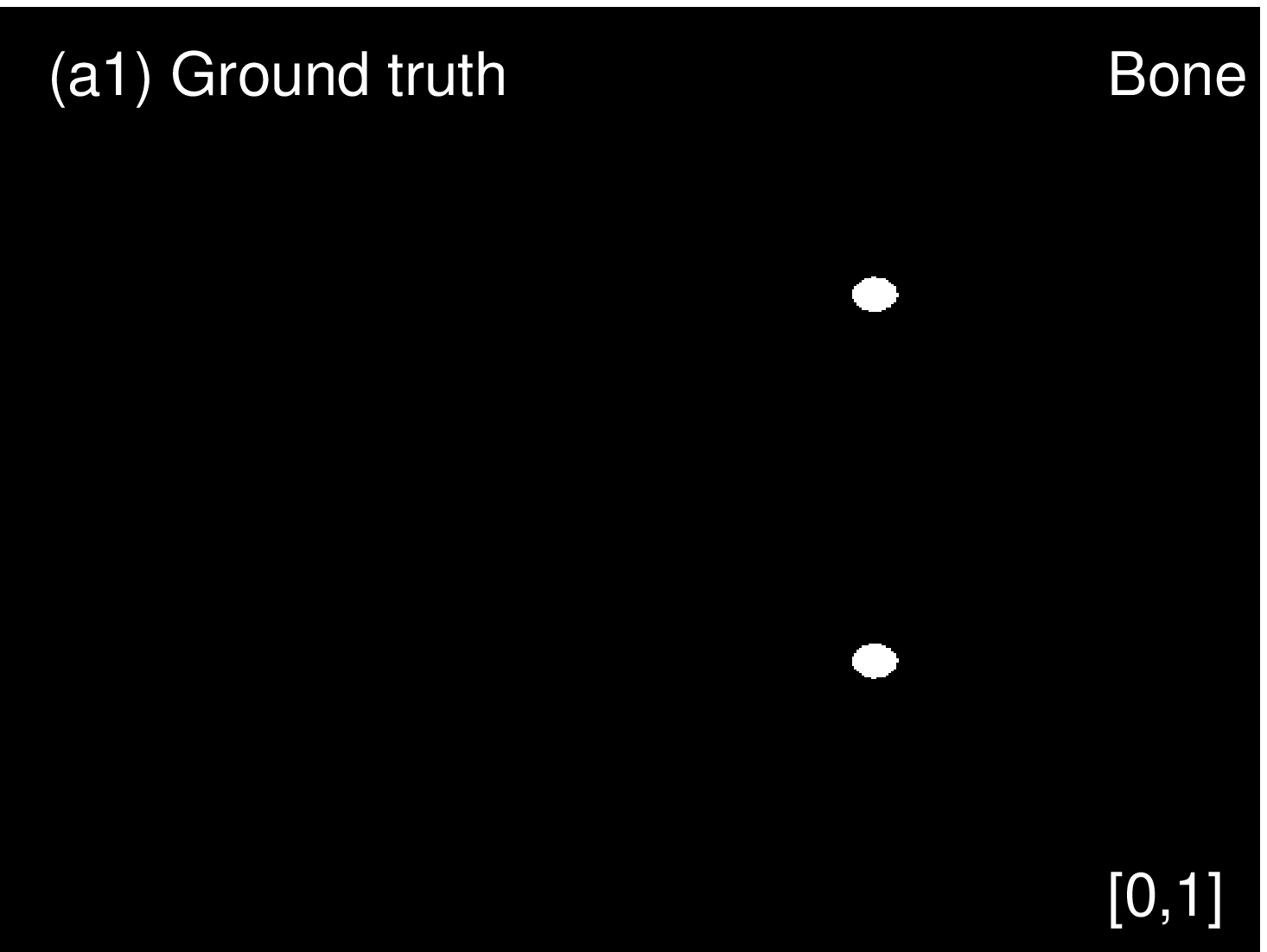}&
\includegraphics[width=.2\linewidth,height=.2\linewidth]{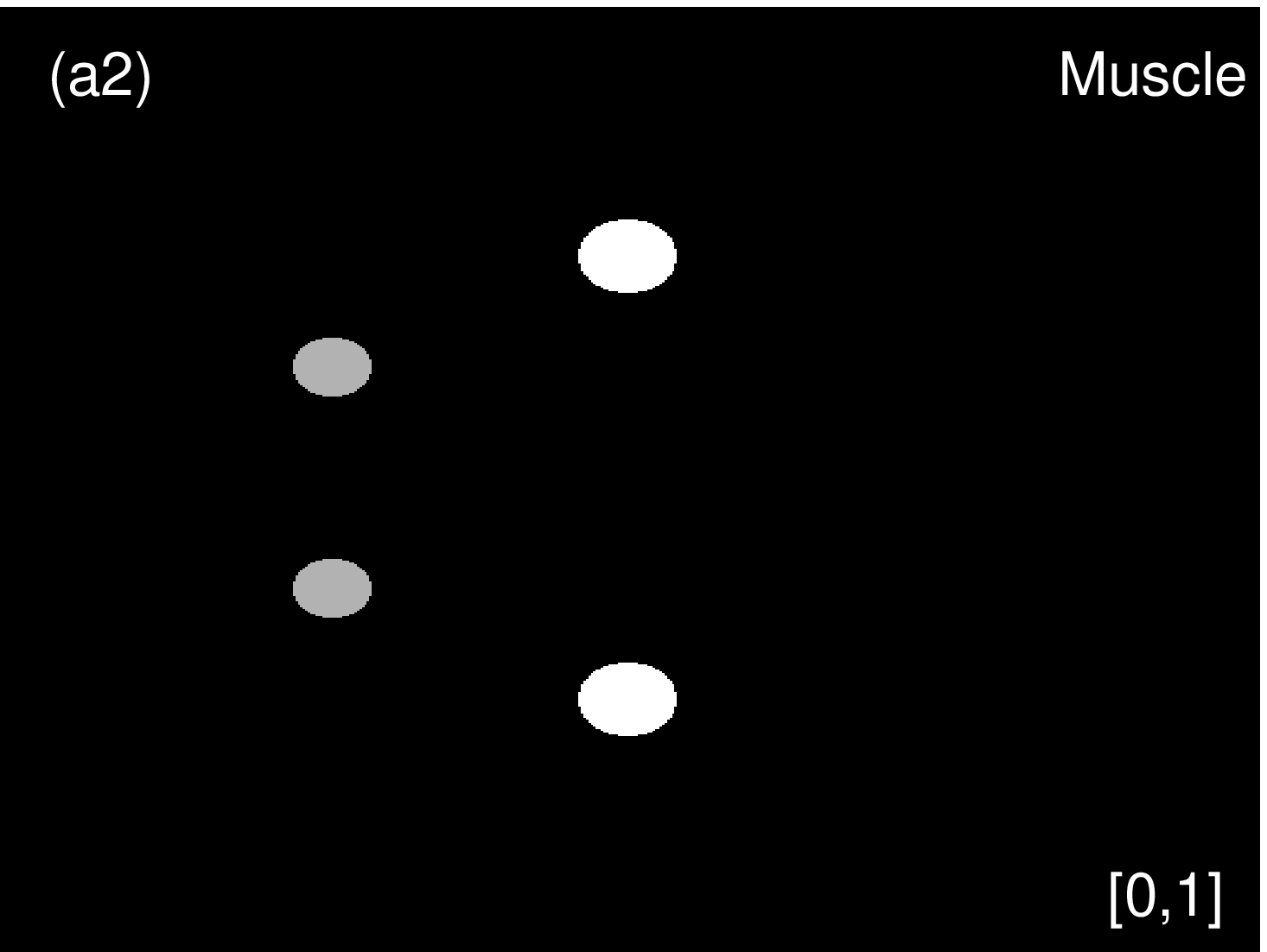}&
\includegraphics[width=.2\linewidth,height=.2\linewidth]{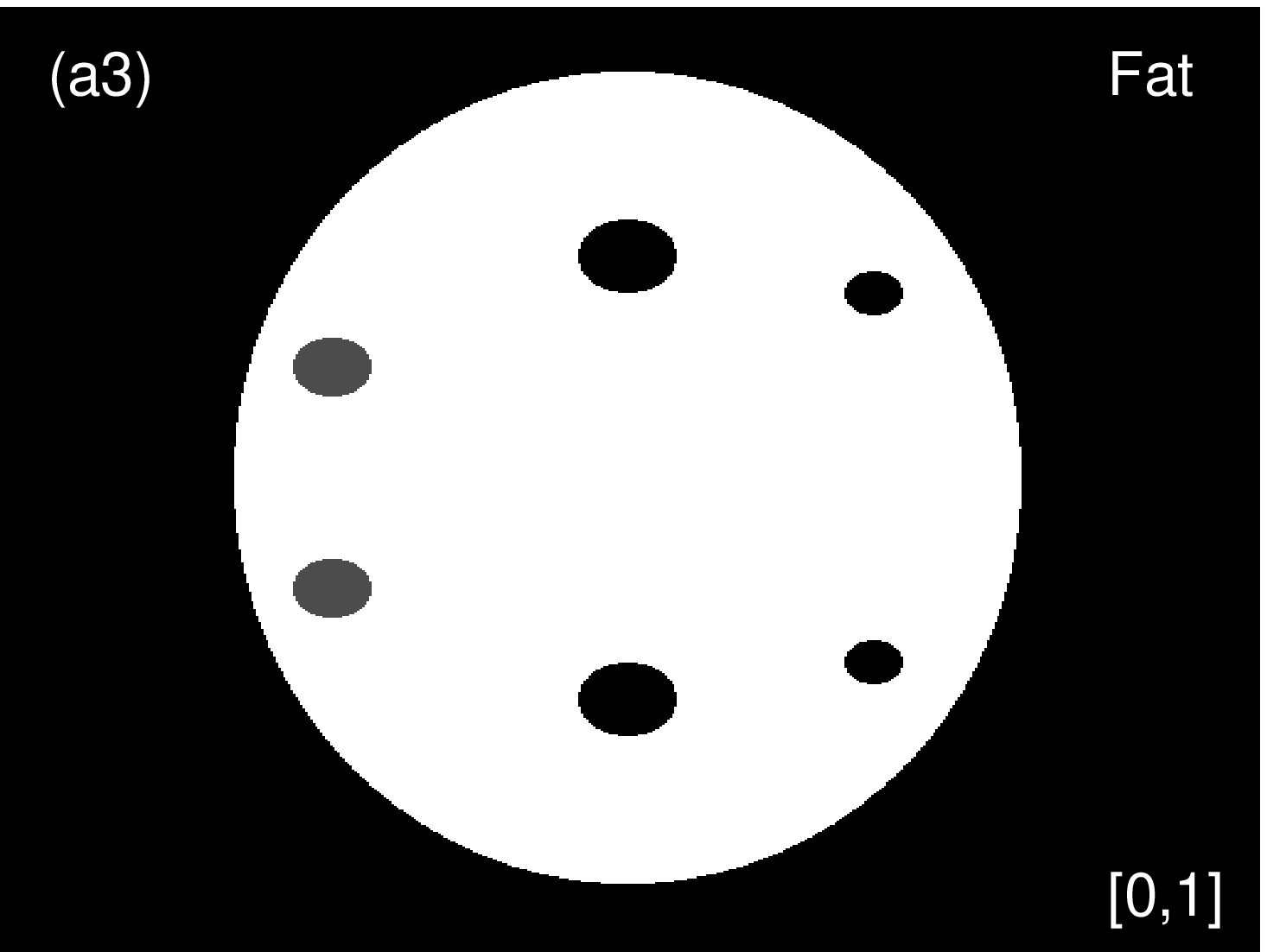}&
\includegraphics[width=.2\linewidth,height=.2\linewidth]{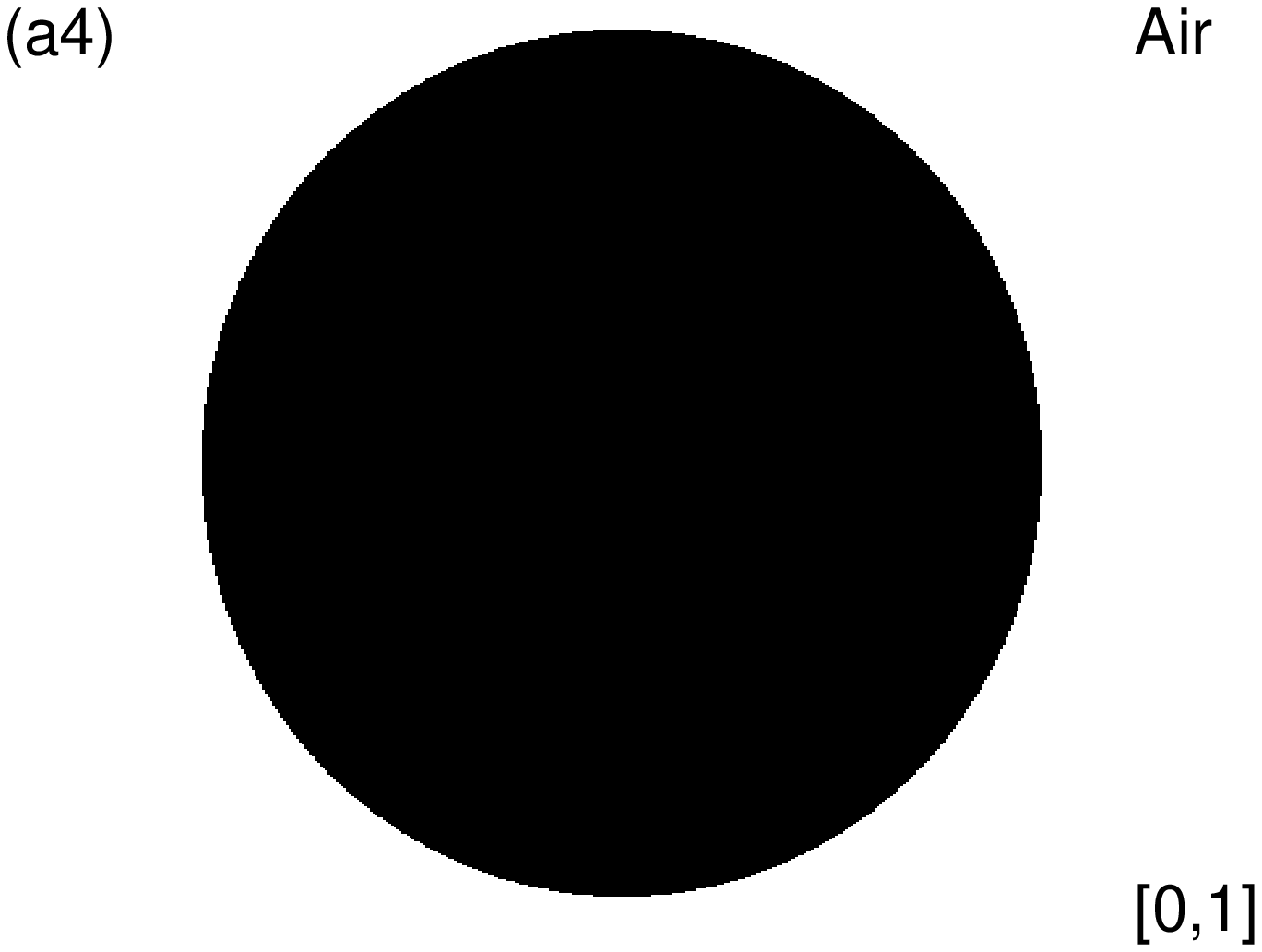}\\
\includegraphics[width=.2\linewidth,height=.2\linewidth]{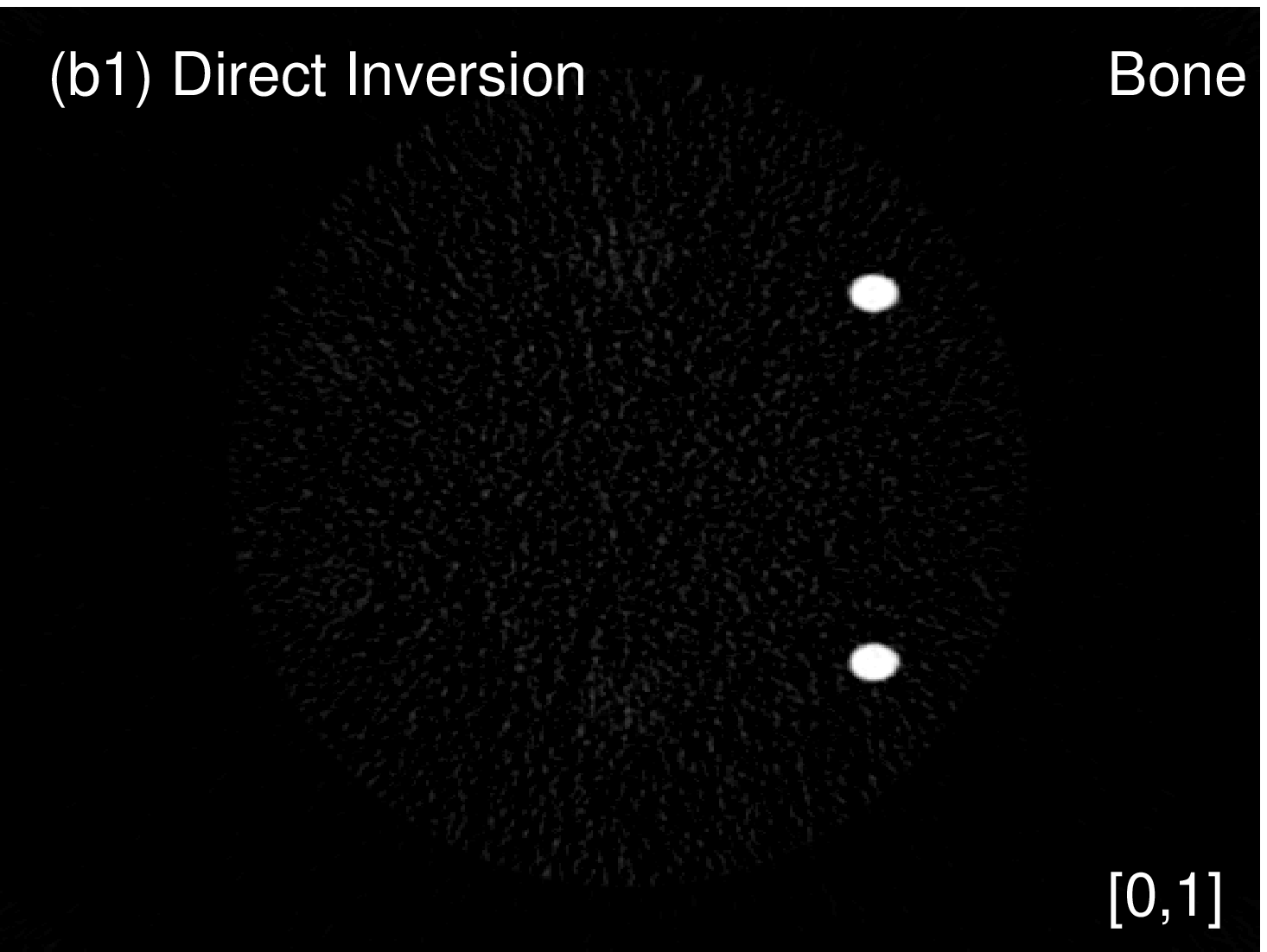}&
\includegraphics[width=.2\linewidth,height=.2\linewidth]{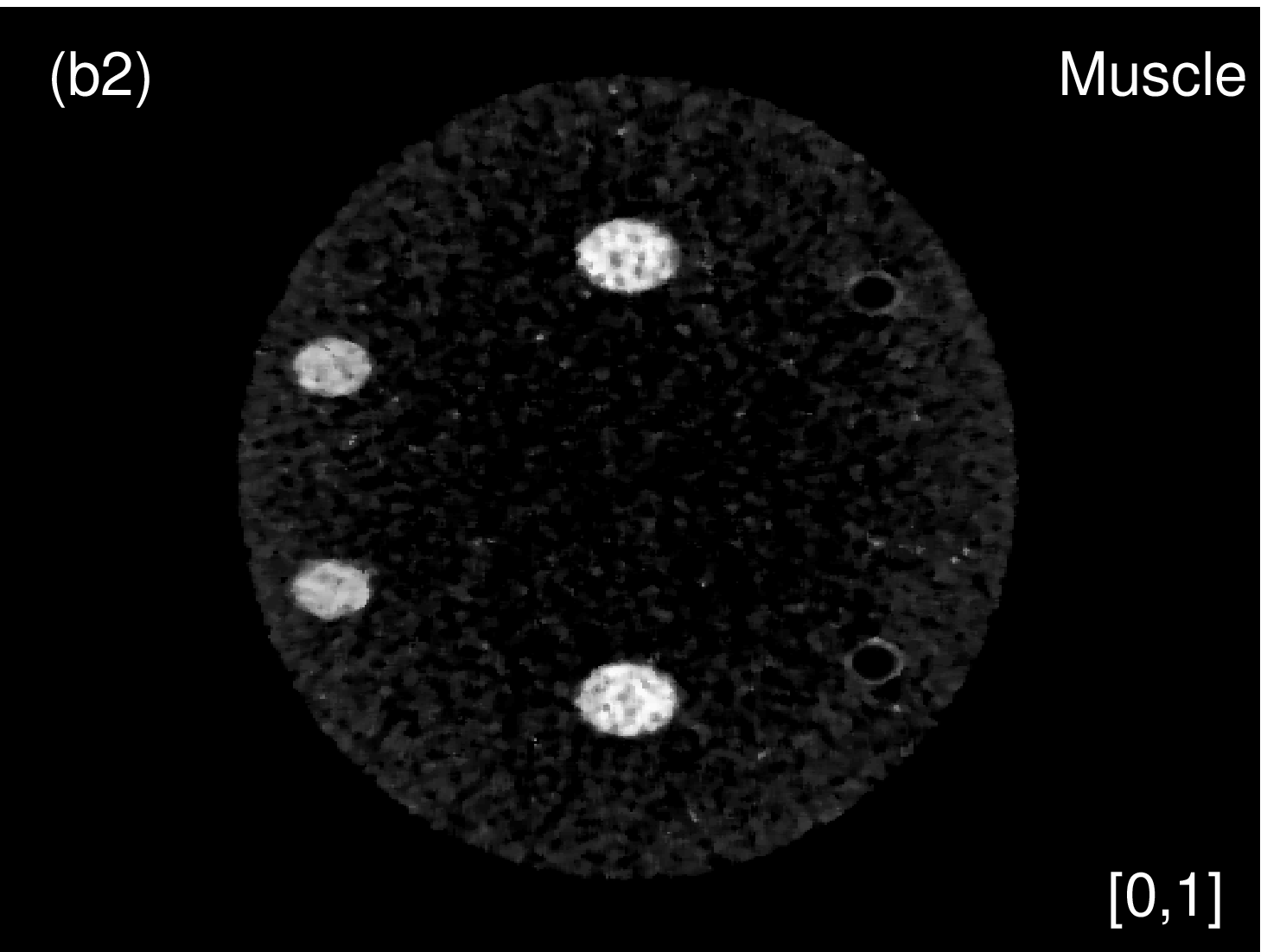}&
\includegraphics[width=.2\linewidth,height=.2\linewidth]{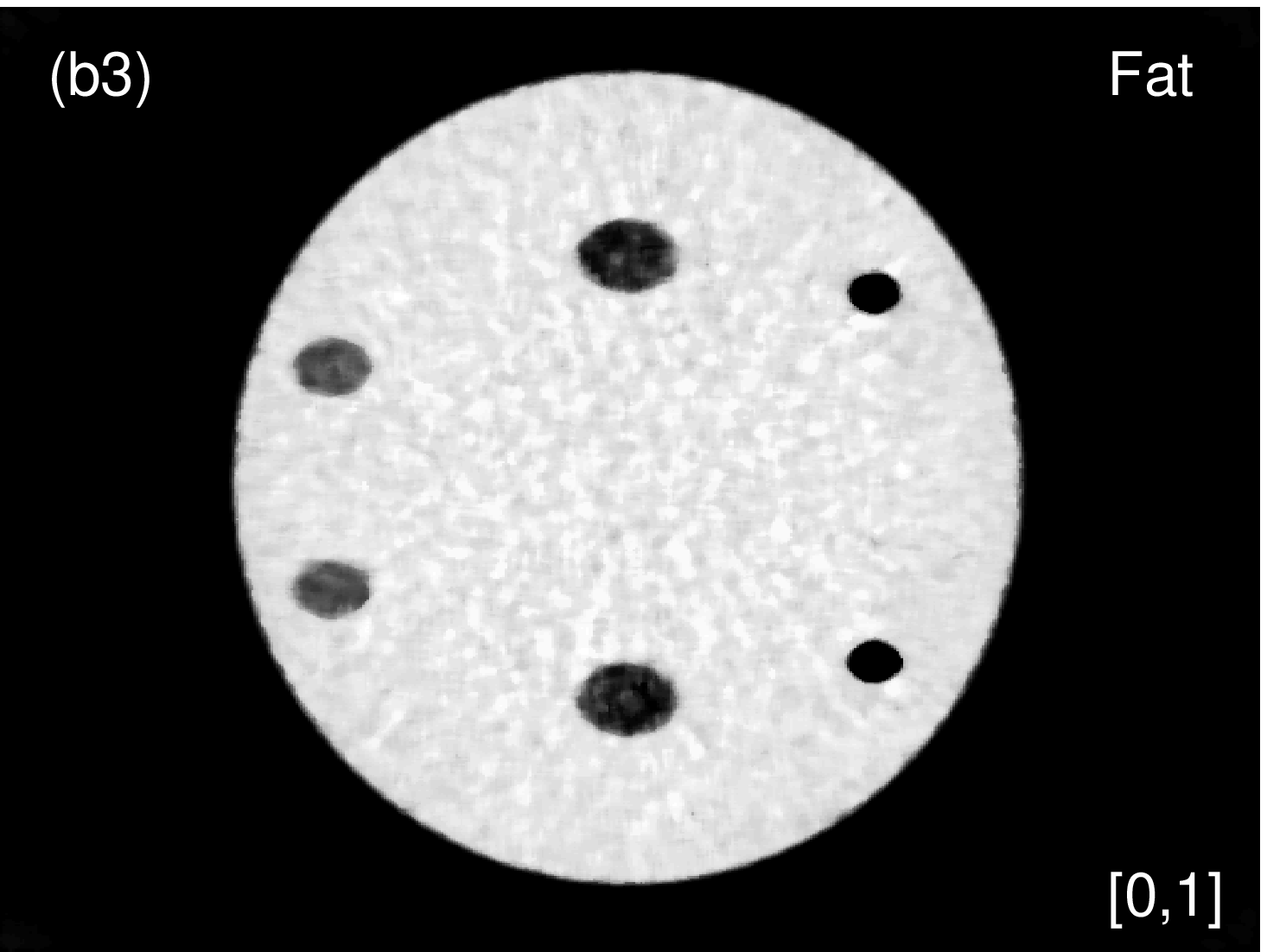}&
\includegraphics[width=.2\linewidth,height=.2\linewidth]{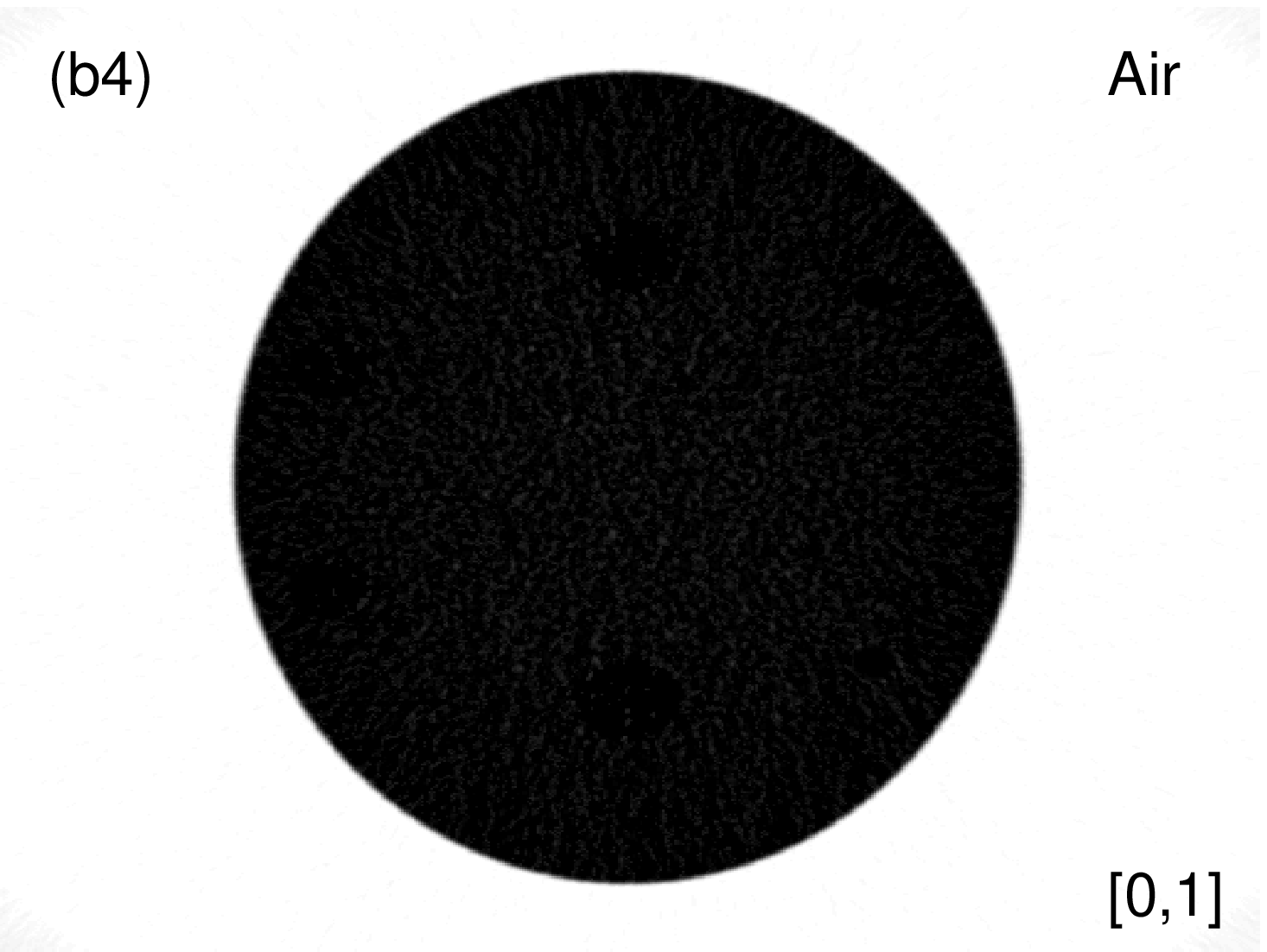}\\
\includegraphics[width=.2\linewidth,height=.2\linewidth]{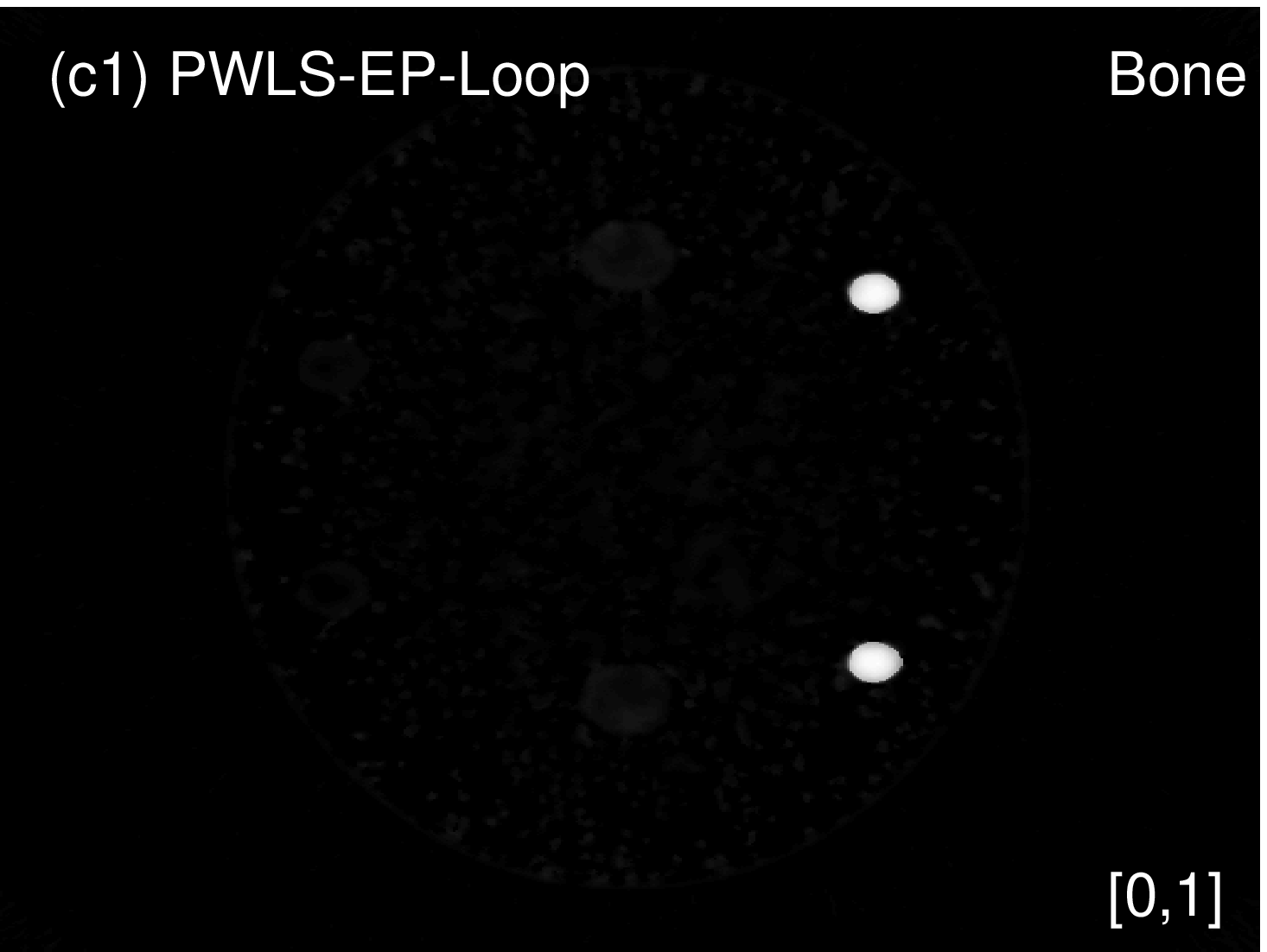}&
\includegraphics[width=.2\linewidth,height=.2\linewidth]{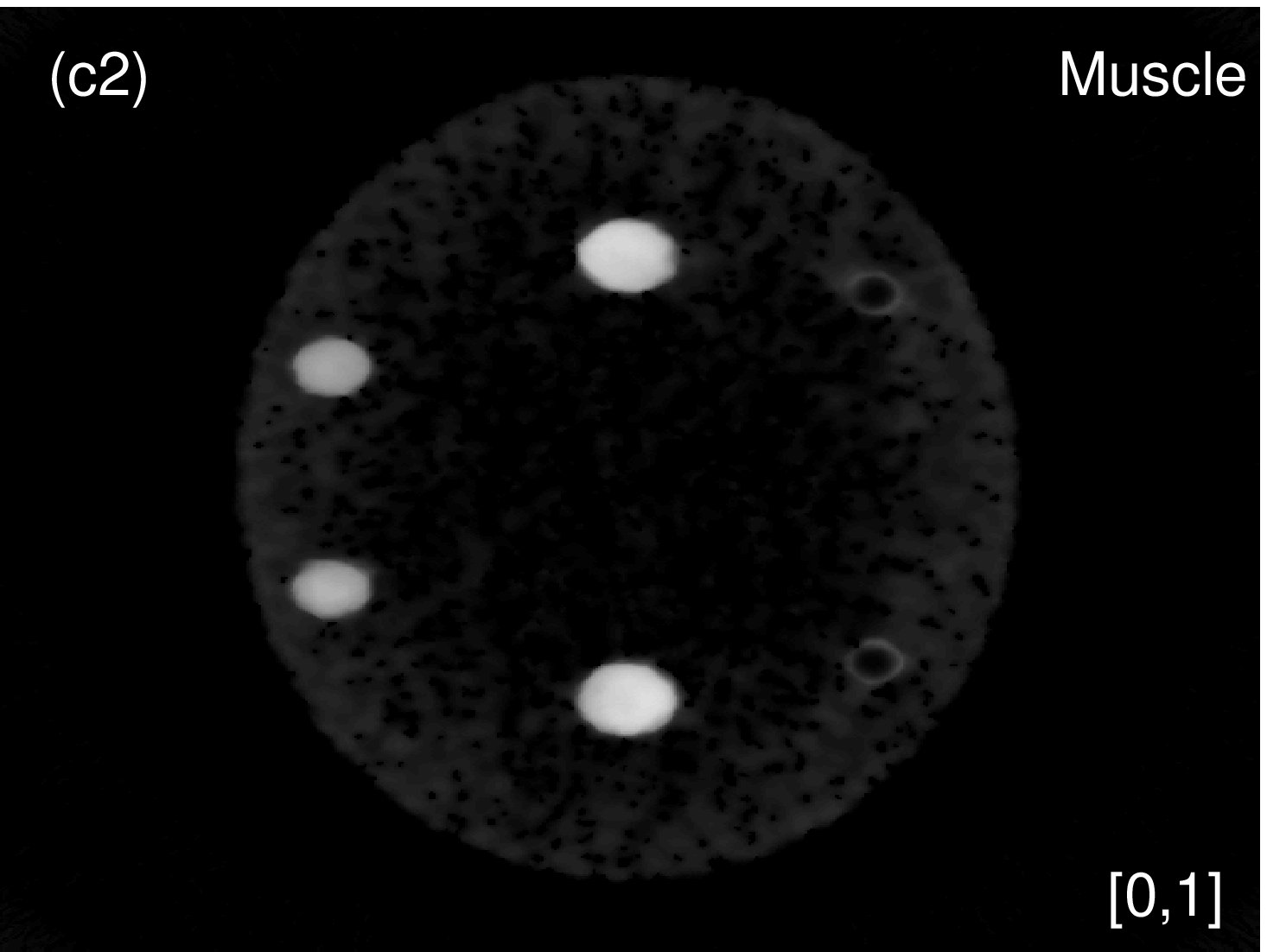}&
\includegraphics[width=.2\linewidth,height=.2\linewidth]{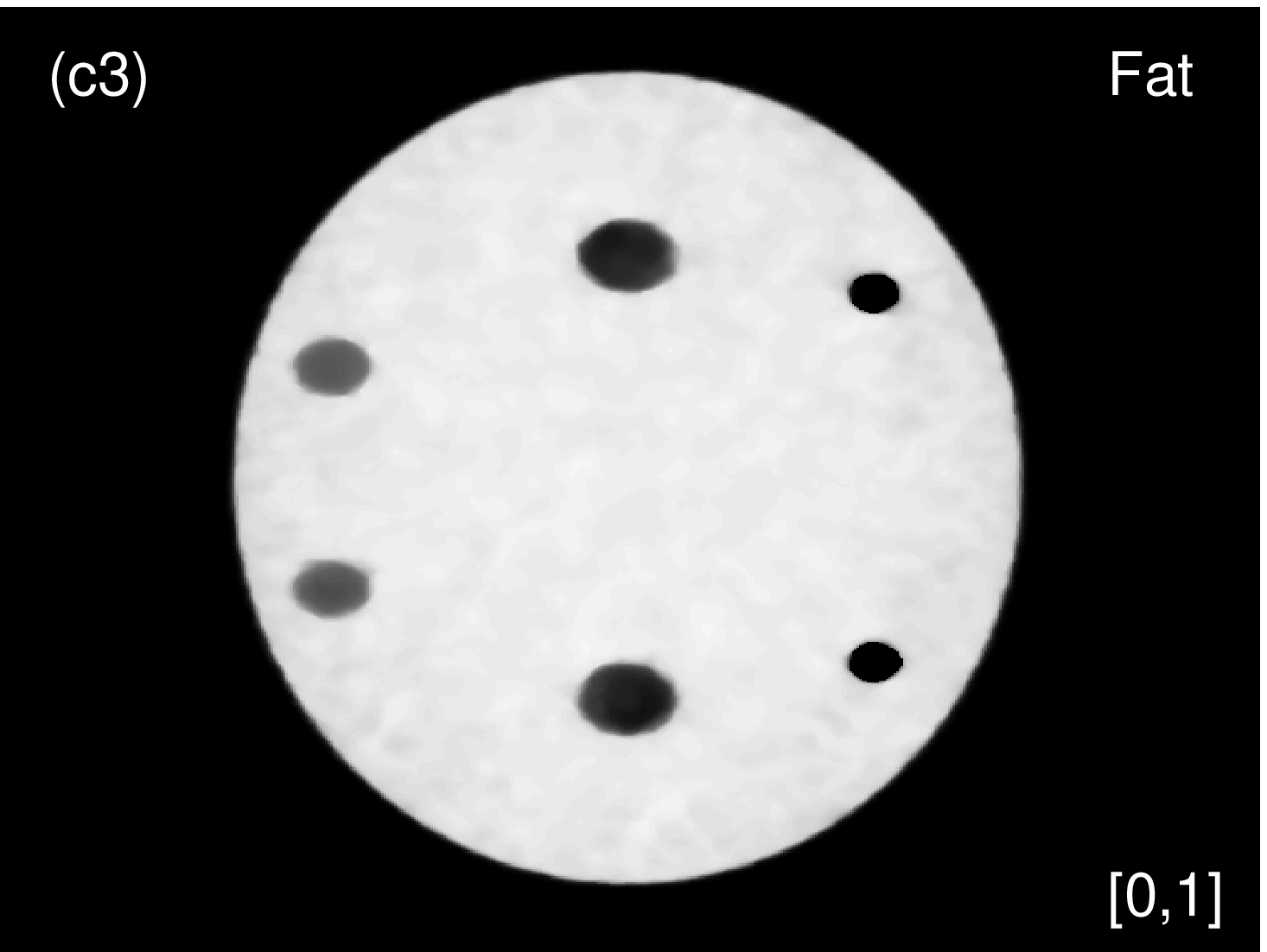}&
\includegraphics[width=.2\linewidth,height=.2\linewidth]{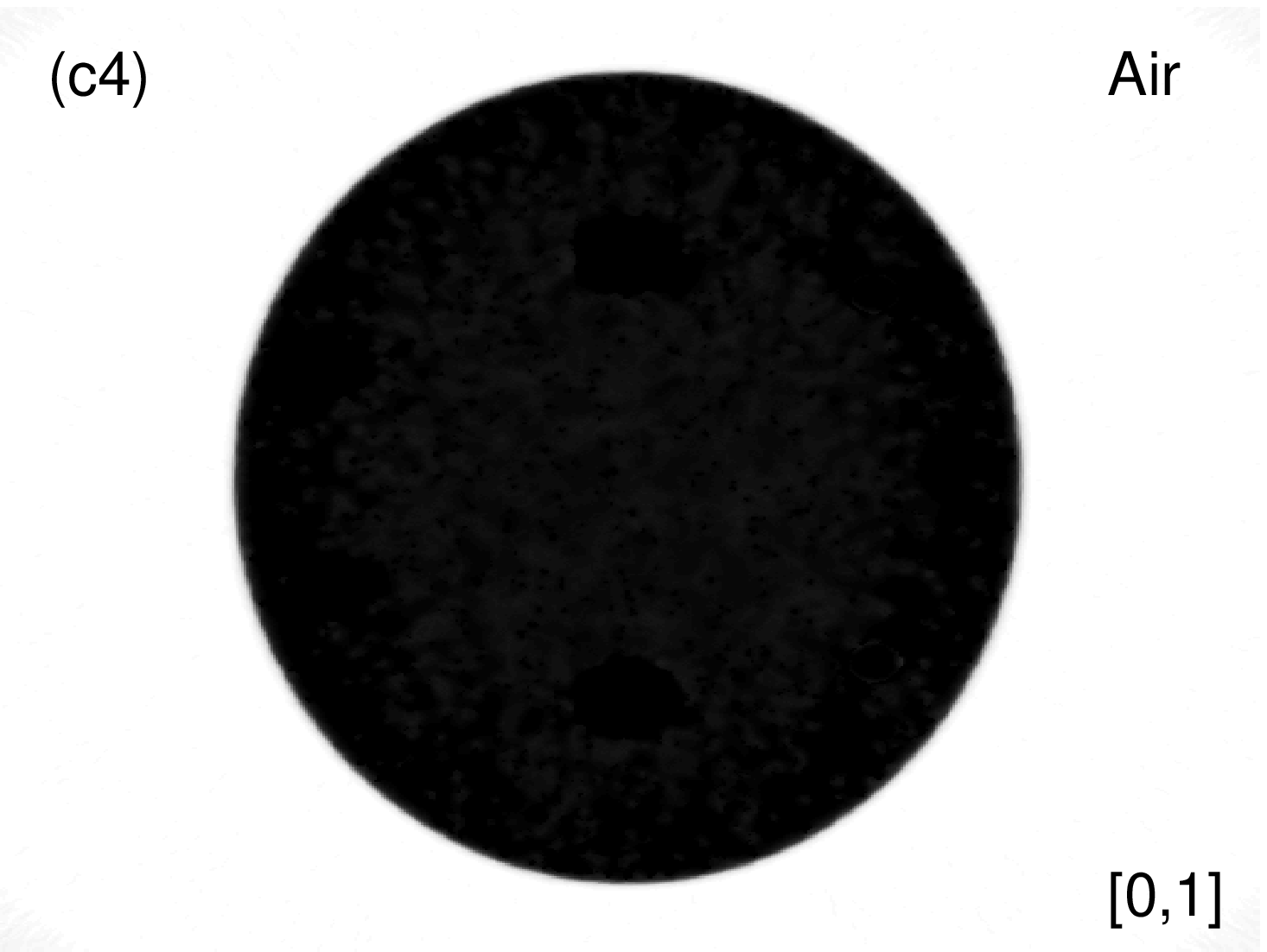}\\
\includegraphics[width=.2\linewidth,height=.2\linewidth]{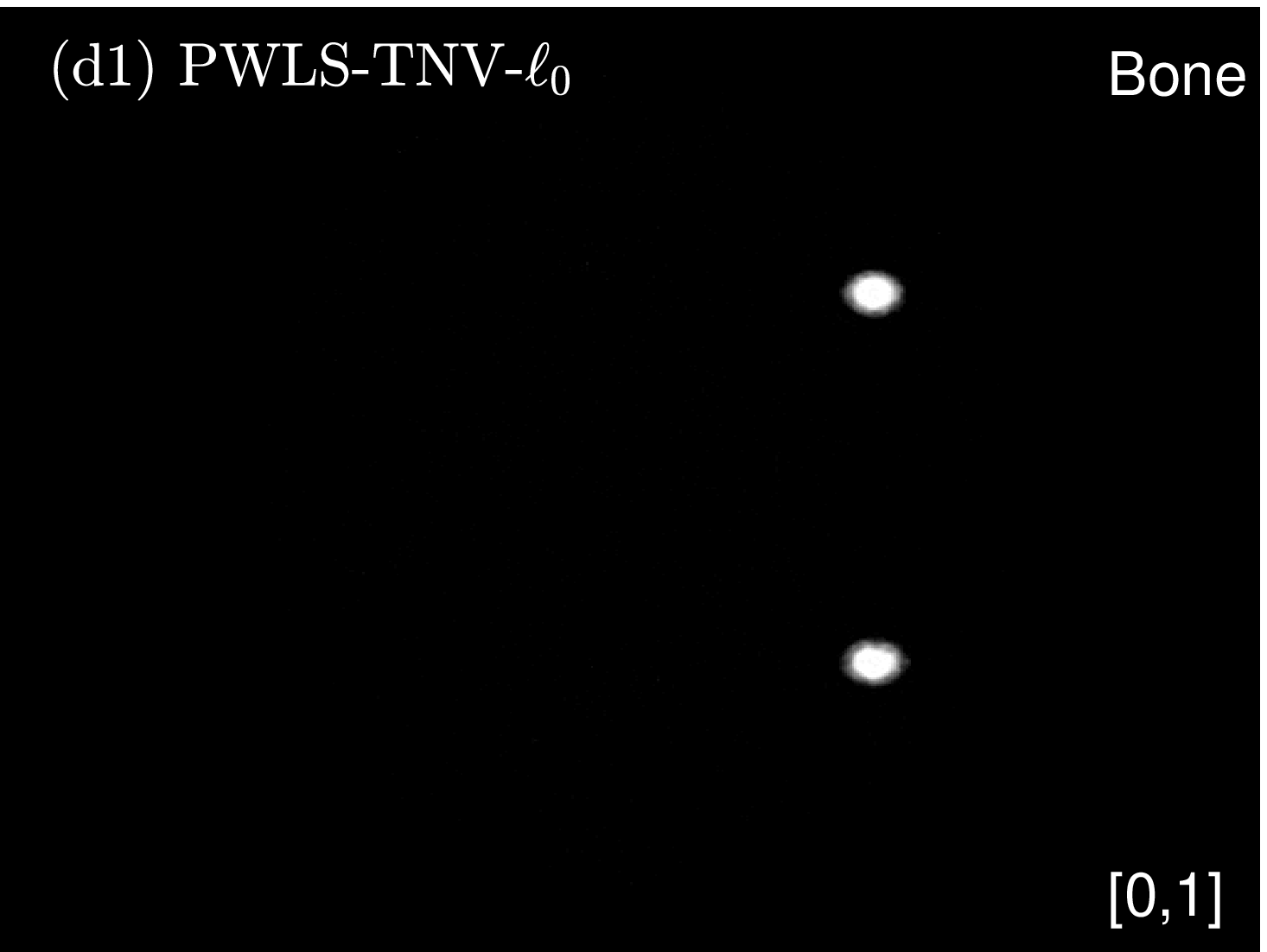}&
\includegraphics[width=.2\linewidth,height=.2\linewidth]{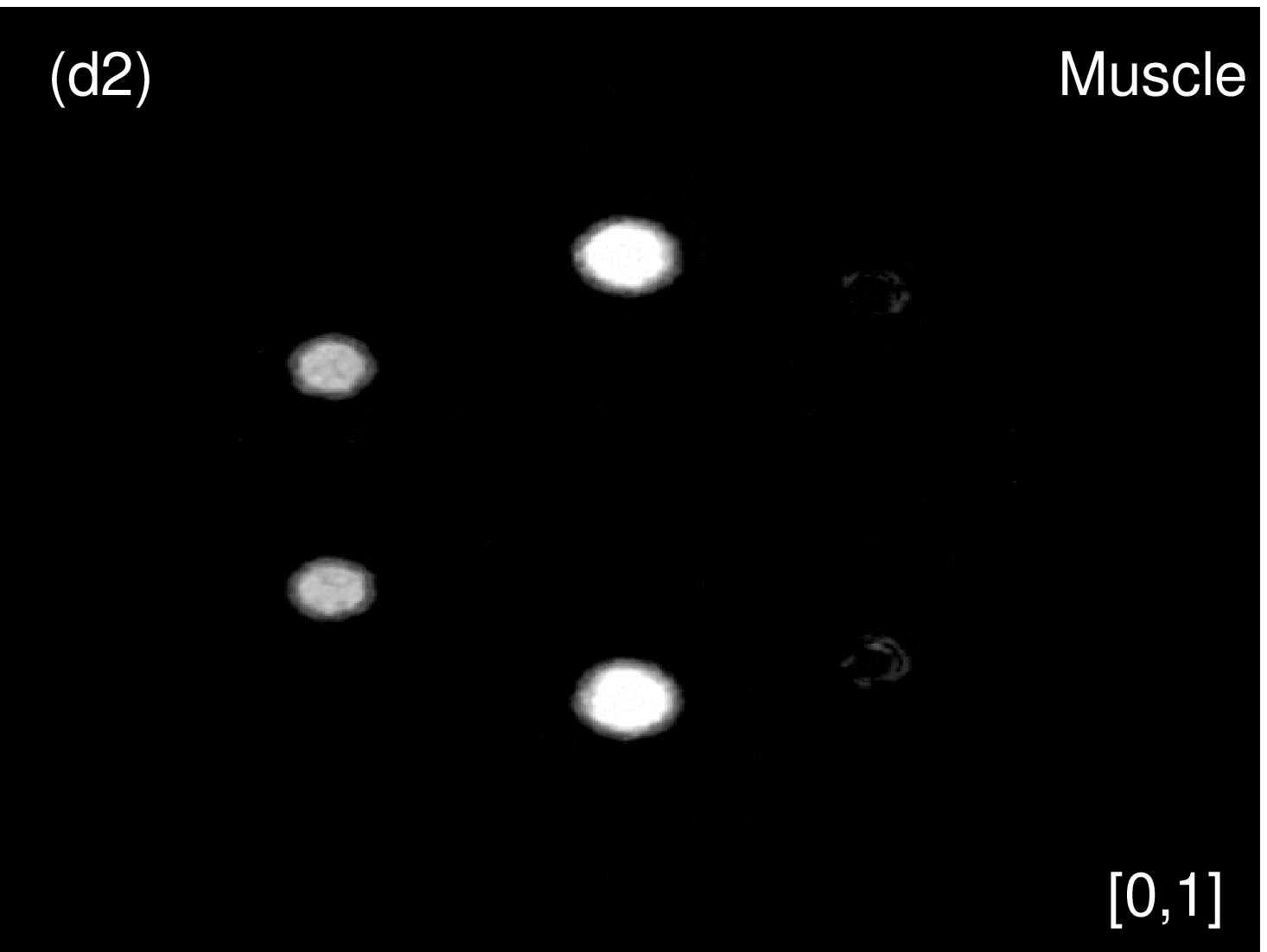}&
\includegraphics[width=.2\linewidth,height=.2\linewidth]{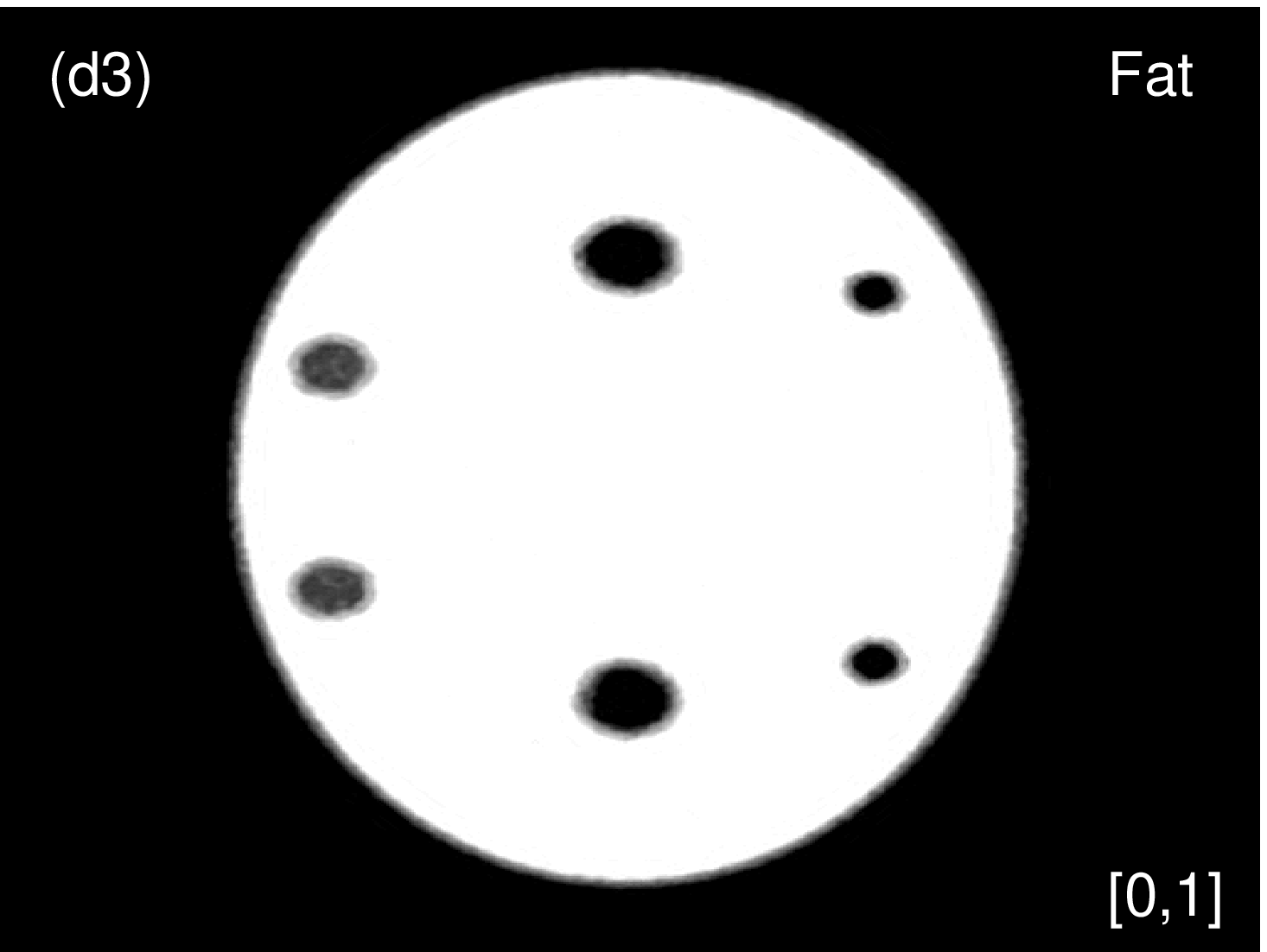}&
\includegraphics[width=.2\linewidth,height=.2\linewidth]{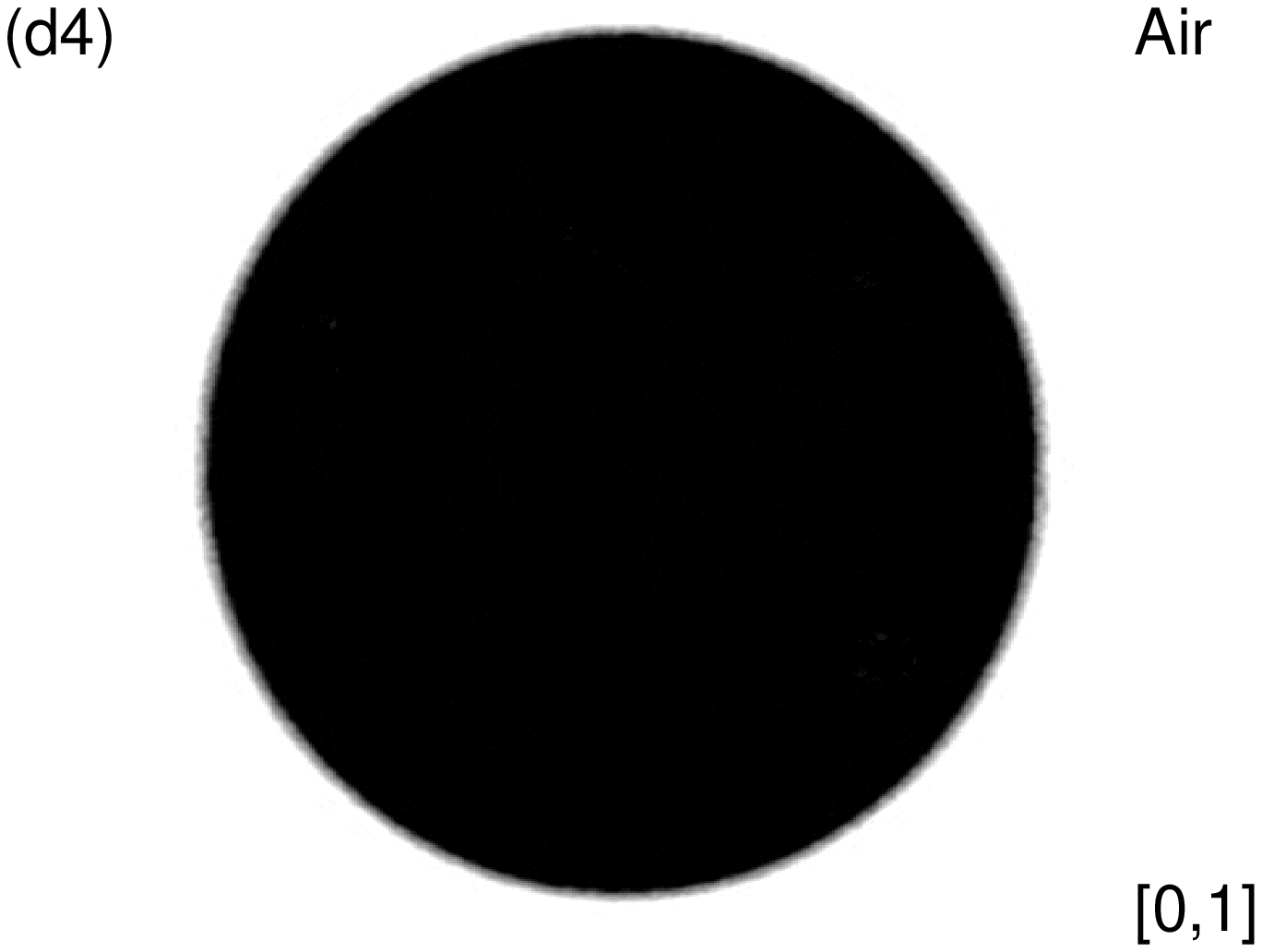}
\end{tabular}
\caption { Material images of ground truth (the $1^{st}$ row), Direct Inversion (the $2^{nd}$ row),
PWLS-DP-LOOP (the $3^{rd}$ row)  and PWLS-TNV-$\ell_0$ (the $4^{th}$ row).
The display windows are shown in the bottom-right corners.
}
\label{ZJUDigTNVMM}
\end{center}
\end{figure*}

We implemented the direct inversion MMD method in \cite{mendonca2014a} and
 used its results as the initialization for the PWLS-EP-LOOP method
 \cite{xue2017statistical} and
 the PWLS-TNV-$\ell_0$ method respectively.
Fig.~\ref{ZJUDigTNVMM}~(a) shows the true material images.
Fig.~\ref{ZJUDigTNVMM}~(b), (c) and (d) show the decomposed
basis material images by the Direct inversion, the
PWLS-EP-LOOP and the PWLS-TNV-$\ell_0$ method respectively.
The PWLS-TNV-$\ell_0$ method reduced noise and crosstalk
in the component images, especially for the muscle image, compared to the
PWLS-EP-LOOP method.
To quantitatively analyze performances of different methods,
we calculated evaluation metrics of decomposed basis material images
in several ROIs located within uniform areas
shown with dashed line circles in Fig.~\ref{DigitalPhantom}~(b).
Table~\ref{DigitalSTD} summarizes the means and noise STDs of the decomposed
basis material images.
For the Direct Inversion, the PWLS-EP-LOOP and the proposed PWLS-TNV-$\ell_0$
method, the volume fraction accuracies were $93.61\%$, $93.27\%$, and $99.31\%$
respectively.
Compared with Direct Inversion and PWLS-EP-LOOP, the proposed method improved
volume fraction accuracy by $5.7\%$ and $6.04\%$ respectively.

\begin{table*}[htbp!]
\caption{\leftline{The means and STDs of decomposed images within ROIs of the
digital phantom.}}
\begin{tabular}{p{2.6cm}p{2.3cm}p{2.3cm}p{2.3cm}p{2.3cm}p{2.3cm}p{2.3cm}}
\hline
\hline
\multirow{2}{*}{Methods}&
        ROI1&  ROI2&    \multicolumn{2}{c}{ROI3}&          ROI4&  ROI5  \cr\cline{4-5}
       &Bone&Muscle&   Muscle&     Fat&                     Fat&  Air\\
\hline
Ground Truth&       $1\pm0$          &  $1\pm0$          &     $0.7\pm0$        &    $0.3\pm0$        &  $1\pm0$          & $1\pm0$\\
Direct Inversion &  $0.9964\pm0.0073$&  $0.7834\pm0.1420$&     $0.6753\pm0.0749$&    $0.3101\pm0.0388$&  $0.9087\pm0.0384$& $0.9970\pm0.0041$\\
PWLS-EP-LOOP&       $0.9588\pm0.0181$&  $0.8107\pm0.0221$&     $0.6756\pm0.0186$&    $0.3187\pm0.0121$&  $0.9261\pm0.0096$& $0.9976\pm0.0038$\\
PWLS-TNV-$\ell_0$&  $0.9989\pm0.0143$&  $0.9995\pm0.0156$&     $0.7071\pm0.0353$&    $0.2919\pm0.0345$&  $0.9983\pm0.0020$& $0.9993\pm0.0011$\\
\hline
\hline
\end{tabular}
\label{DigitalSTD}
\end{table*}

\subsection{Catphan\copyright600 phantom study}
\begin{figure*}[htbp!]
\centering
\subfigure{\includegraphics[width=0.3\linewidth,height=0.3\linewidth]{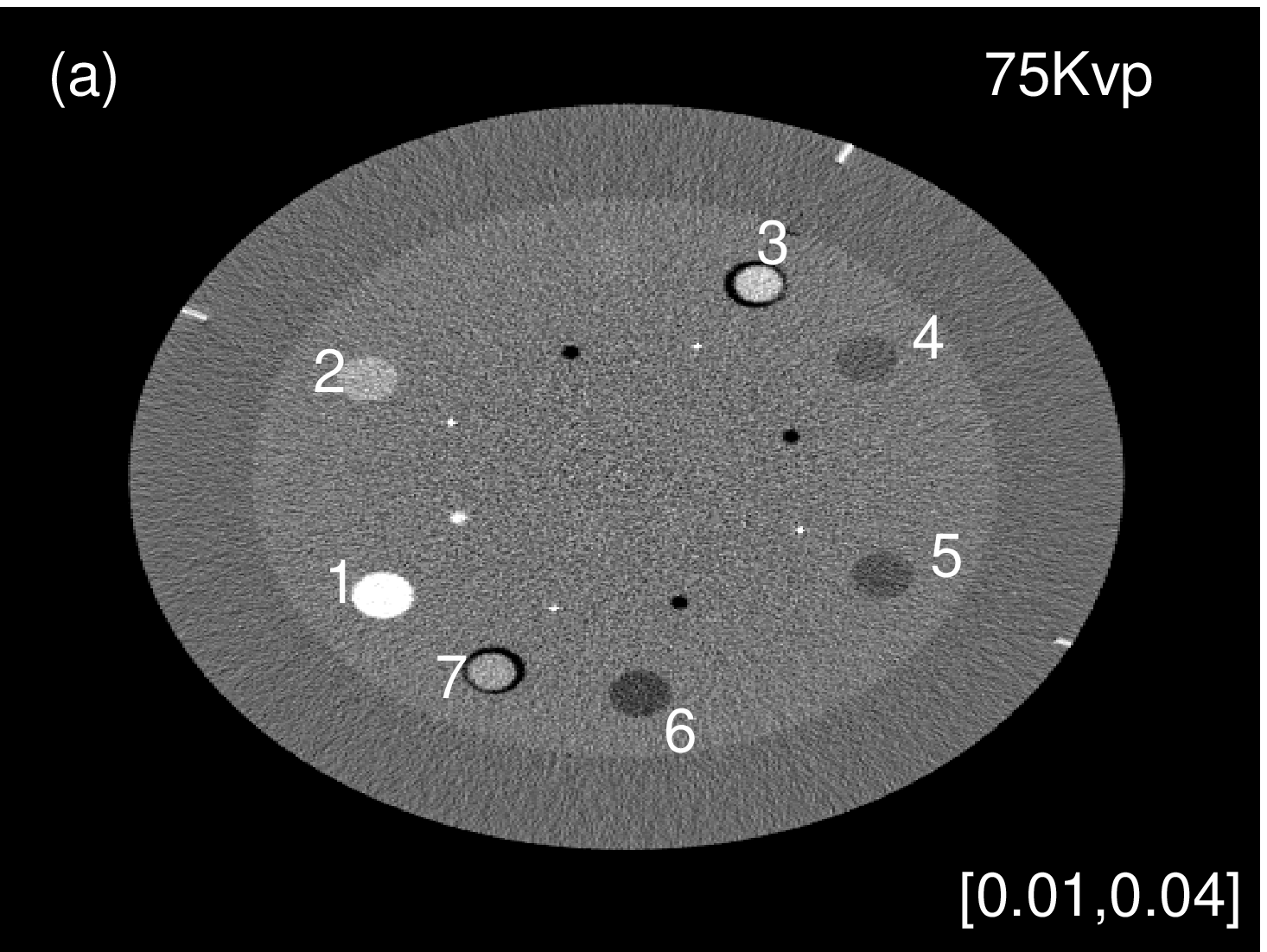}}
\subfigure{\includegraphics[width=0.3\linewidth,height=0.3\linewidth]{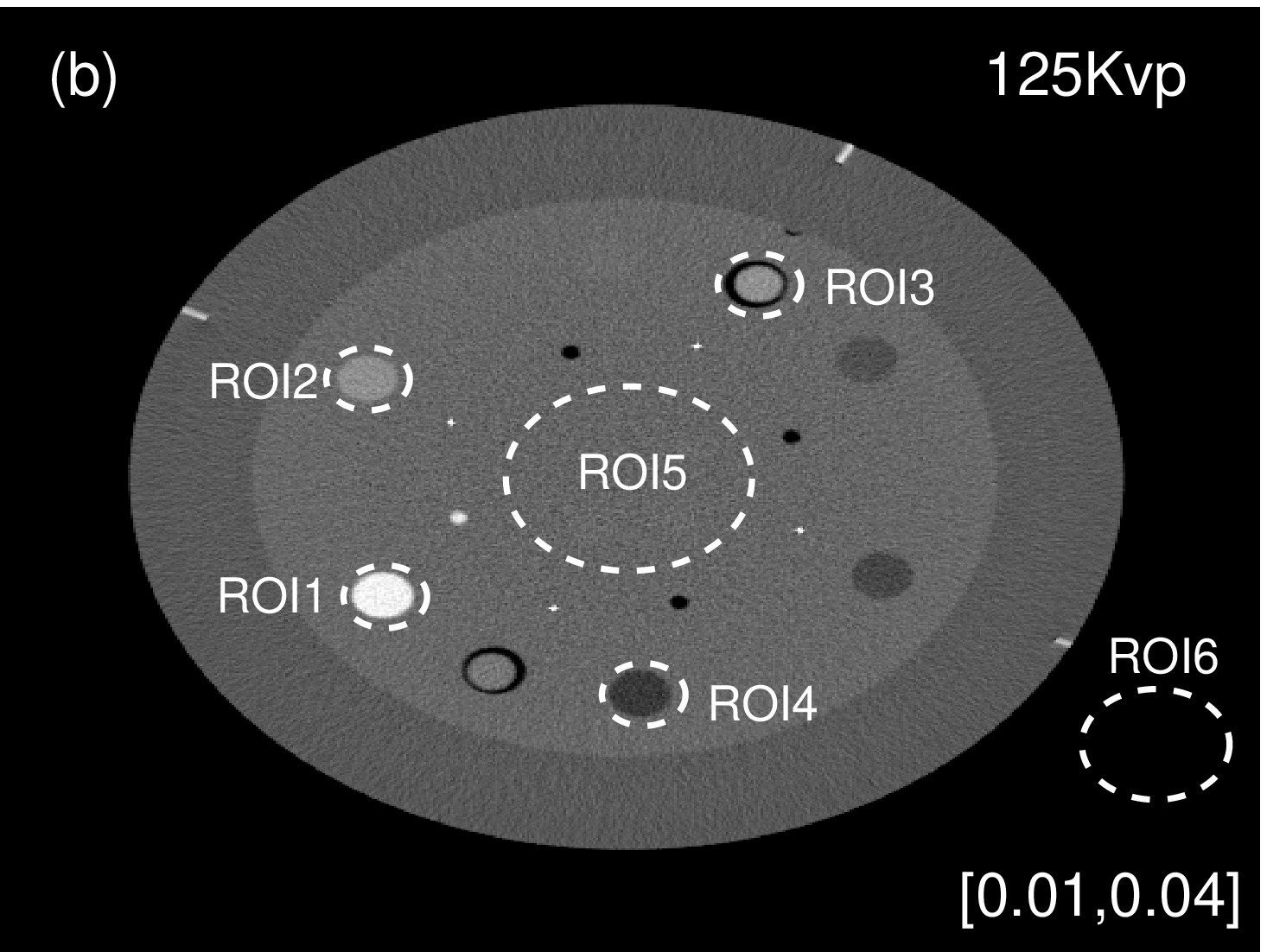}}
\caption { CT images of the Catphan\copyright600 phantom on the contrast rods slice: (a) The low-energy: 75 kVp and (b) The high-energy: 125 kVp. Display window is
$[0.01, 0.04]\mathrm{mm}^{-1}$. The components of ROIs are Teflon (ROI1), Delrin (ROI2), Iodine of $10~\mathrm{mg}/\mathrm{ml}$ (ROI3), PMP (ROI4), Inner soft tissue (ROI5) and
Air (ROI6).}
\label{CatphanPhantom}
\end{figure*}

We acquired the Catphan\copyright600 phantom data on
a tabletop cone-beam CT (CBCT) system
whose geometry matched that of a Varian On-Board Imager (OBI)
on the Trilogy radiation therapy machine.
We inserted iodine solutions with nominal concentrations of
$10~\mathrm{mg}/\mathrm{ml}$ and $5~\mathrm{mg}/\mathrm{ml}$ into
the phantom.
There were $1024\times768$ pixels with a physical size of
$0.388~ \mathrm{mm} \times 0.388~ \mathrm{mm}$ per pixel
on the CB4030 flat-panel detector (Varian Medical Systems).
The DECT measurements were obtained at $75$~kVp and $125$~kVp
with a tube current of $80$~mA and a pulse width of $13$~ms.
We acquired $655$ projections over $[0^\circ, 360^\circ)$
in each scan.
Using a fan-beam geometry with a longitudinal beam width of $15~\mathrm{mm}$
on the detector \cite{niu2010shading},
We acquired projections with scatter contamination inherently suppressed.
We used a contrast rod slice of the Catphan\copyright600 phantom
to evaluated the proposed method.
We reconstructed attenuation images of size $512\times512$ with a pixel size of
$0.5\mathrm{ mm} \times 0.5~\mathrm{mm}$.
Fig.~\ref{CatphanPhantom} shows the low- and high-energy CT images.
Fig.~\ref{CatphanPhantom}(a) identifies the rods with labels:
Teflon (labeled as $\#1$),
Delrin (labeled as $\#2$),
Iodine solution of $10~\mathrm{mg}/\mathrm{ml}$ (labeled as $\# 3$),
Polystyrene (labeled as $\#4$),
low-density Polyethylene (LDPE) (labeled as $\#5$),
Polymethylpentene (PMP) (labeled as $\#6$),
Iodine solution of $5~\mathrm{mg}/\mathrm{ml}$ (labeled as $\#7$).
Fig.~\ref{CatphanPhantom}(b) shows selected basis materials and ROIs
in white dashed line circles: Teflon (ROI1),
Delrin (ROI2), Iodine solution of
$10~\mathrm{mg}/\mathrm{ml}$ (ROI3), PMP (ROI4), Inner soft tissue (ROI5) and
Air (ROI6).

\begin{figure*}[htbp!]
\begin{center}
\begin{tabular}{c@{\hspace{0pt}}c@{\hspace{0pt}}c@{\hspace{0pt}}c@{\hspace{0pt}}c@{\hspace{0pt}}c@{\hspace{0pt}}c@{\hspace{0pt}}c}
\includegraphics[width=.17\linewidth,height=.17\linewidth]{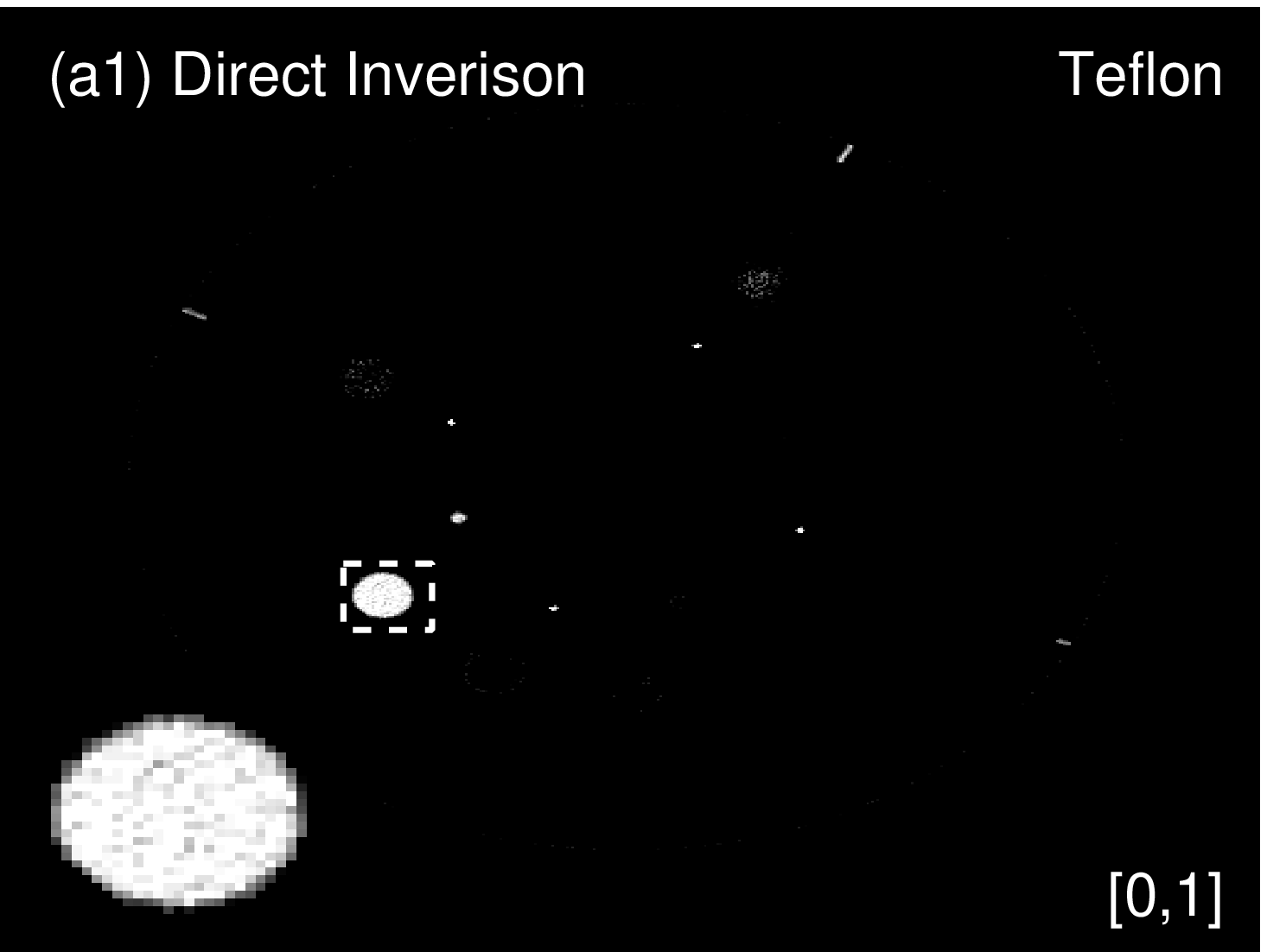}&
\includegraphics[width=.17\linewidth,height=.17\linewidth]{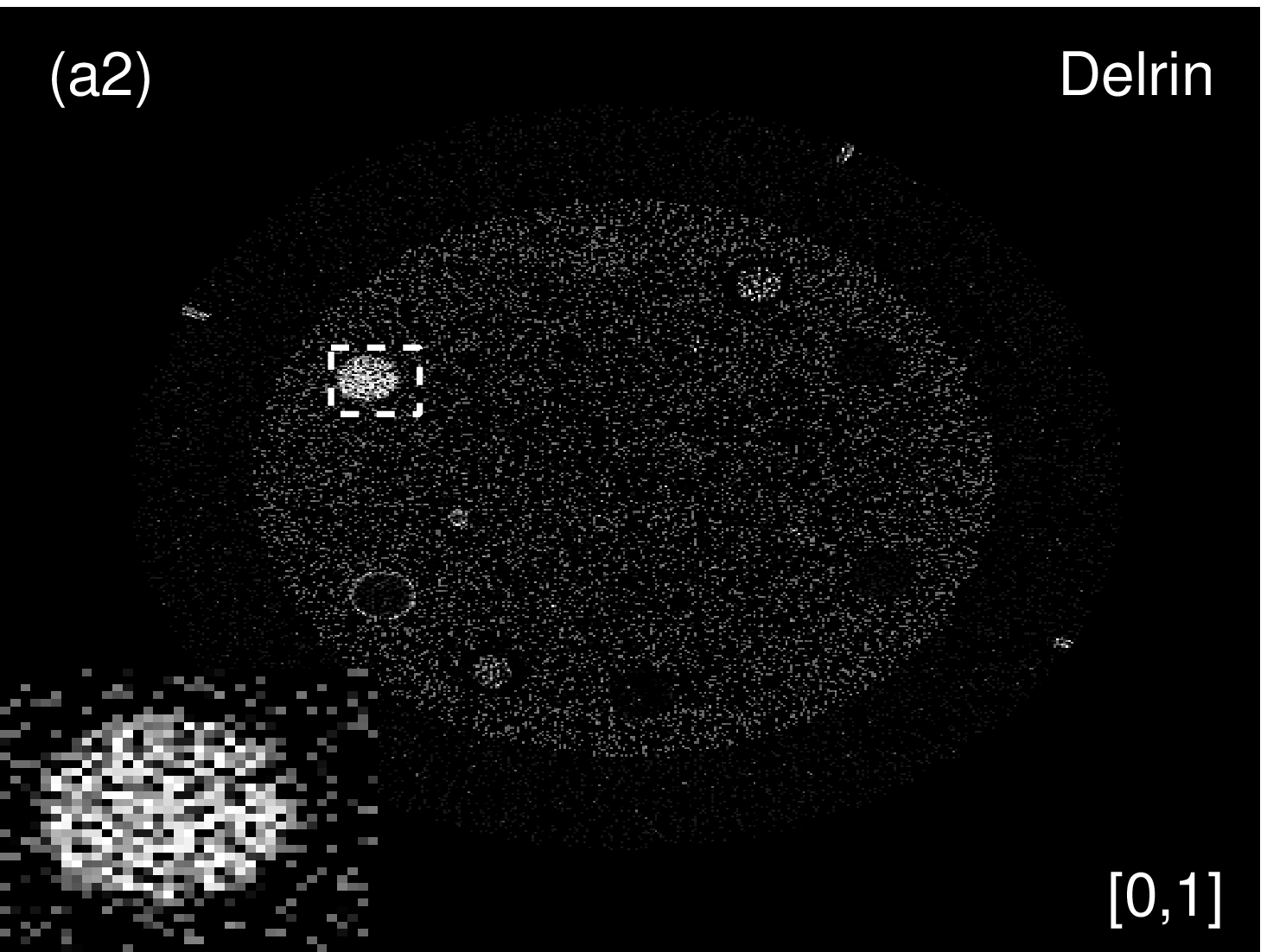}&
\includegraphics[width=.17\linewidth,height=.17\linewidth]{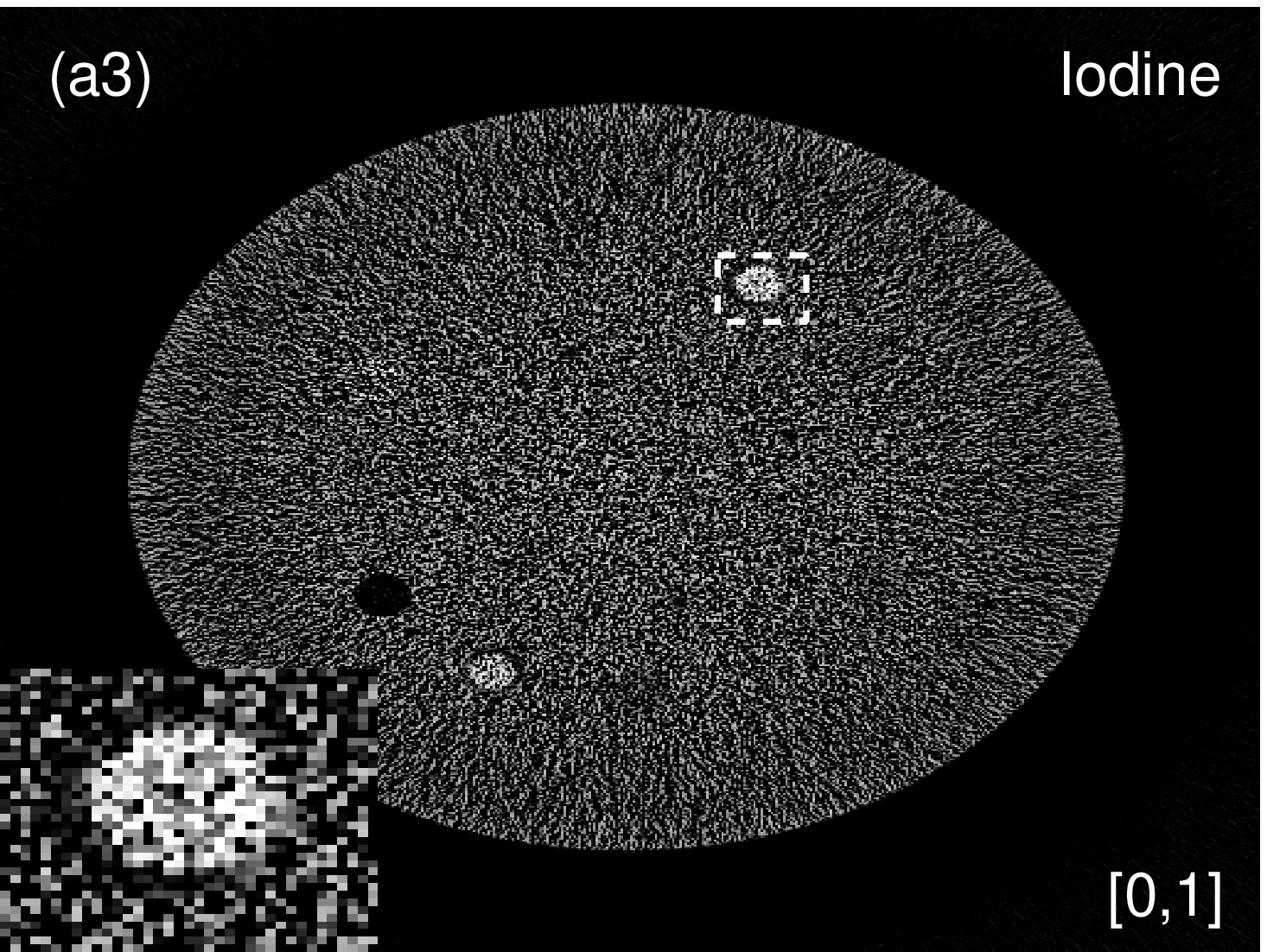}&
\includegraphics[width=.17\linewidth,height=.17\linewidth]{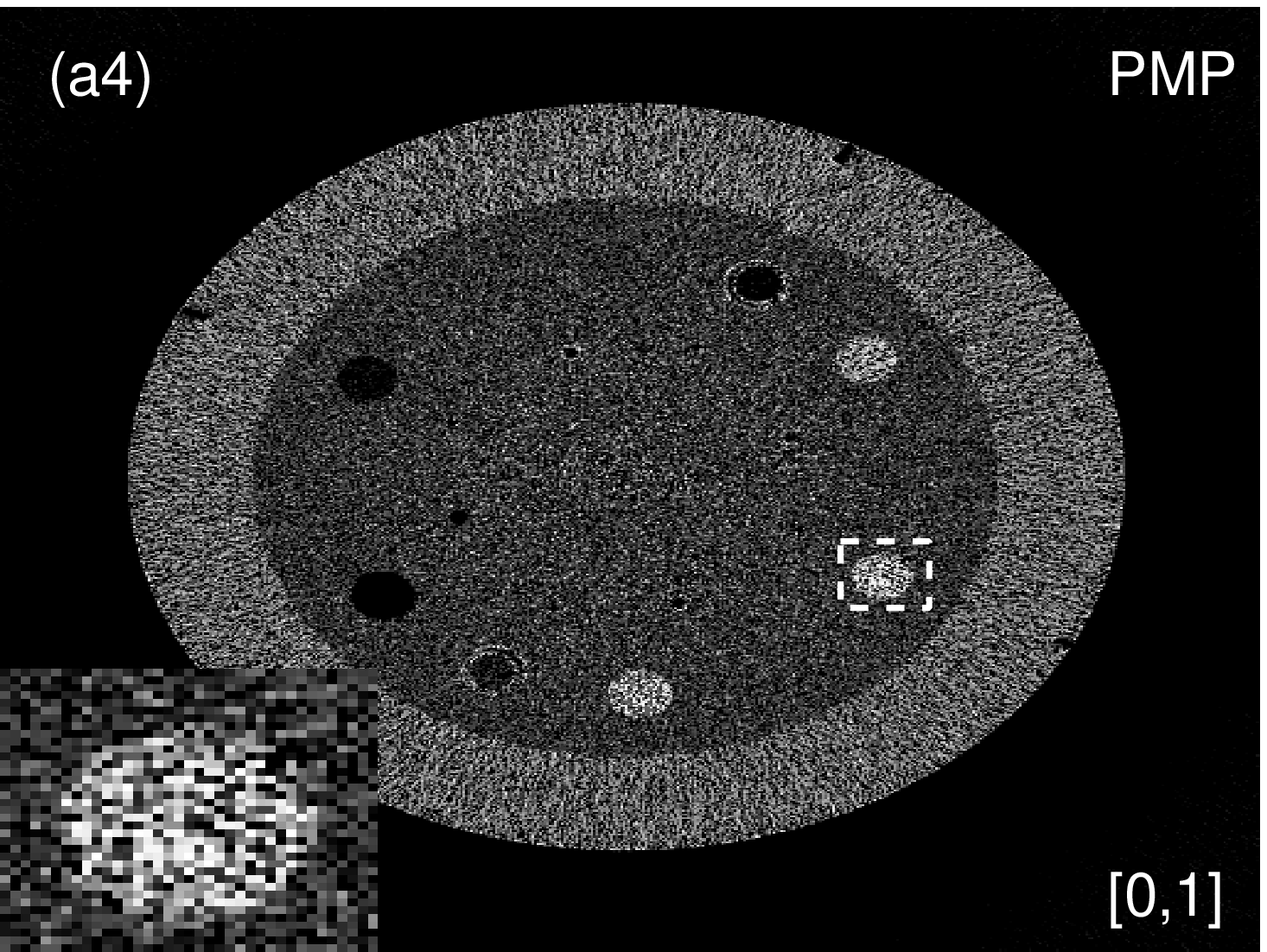}&
\includegraphics[width=.17\linewidth,height=.17\linewidth]{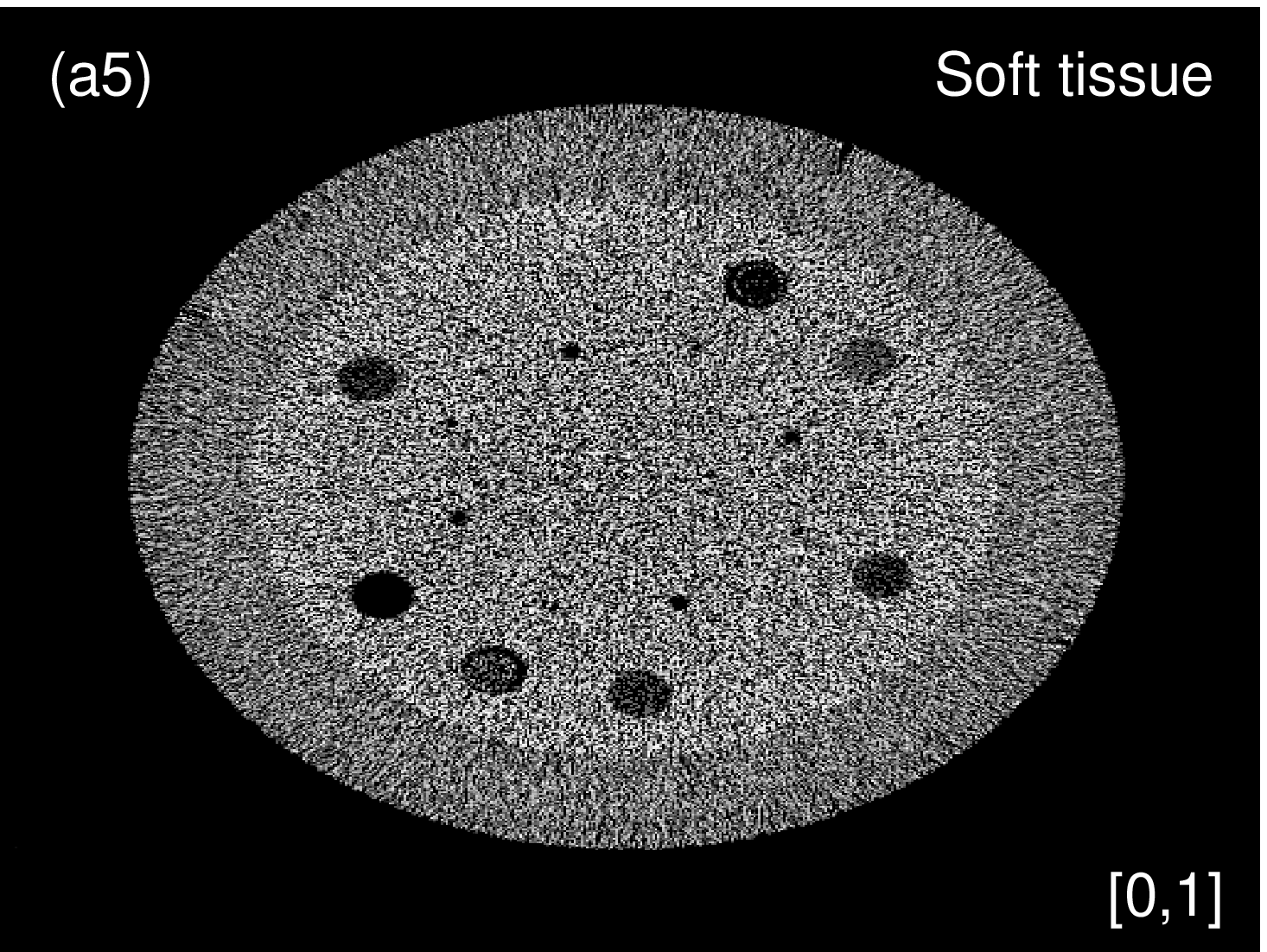}&
\includegraphics[width=.17\linewidth,height=.17\linewidth]{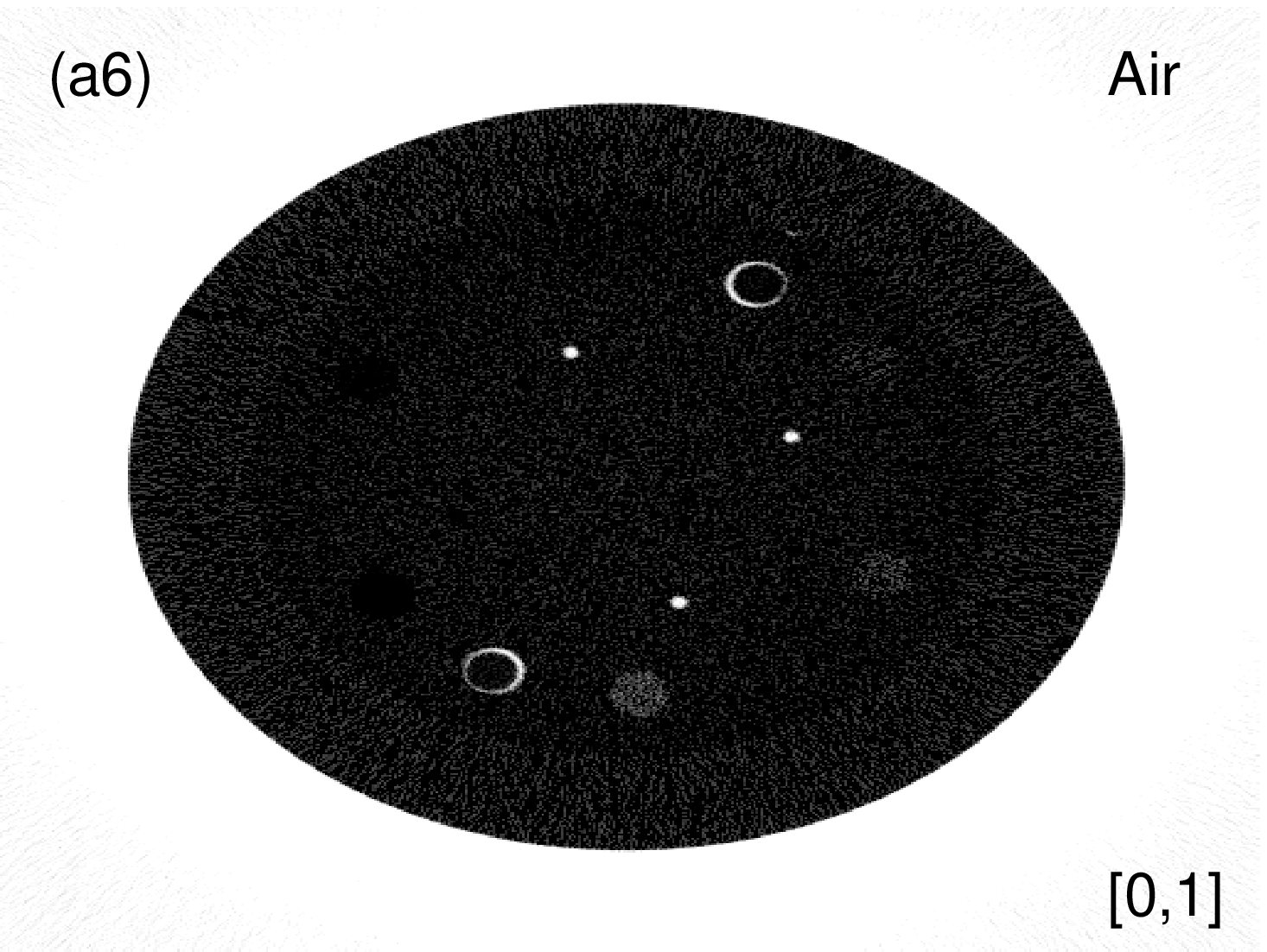}\\
\includegraphics[width=.17\linewidth,height=.17\linewidth]{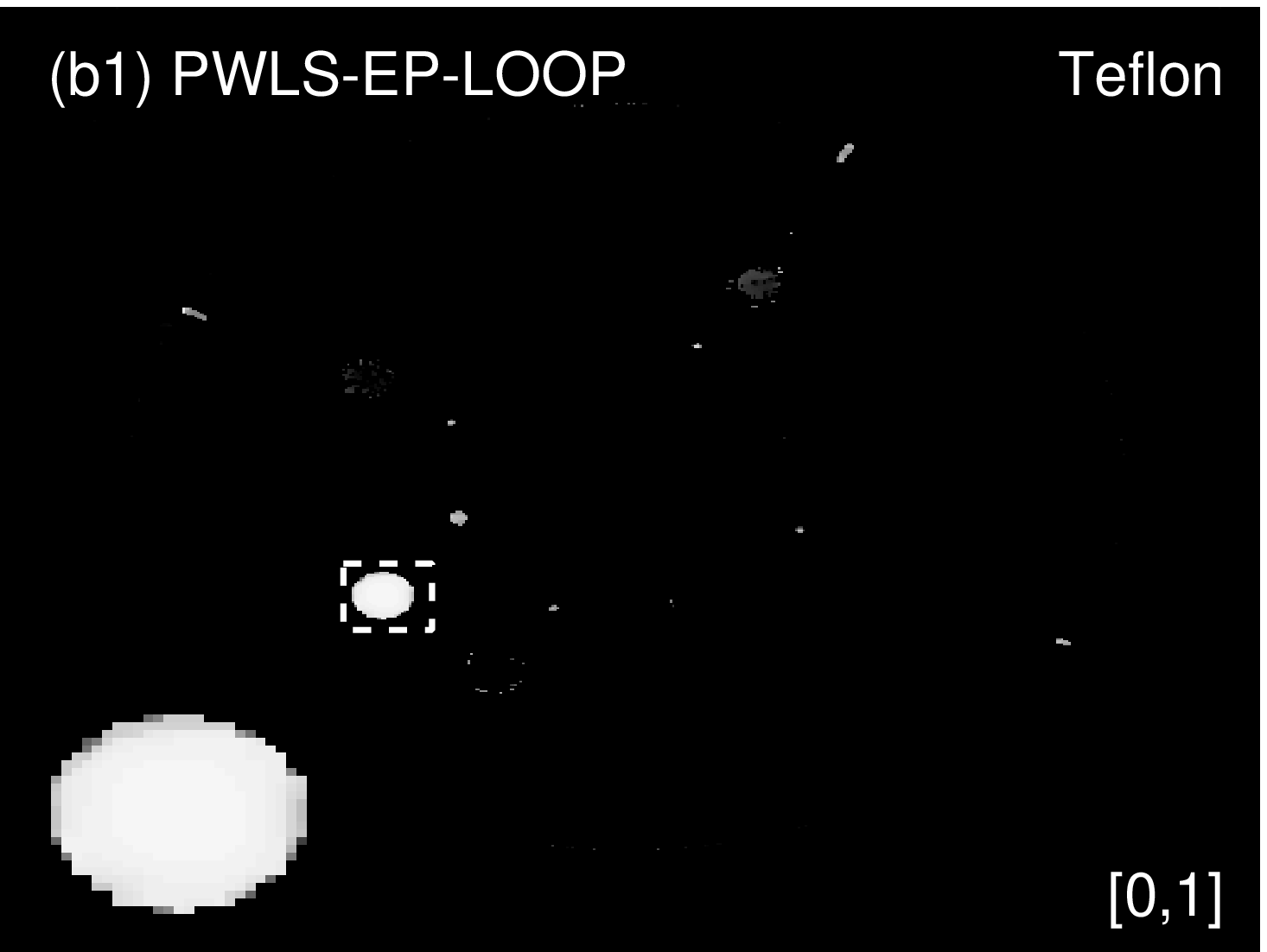}&
\includegraphics[width=.17\linewidth,height=.17\linewidth]{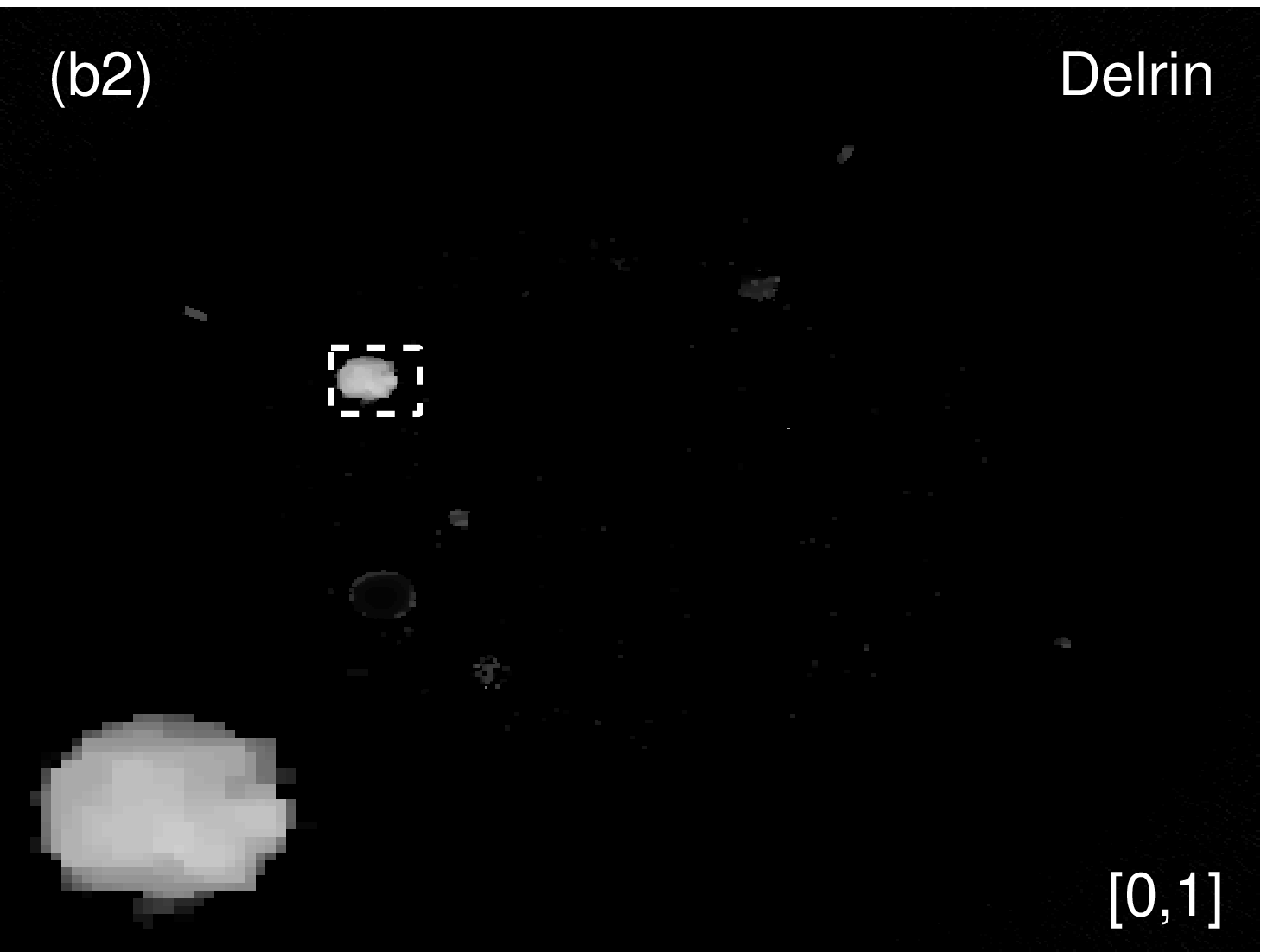}&
\includegraphics[width=.17\linewidth,height=.17\linewidth]{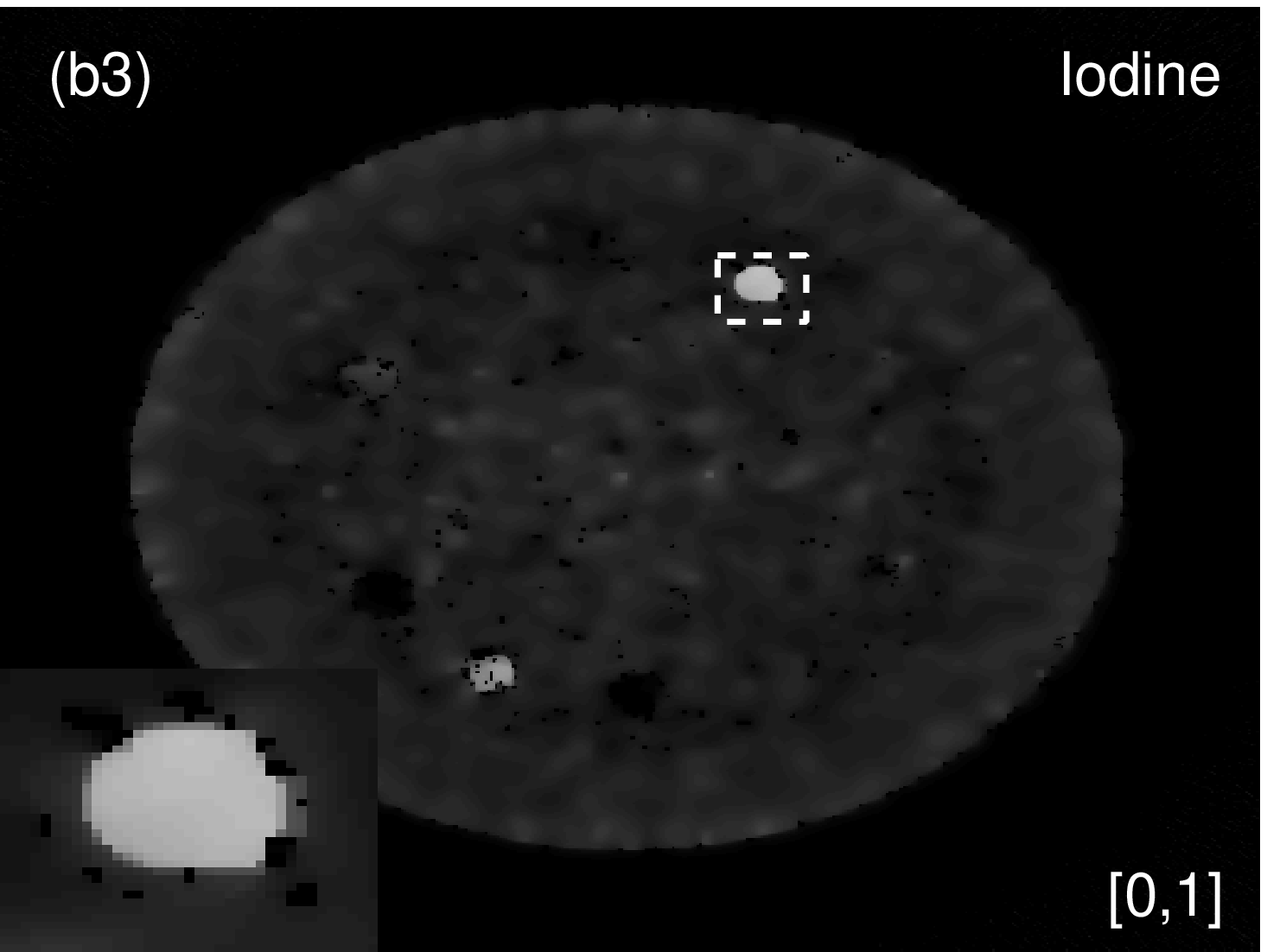}&
\includegraphics[width=.17\linewidth,height=.17\linewidth]{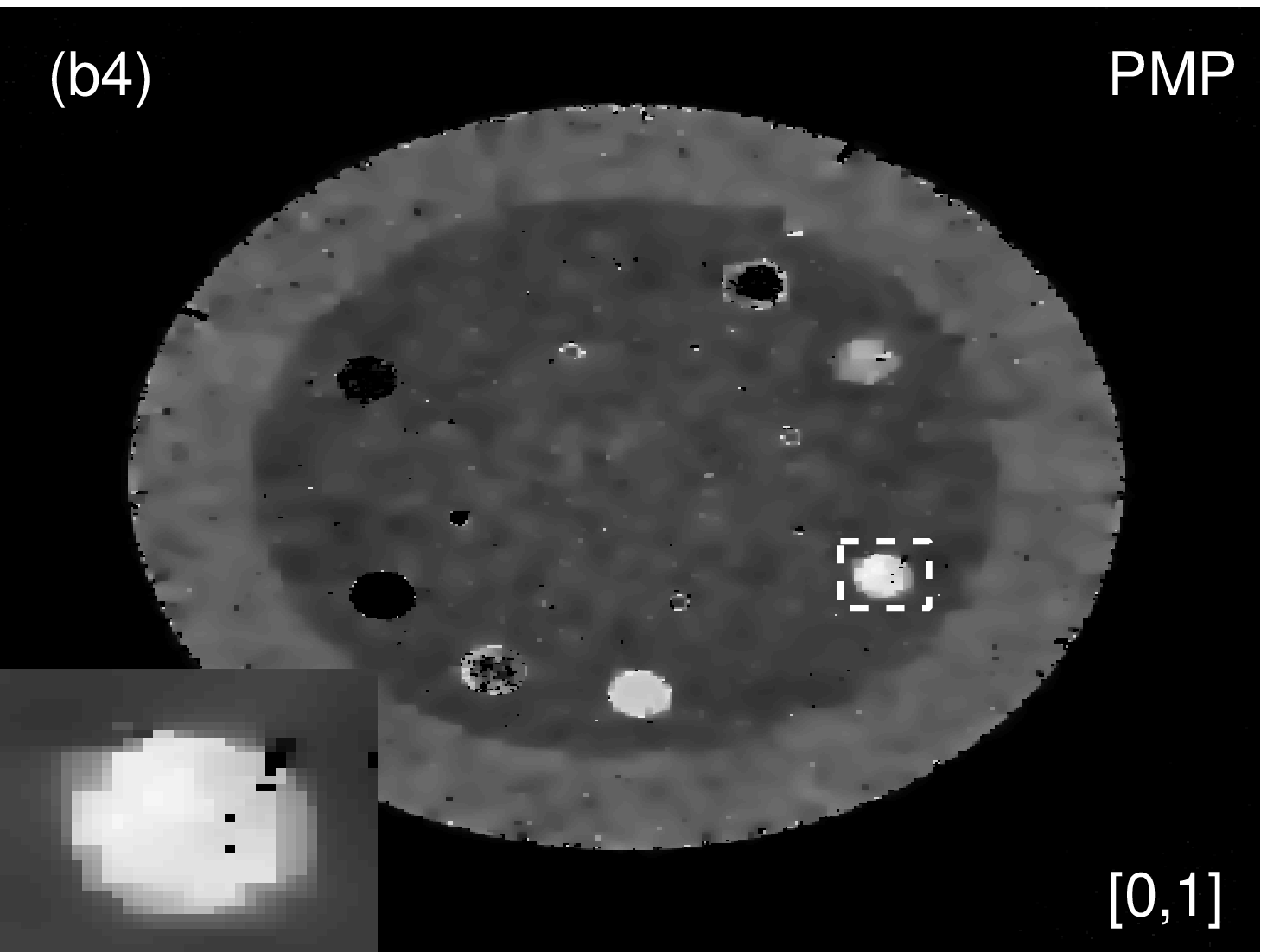}&
\includegraphics[width=.17\linewidth,height=.17\linewidth]{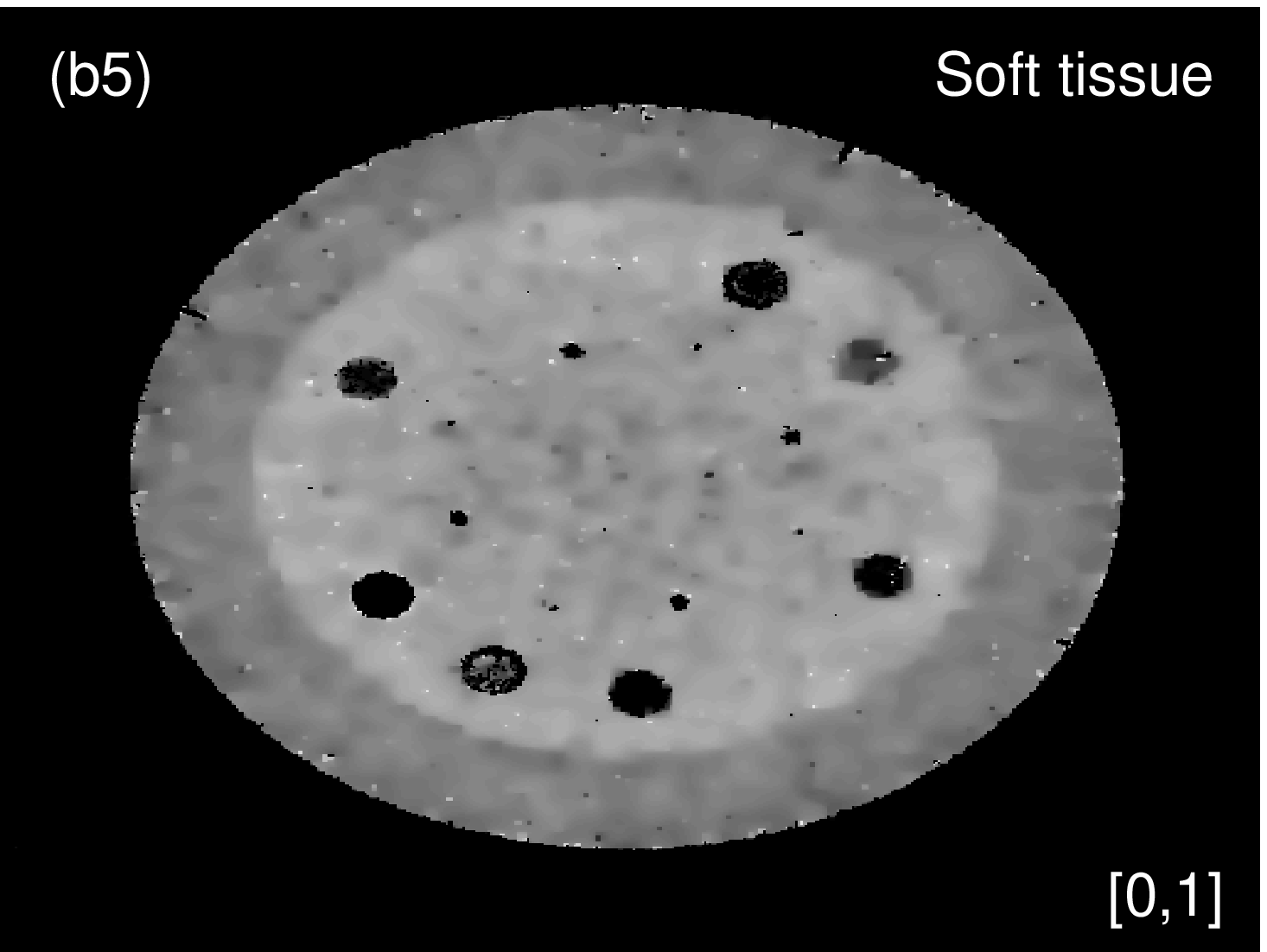}&
\includegraphics[width=.17\linewidth,height=.17\linewidth]{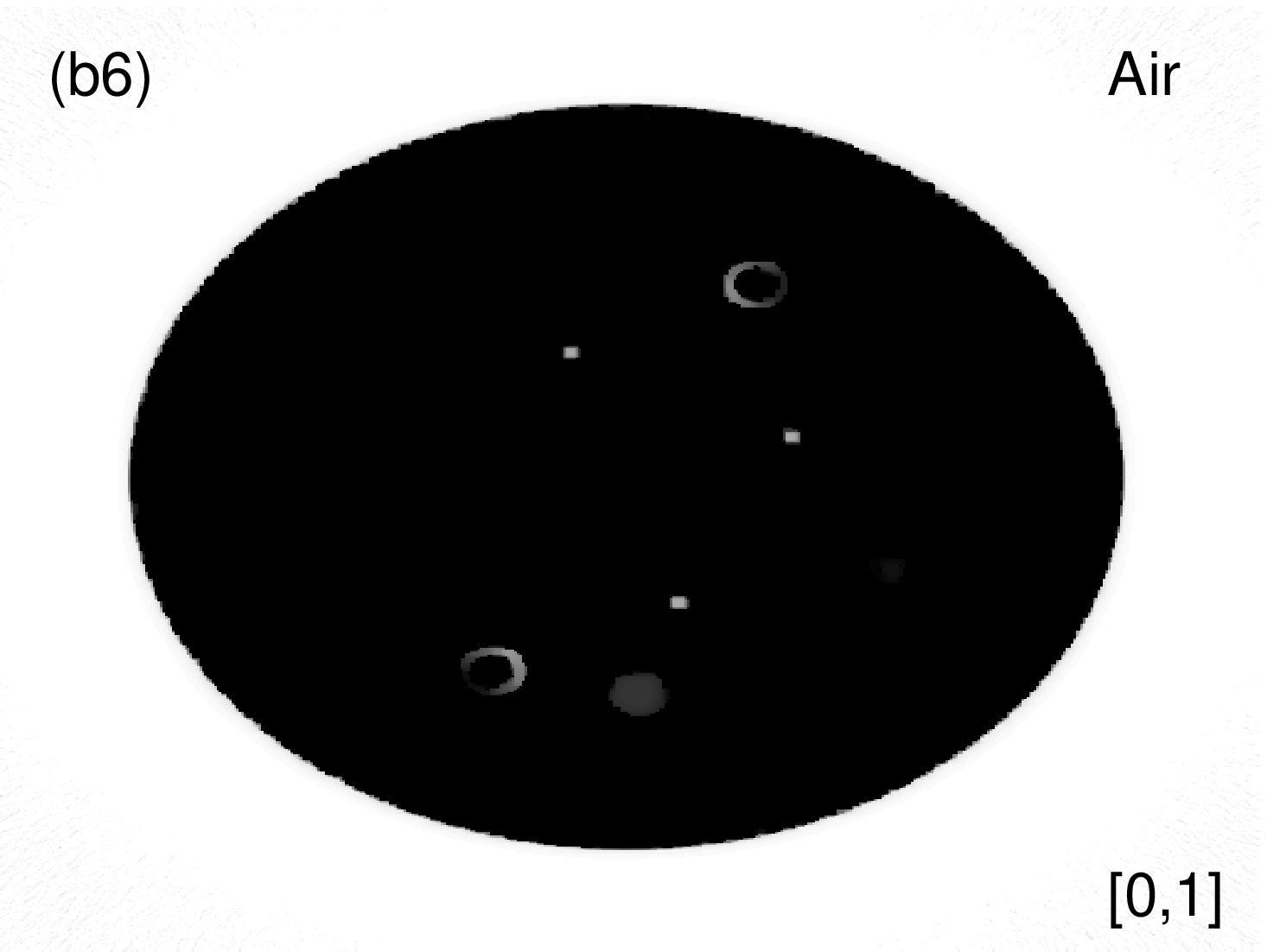}\\
\includegraphics[width=.17\linewidth,height=.17\linewidth]{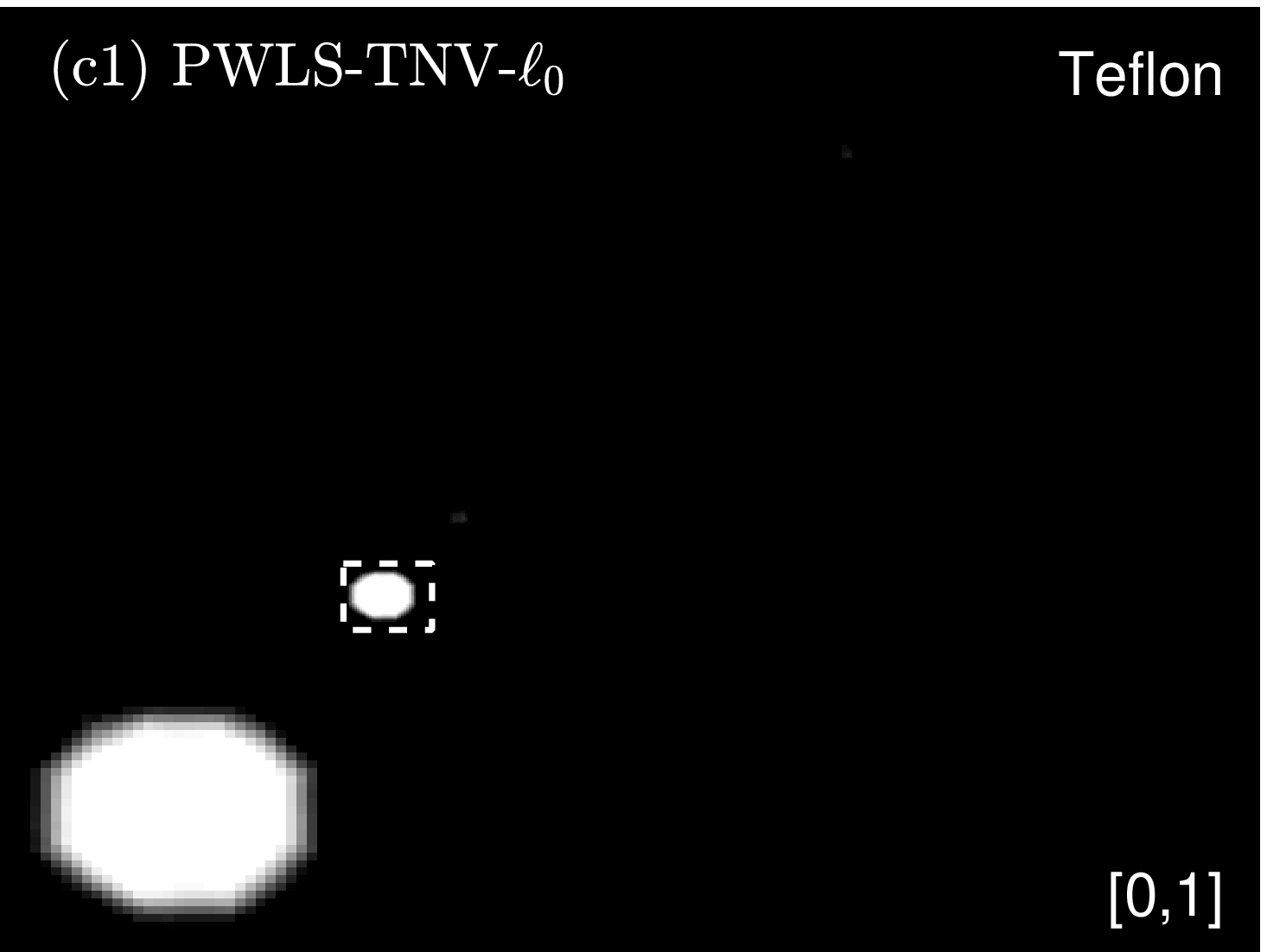}&
\includegraphics[width=.17\linewidth,height=.17\linewidth]{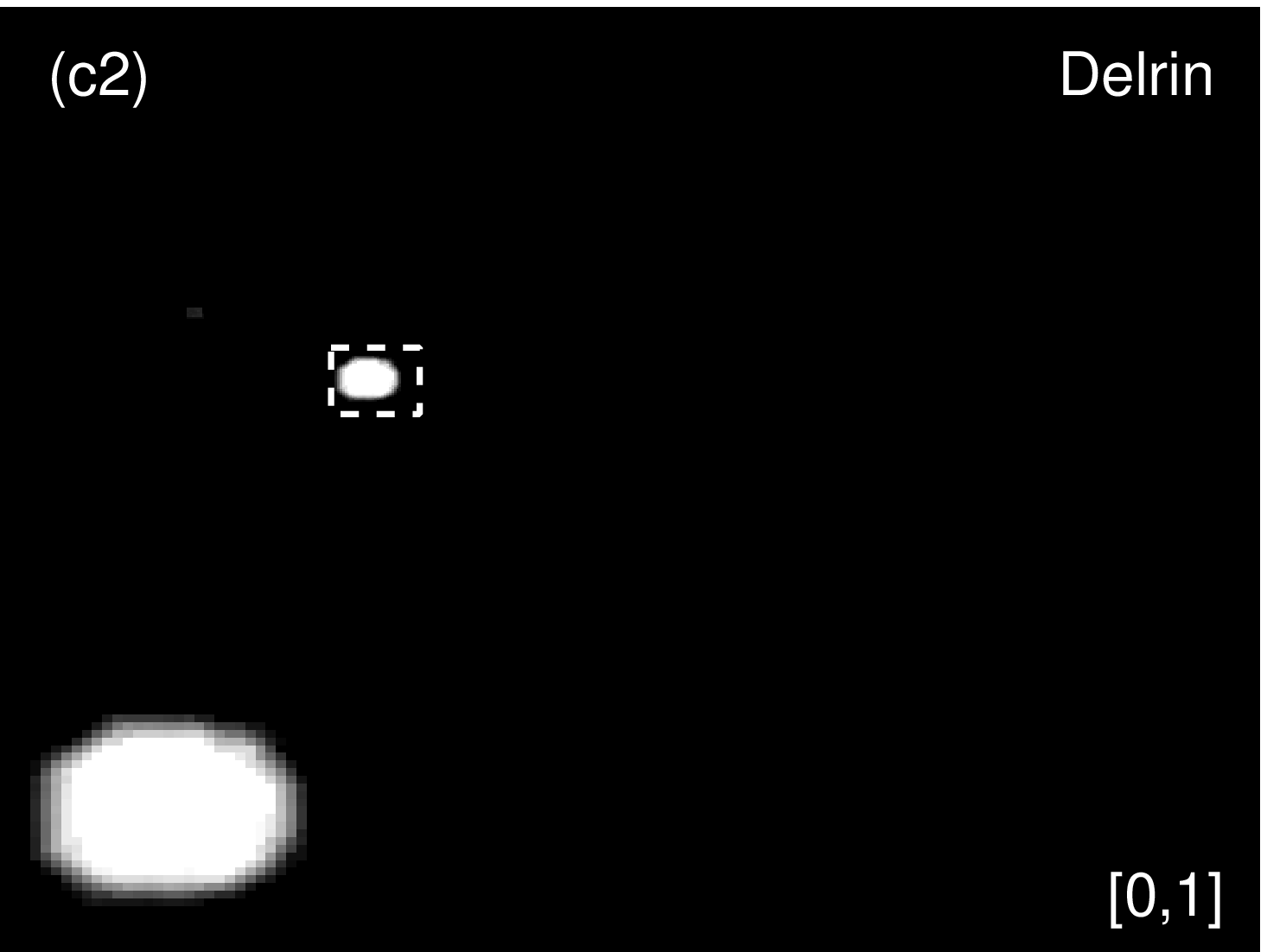}&
\includegraphics[width=.17\linewidth,height=.17\linewidth]{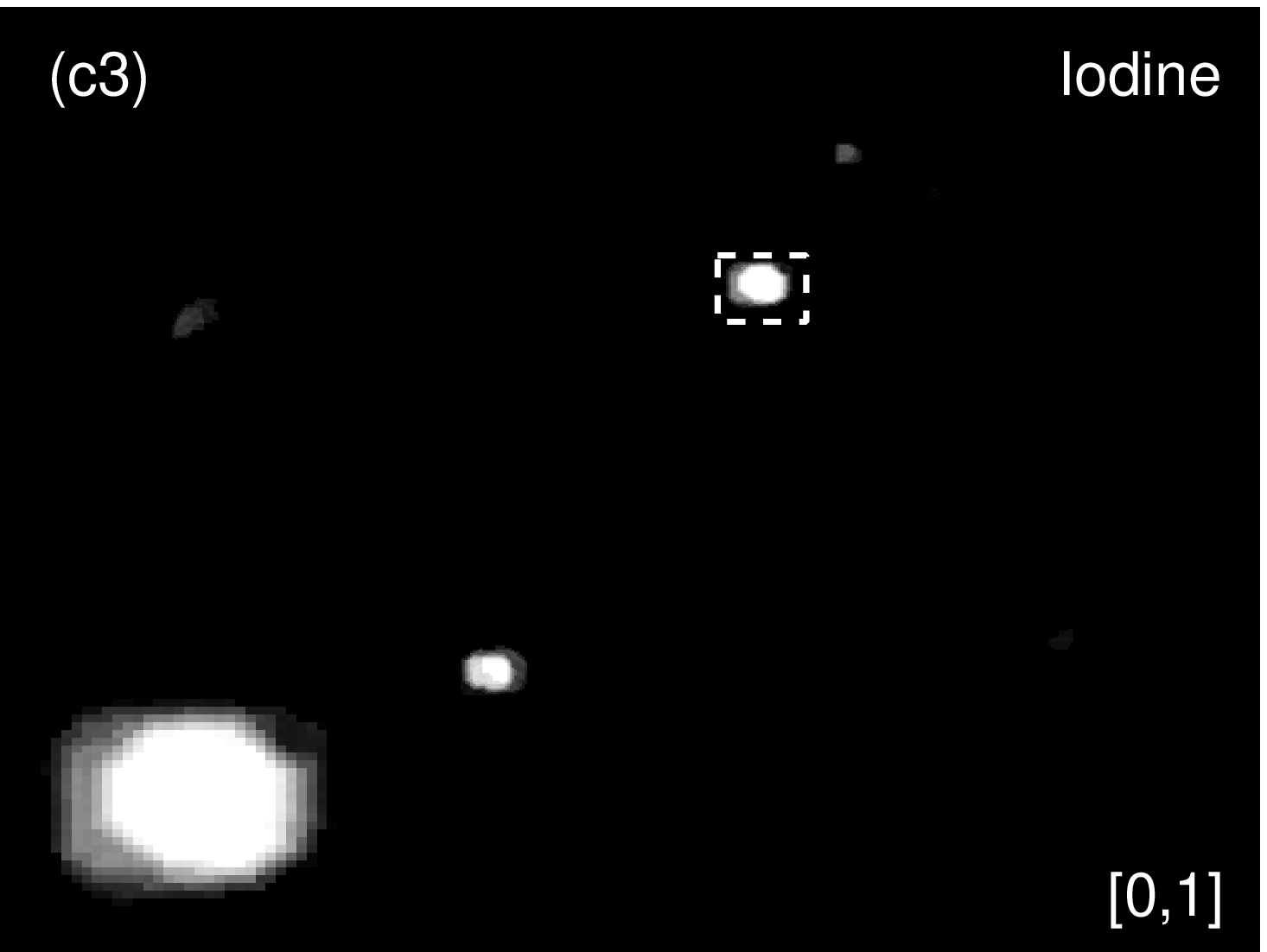}&
\includegraphics[width=.17\linewidth,height=.17\linewidth]{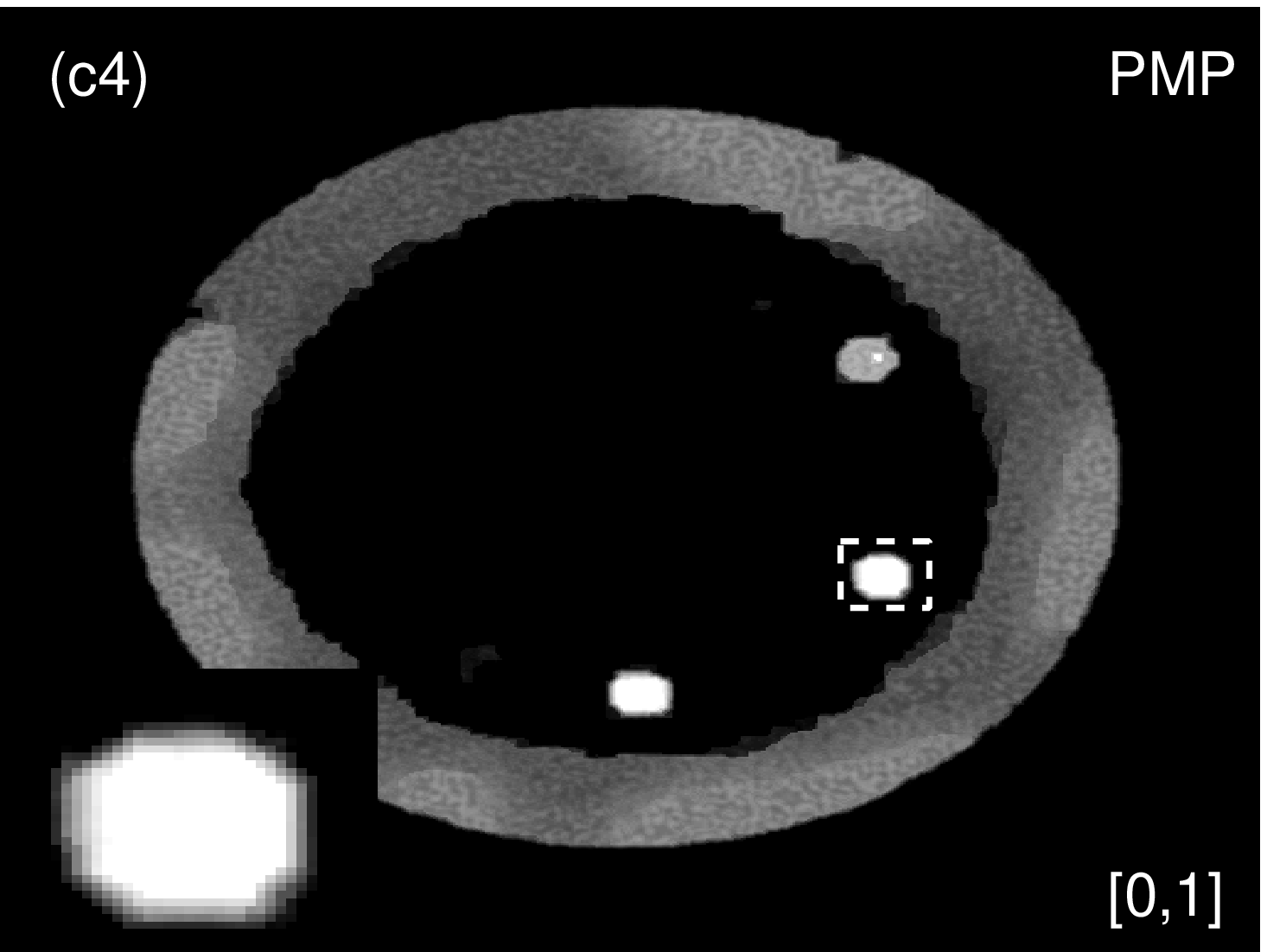}&
\includegraphics[width=.17\linewidth,height=.17\linewidth]{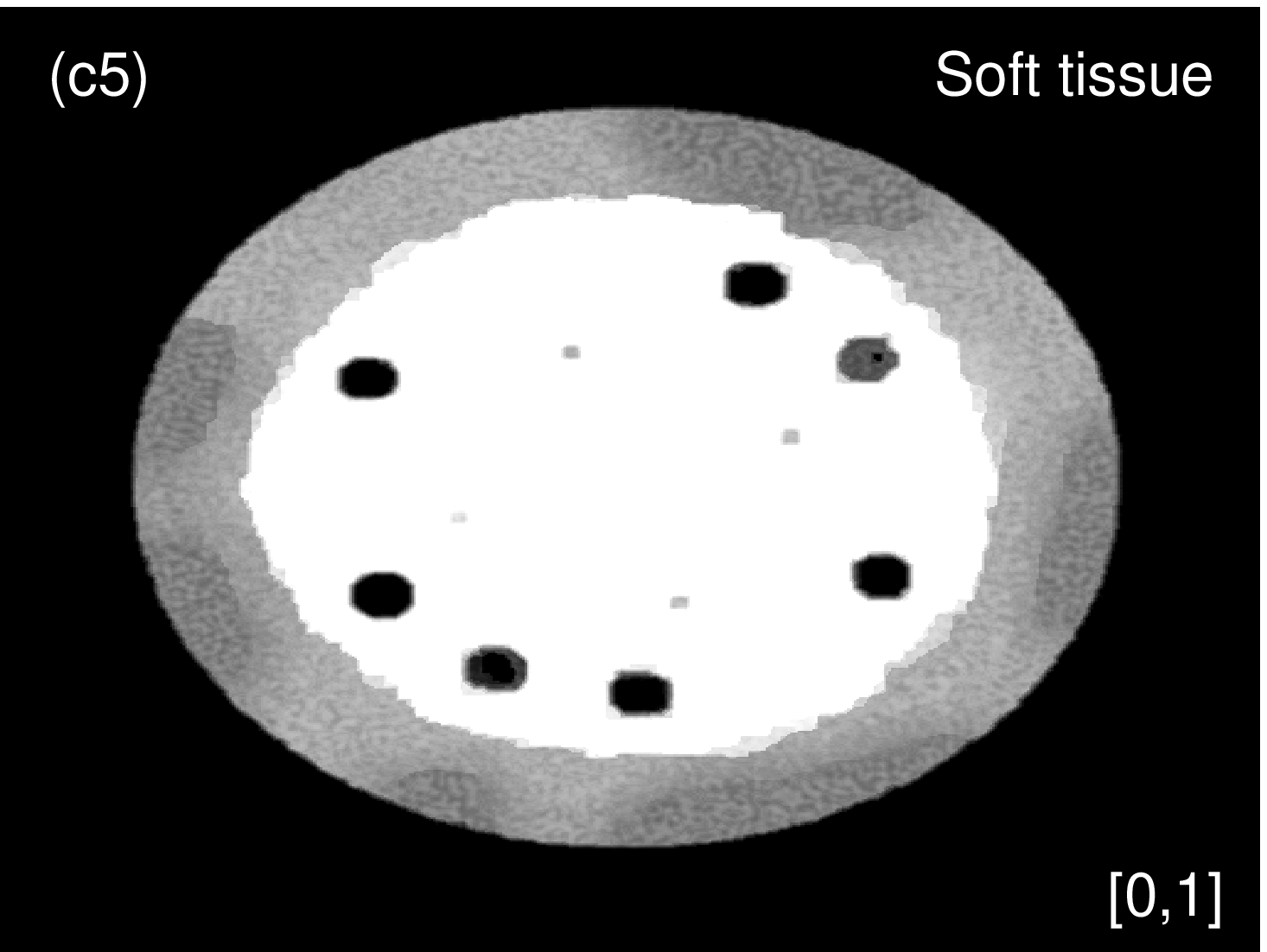}&
\includegraphics[width=.17\linewidth,height=.17\linewidth]{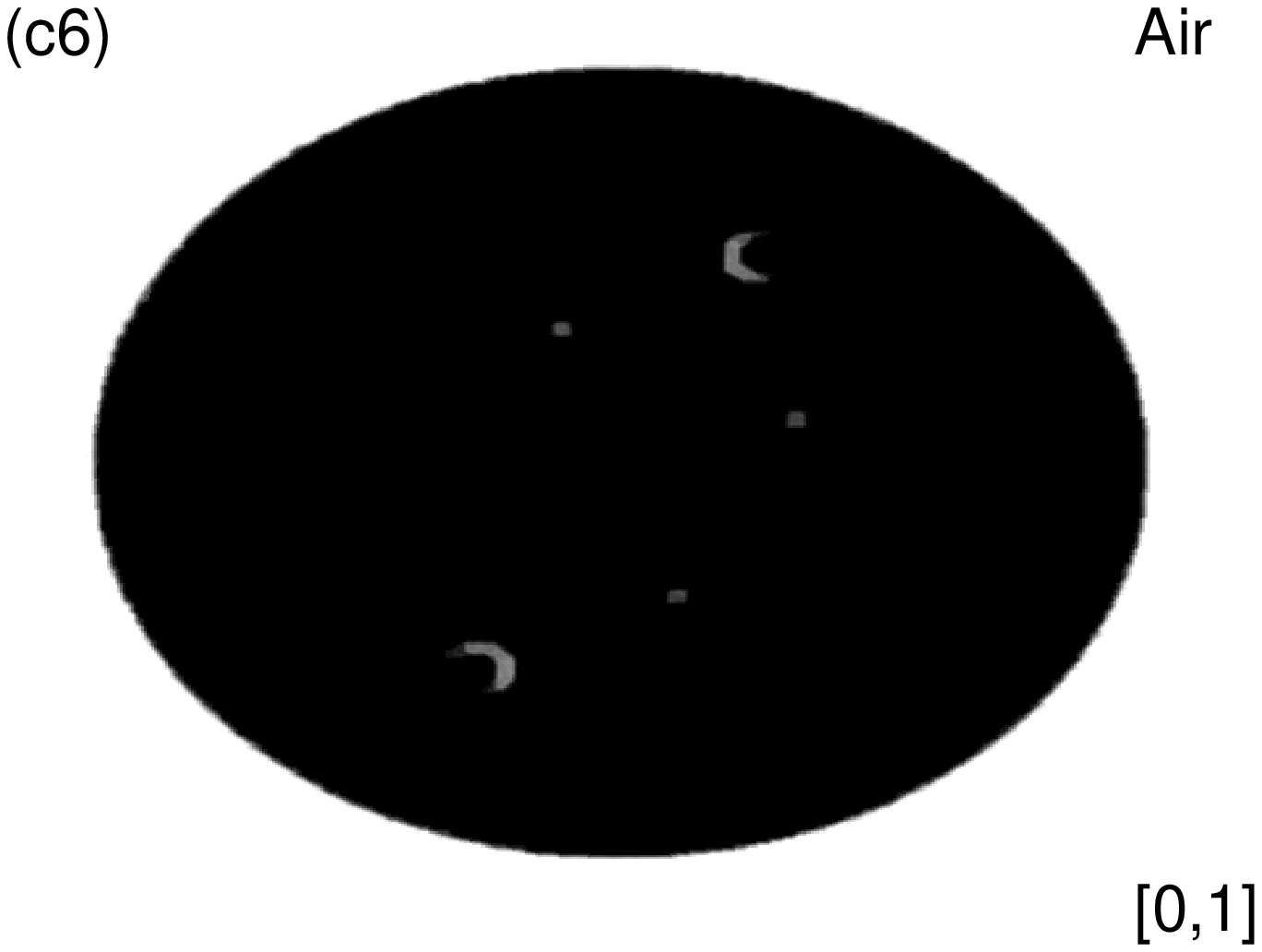}
\end{tabular}
\caption {  Material images of  Direct Inversion (the $1^{st}$ row), PWLS-DP-LOOP (the $2^{rd}$ row) and PWLS-TNV-$\ell_0$ (the $3^{nd}$ row).
The decomposed Teflon (the $1^{st}$  column),  delrin (the $2^{ed}$  column), iodine  solution(the $3^{nd}$ column), PMP (the $4^{th}$ column),  soft tissue (the $5^{th}$  column) and air (the $6^{th}$  column) images of the Catphan\copyright600 phantom on the contrast rods slice .
 The display windows are shown in the bottom-right corners.}
\label{CatPhanMM}
\end{center}
\end{figure*}

Fig.~\ref{CatPhanMM} shows the decomposed material images by the Direct
Inversion, the PWLS-EP-LOOP and the PWLS-TNV-$\ell_0$ method.
The left corners of the $1^{st}$ to the $4^{th}$ column of Fig.~\ref{CatPhanMM}
show enlarged rods that are highlighted with white dashed boxes in
decomposed material images.
Table~\ref{CatphanSTD} summarizes the means and noise STDs of ROIs of
decomposed basis material images.
The volume fraction (VF) accuracies were $68.62\%$, $79.33\%$, and $99.88\%$
for the Direction Inversion, the PWLS-EP-LOOP and the PWLS-TNV-$\ell_0$ method,
respectively.
Compared with the Direct Inversion and the PWLS-EP-LOOP method, the proposed
PWLS-TNV-$\ell_0$ method increases the VF accuracy by $31.18\%$
and $20.45\%$ respectively.
\begin{table*}[htbp!]
\caption{\leftline{The means and STDs of decomposed images within each ROI of Catphan\copyright600.}}
\scalebox{1}{
\begin{tabular}{p{2.7cm}p{2.4cm}p{2.4cm}p{2.4cm}p{2.4cm}p{2.4cm}p{2.4cm}p{2.3cm}}
\hline
\hline
\multirow{2}{*}{Methods}&  ROI1&  ROI2&     ROI3&    ROI4&          ROI5&  ROI6\\
                        &Teflon&Delrin&   Iodine&     PMP&   Soft Tissue&  Air\\
\hline
Ground Truth &           $1\pm0$          &   $1\pm0$         &     $1\pm0$          &    $1\pm0$          &  $1\pm0$          & $1\pm0$\\
Direct Inversion &       $0.9578\pm0.0642$&  $0.5852\pm0.3340$&     $0.6190\pm0.3290$&    $0.5067\pm0.3088$&  $0.4265\pm0.3309$& $0.9995\pm0.0037$\\
PWLS-EP-LOOP&            $0.9615\pm0.0043$&  $0.7306\pm0.0367$&     $0.7112\pm0.0188$&    $0.7788\pm0.0071$&  $0.5779\pm0.0277$& $0.9999\pm0.0018$\\
PWLS-TNV-$\ell_0$&       $1.0000\pm0.0025$&  $0.9971\pm0.0037$&     $1.0026\pm0.0074$&    $0.9989\pm0.0093$&  $1.0002\pm0.0001$& $1.0001\pm0.0001$\\
\hline
\hline
\end{tabular}
}
\label{CatphanSTD}
\end{table*}
Table~\ref{CatphanRMSE} summarizes the average electron densities of contrast
rods and RMSE($\%$) of electron density for the three MMD methods.
The RMSE($\%$) was $12.27\%$, $11.81\%$ and $4.42\%$ for the Direct Inversion
method, the PWLS-EP-LOOP method and the proposed PWLS-TNV-$\ell_0$ method,
respectively.
The proposed PWLS-TNV-$\ell_0$ method suppressed noise,
decreases crosstalk
and increased decomposition accuracy
in the material images,
while maintaining high image quality.

\begin{table*}[htbp!]
\caption{Electron densities inside the Catphan\copyright600 contrast rods. The numbers of the rods are marked in Fig. \ref{CatphanPhantom}(a). The last column is RMSE$(\%)$ of the seven rods. The electron density of iodine solutions is calculated based on iodine concentrations. The unit of the electron density is $10^{23} \mathrm{e}/\mathrm{cm}^3.$}
\begin{tabular}{lccccccccccccccc}
\hline
\hline
\multirow{2}{*}{Rods } &  1   &  2   &     3           &      4    &         5&    6&                 7&    \\
                       &Teflon&Delrin& Iodine(10 mg/ml)&Polystyrene&LDPE      &  PMP&   Iodine(5 mg/ml)&  RMSE$(\%)$\\
\hline
Ground truth&                       $6.240$&  $4.525$&    $3.368$&   $3.400$&   $3.155$&    $2.851$&    $3.356$&\\
\hline
Direct Inversion&                   $6.158$&  $4.127$&    $3.882$&   $2.984$&   $2.729$&    $2.274$&    $3.370$&\\
Average Percentage Errors E$(\%)$ & $1.32\%$& $8.80\%$&   $15.25\%$& $12.24\%$& $13.49\%$&  $20.23\%$&  $0.42\%$& $12.27\%$\\
\hline
PWLS-EP-LOOP&                       $6.171$&  $4.288$&    $3.936$&   $3.140$&   $2.769$&    $2.243$&    $3.348$&\\
Average Percentage Errors E$(\%)$ & $1.10\%$& $5.23\%$&   $16.85\%$& $7.65\%$&  $12.23\%$&  $21.32\%$&  $0.25\%$& $11.81\%$\\
\hline
PWLS-TNV-$\ell_0$&                  $6.242$&  $4.525$&    $3.390$&   $3.173$&   $2.854$&    $2.854$&    $3.375$&\\
Average Percentage Errors E$(\%)$ & $0.02\%$& $0.00\%$&   $0.66\%$&  $6.68\%$&  $9.54\%$&   $0.11\%$&   $0.56\%$& $4.42\%$\\
\hline
\hline
\end{tabular}
\label{CatphanRMSE}
\end{table*}

\subsection{Pelvis Data Study}

\begin{table*}[htbp!]
\caption{\leftline{Data acquisition parameters applied in pelvis data acquisition.}}
\begin{tabular}{|p{3.3cm}|p{1.8cm}|p{1.6cm}|p{1.3cm}|p{2.5cm}|p{1.6cm}|p{1.0cm}|p{3.3cm}|}
\hline
 Siemens SOMATOM Definition flash CT  &Peak voltage (kVp)&X-ray Tube Current (mA)&Exposure Time(s)&Current-exposure Time Product (mAs)&Noise STD ($\mathrm{mm}^{-1}$)&Helical Pitch&Gantry Rotation Speed (circle/second) \cr
\hline
High-energy CT image&$140$&$146$                &    $0.500$     &    $73.0$                         &  $4.09\mathrm{e}-04$     & $0.7$       &$0.28$\\
\hline
Low-energy CT image &$100$&$186$                &    $0.500$     &    $93.0$                         &  $7.27\mathrm{e}-04$     &$0.7$        &$0.28$\\
\hline
\end{tabular}
\label{DataAcquire}
\end{table*}

\begin{figure*}[htbp!]
\centering
\subfigure{\includegraphics[width=0.3\linewidth, height=0.3\linewidth ]{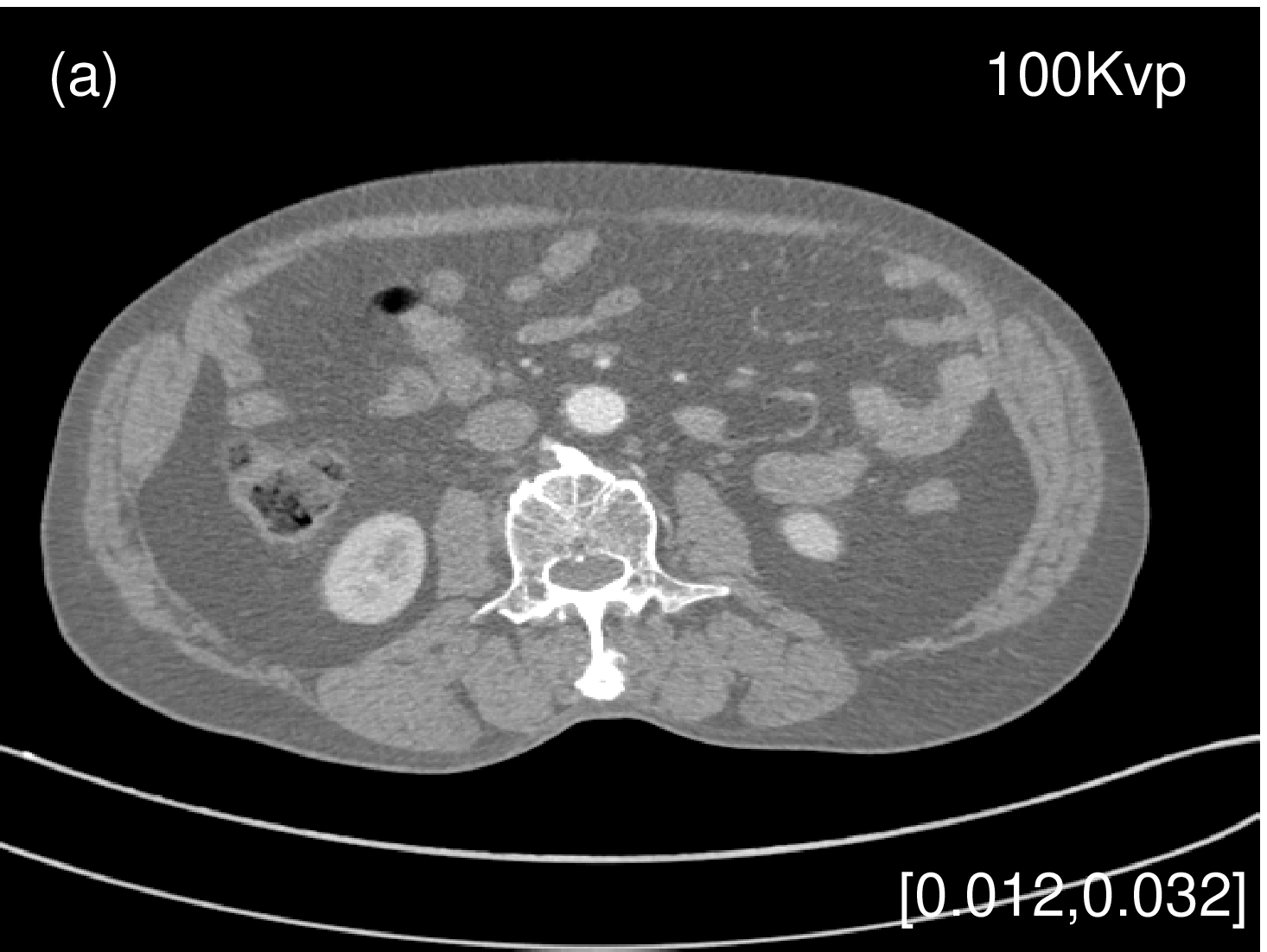}}
\subfigure{\includegraphics[width=0.3\linewidth, height=0.3\linewidth ]{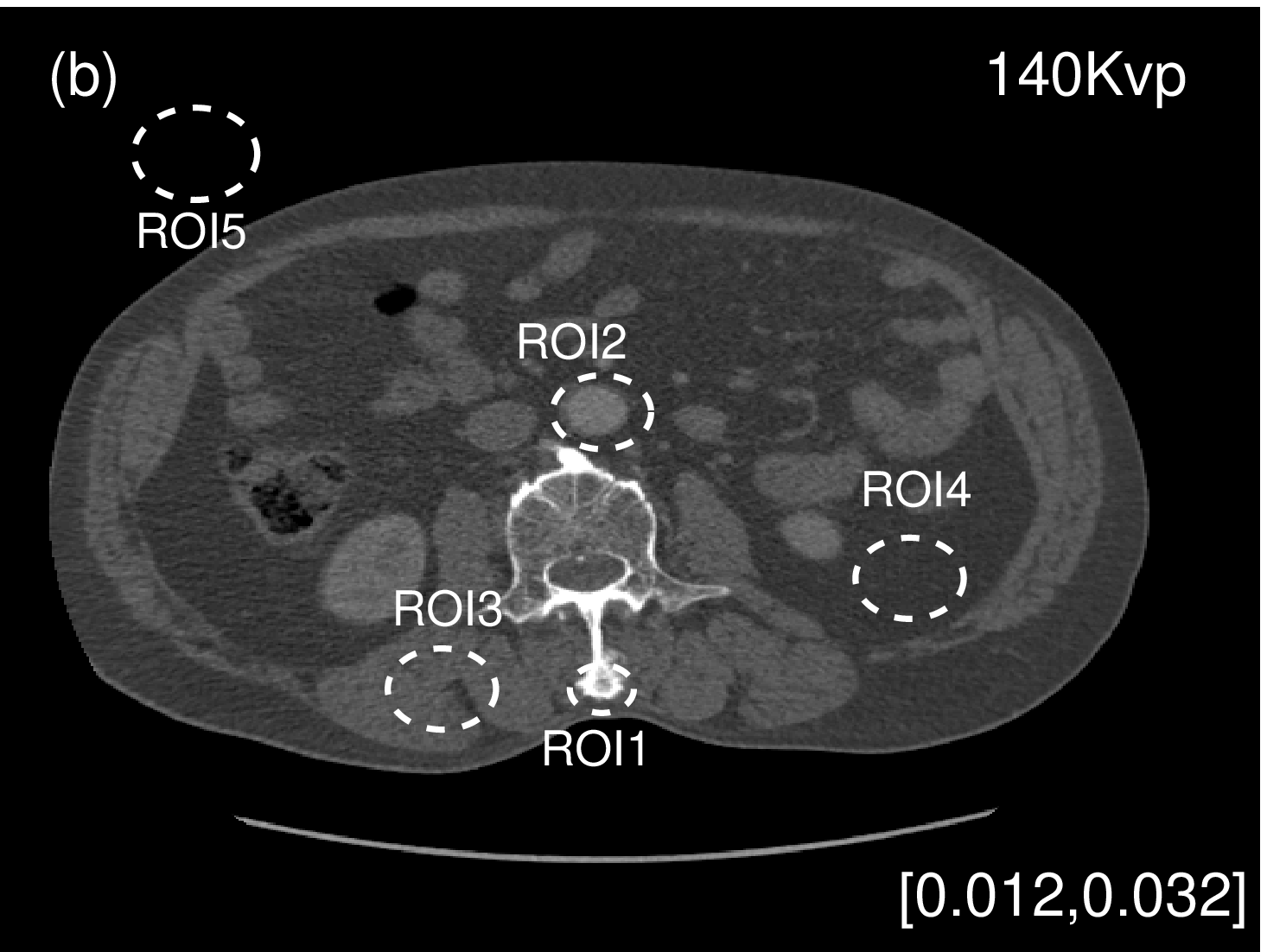}}
\caption { CT images of a pelvis patient. (a) The low-energy: 100 kVp and (b) The high-energy: 140 kVp. Display window is [0.012, 0.032] $\mathrm{mm}^{-1}$. The major components
of ROIs are bone (ROI1), iodine solution (ROI2), muscle (ROI3), fat (ROI4) and air (ROI5).}
\label{fig:pelvis}
\end{figure*}

\begin{figure*}[htbp!]
\begin{center}
\begin{tabular}{c@{\hspace{0pt}}c@{\hspace{0pt}}c@{\hspace{0pt}}c@{\hspace{0pt}}c@{\hspace{0pt}}c@{\hspace{0pt}}c@{\hspace{0pt}}c}
\includegraphics[width=.2\linewidth,height=.2\linewidth]{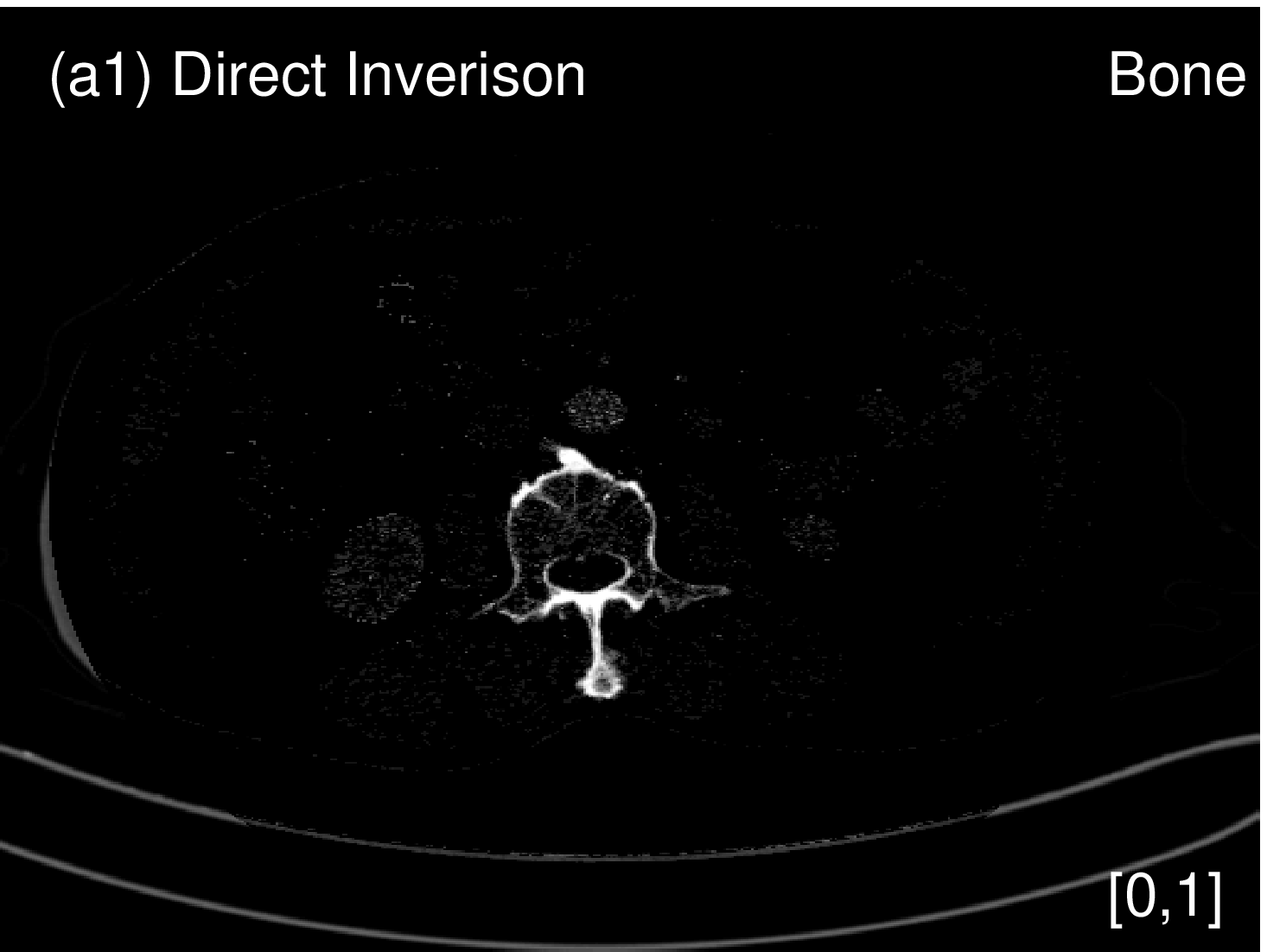}&
\includegraphics[width=.2\linewidth,height=.2\linewidth]{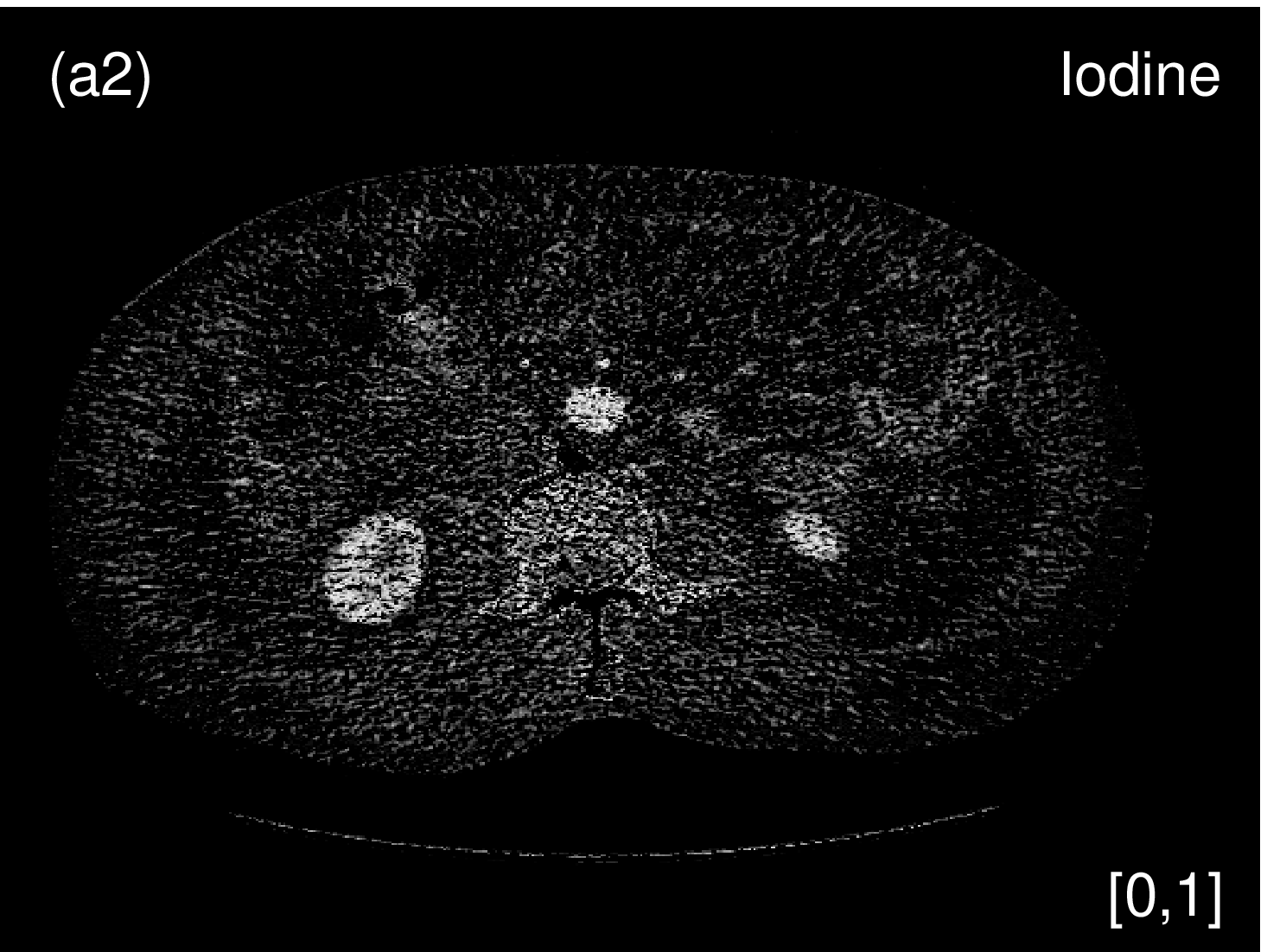}&
\includegraphics[width=.2\linewidth,height=.2\linewidth]{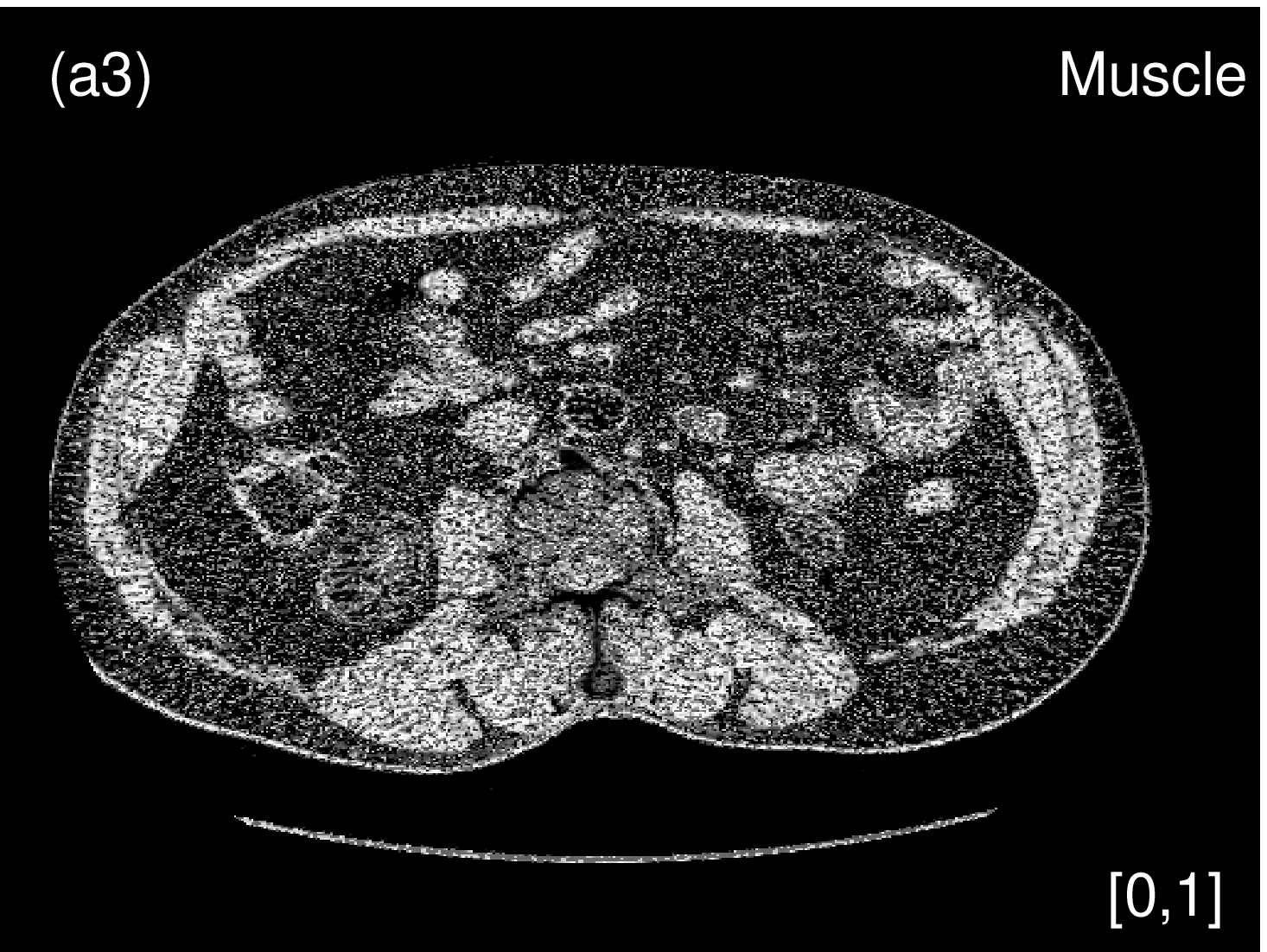}&
\includegraphics[width=.2\linewidth,height=.2\linewidth]{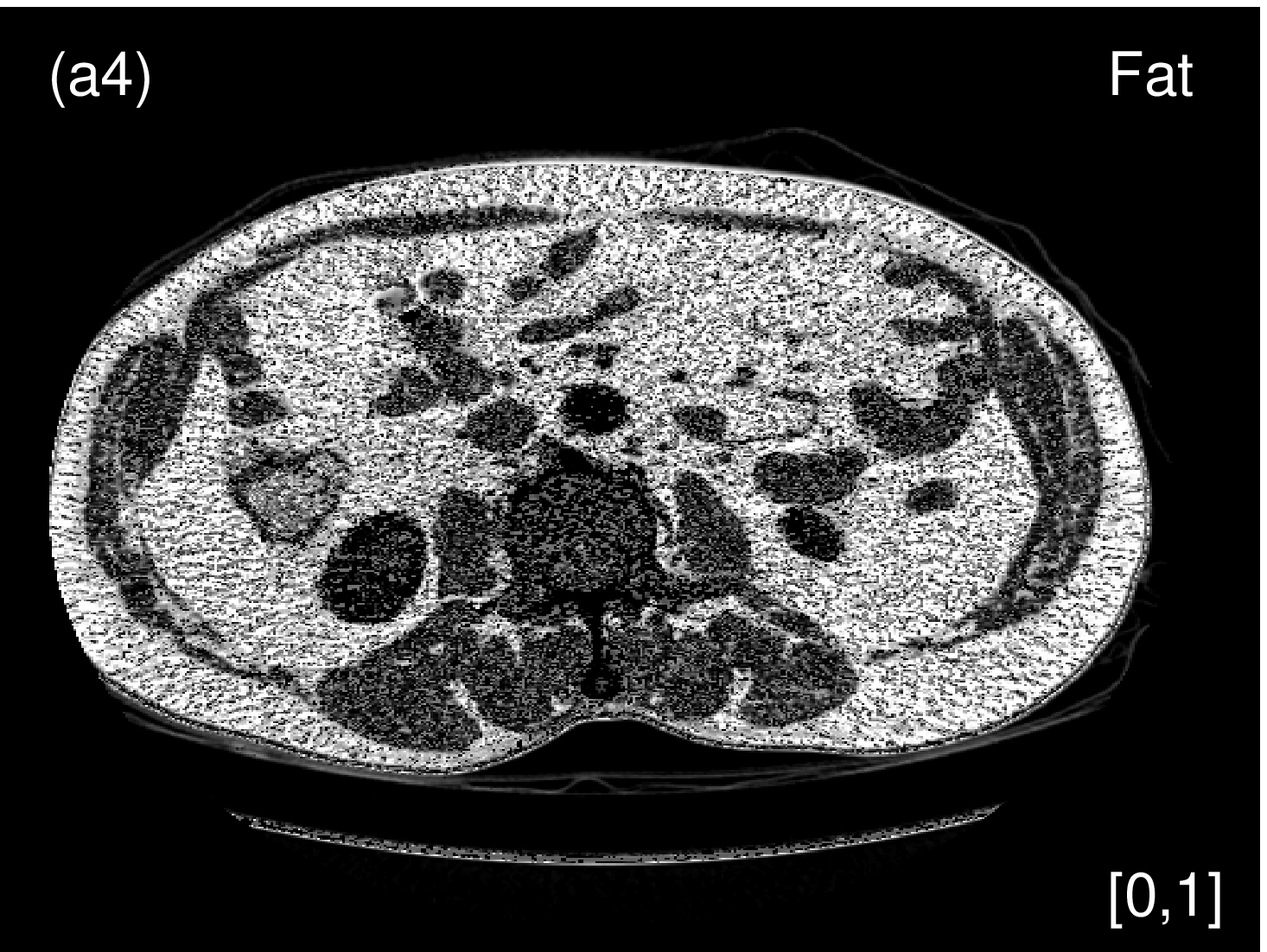}&
\includegraphics[width=.2\linewidth,height=.2\linewidth]{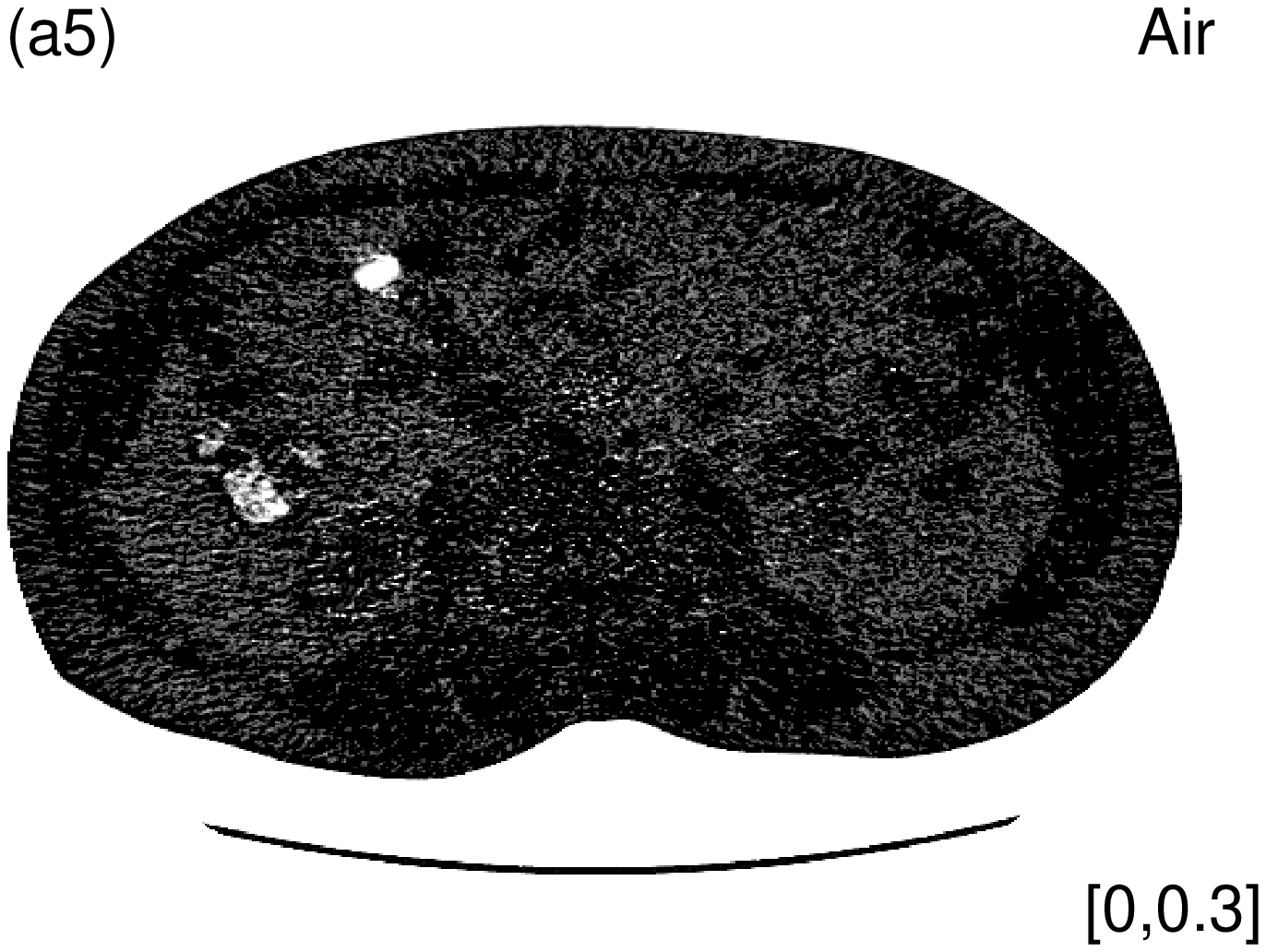}\\
\includegraphics[width=.2\linewidth,height=.2\linewidth]{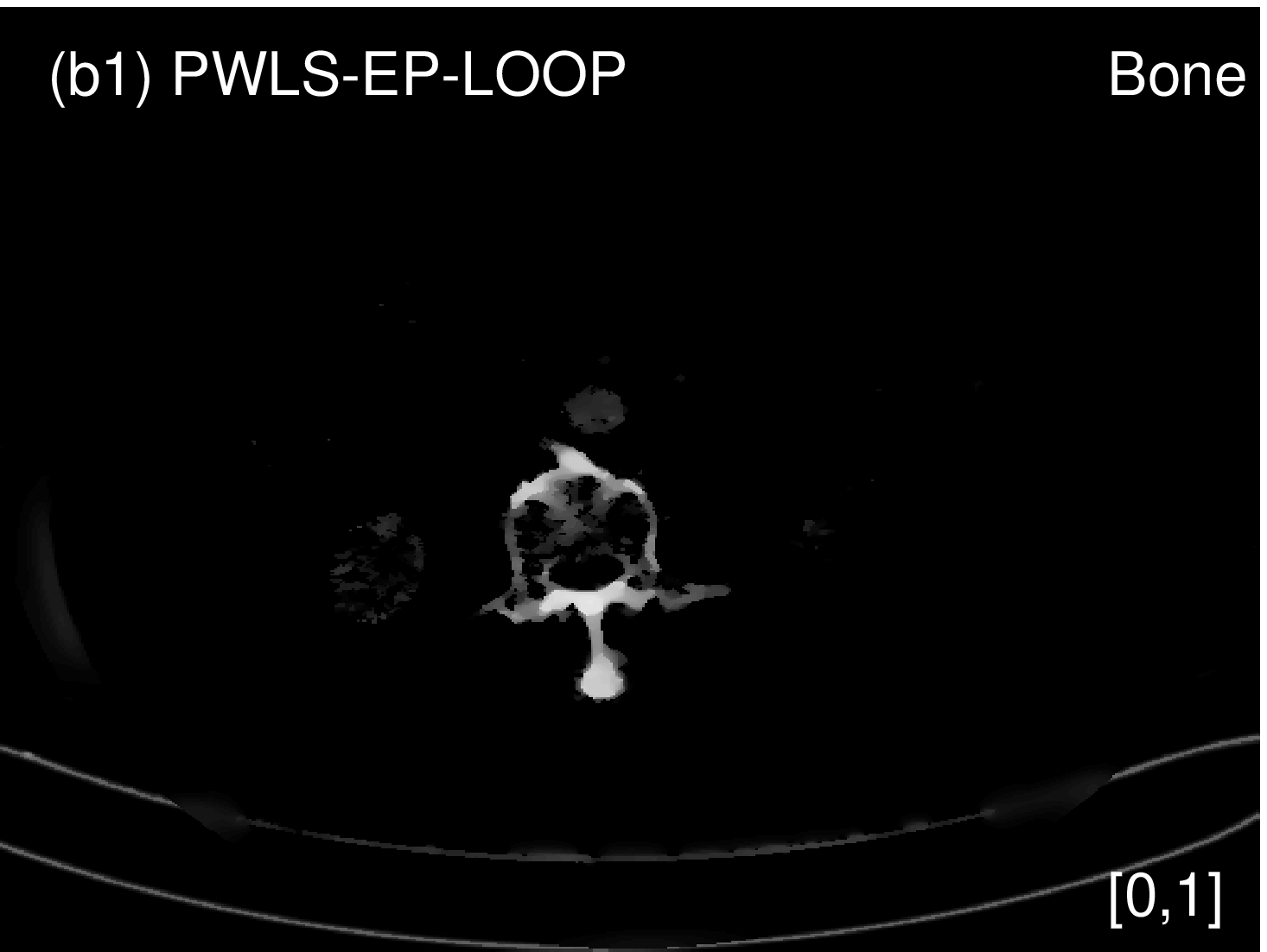}&
\includegraphics[width=.2\linewidth,height=.2\linewidth]{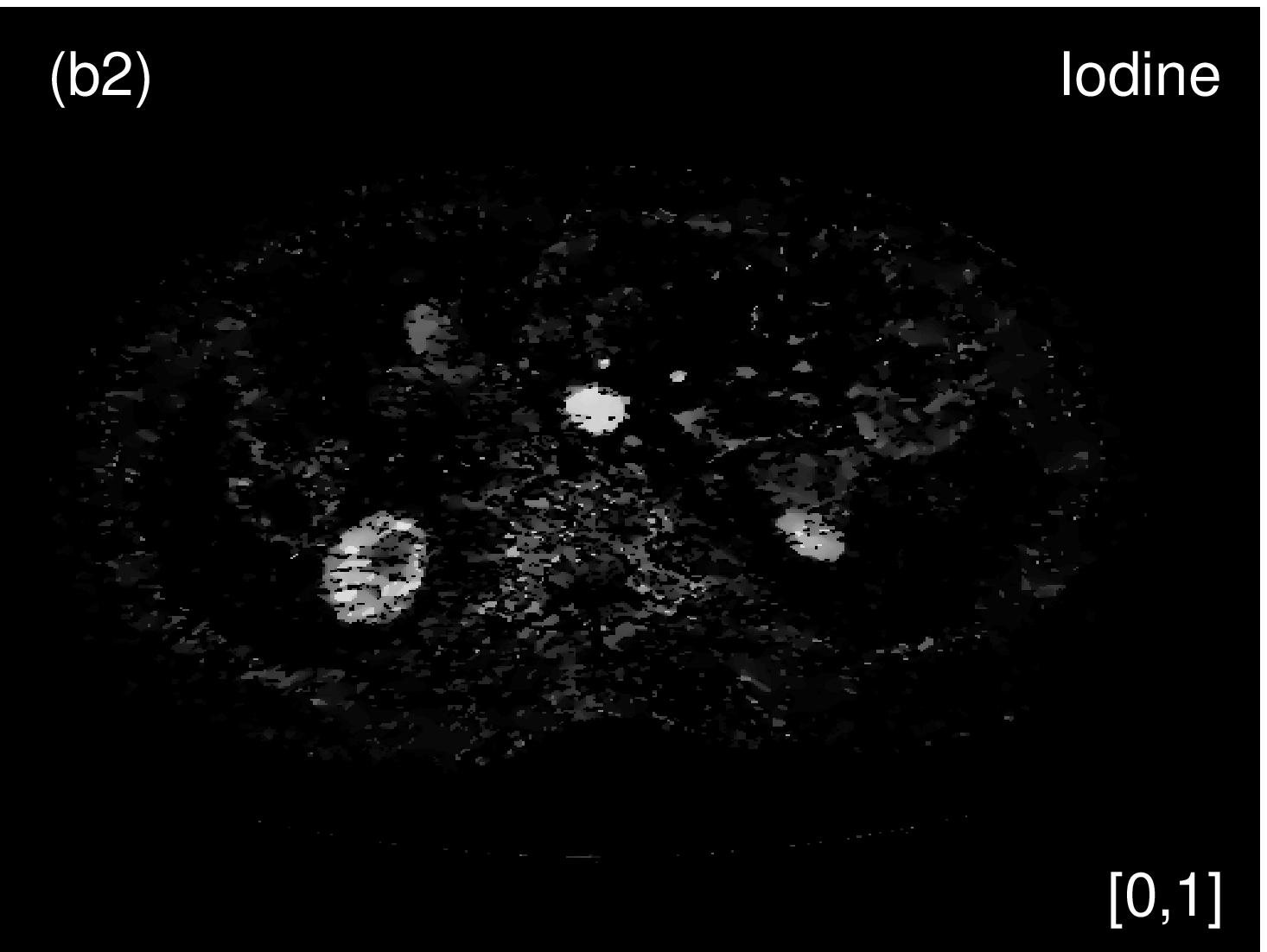}&
\includegraphics[width=.2\linewidth,height=.2\linewidth]{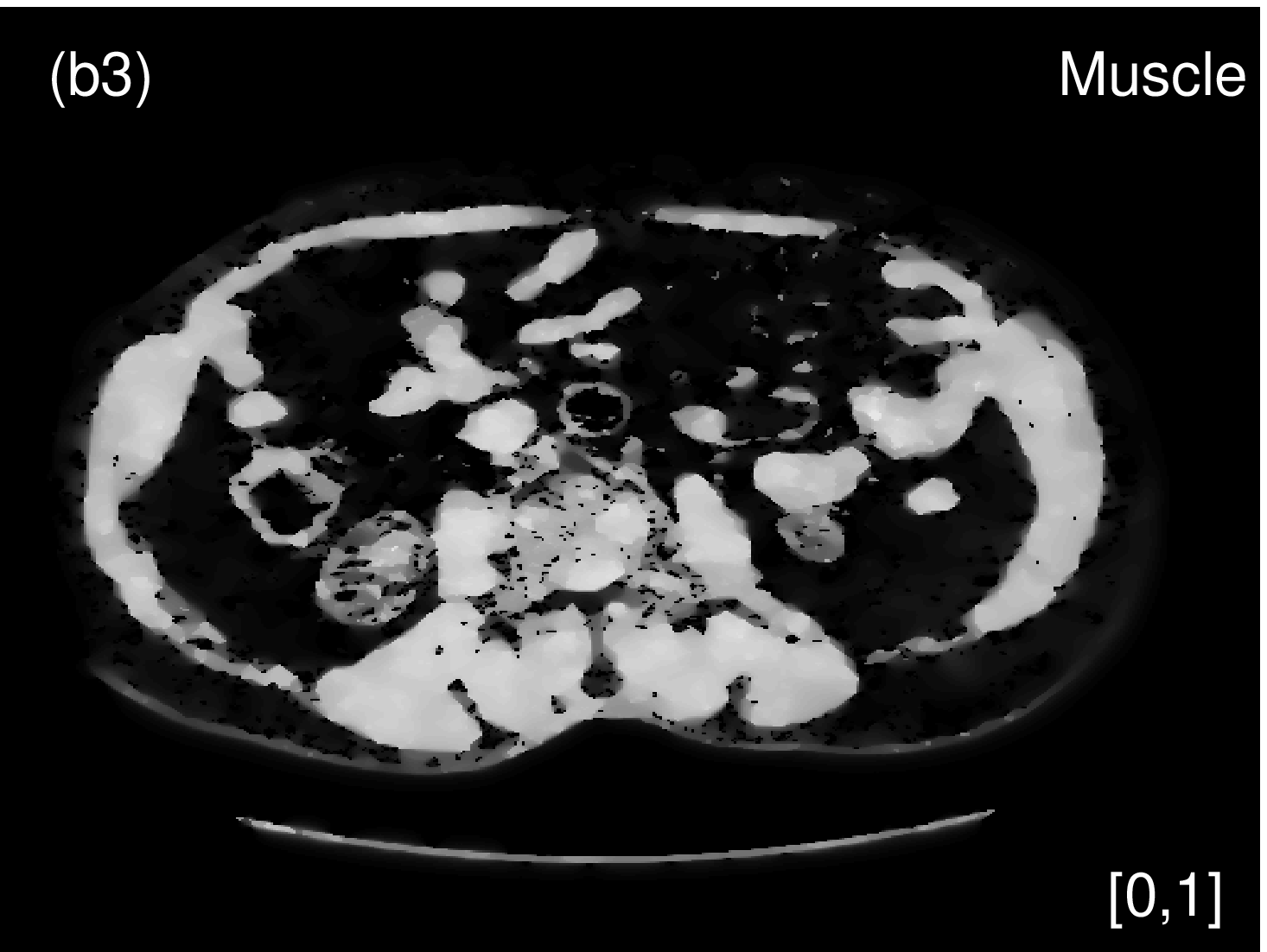}&
\includegraphics[width=.2\linewidth,height=.2\linewidth]{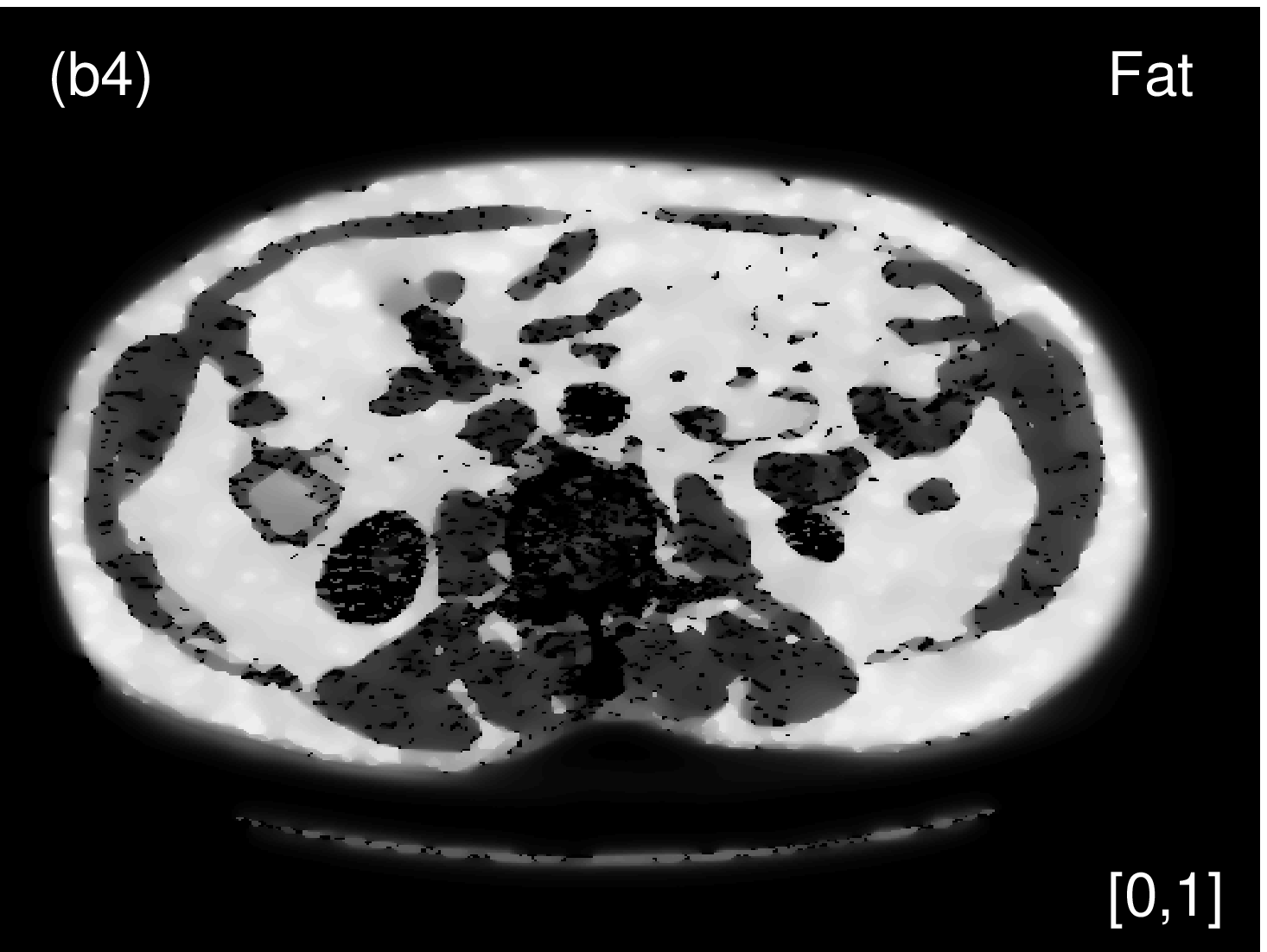}&
\includegraphics[width=.2\linewidth,height=.2\linewidth]{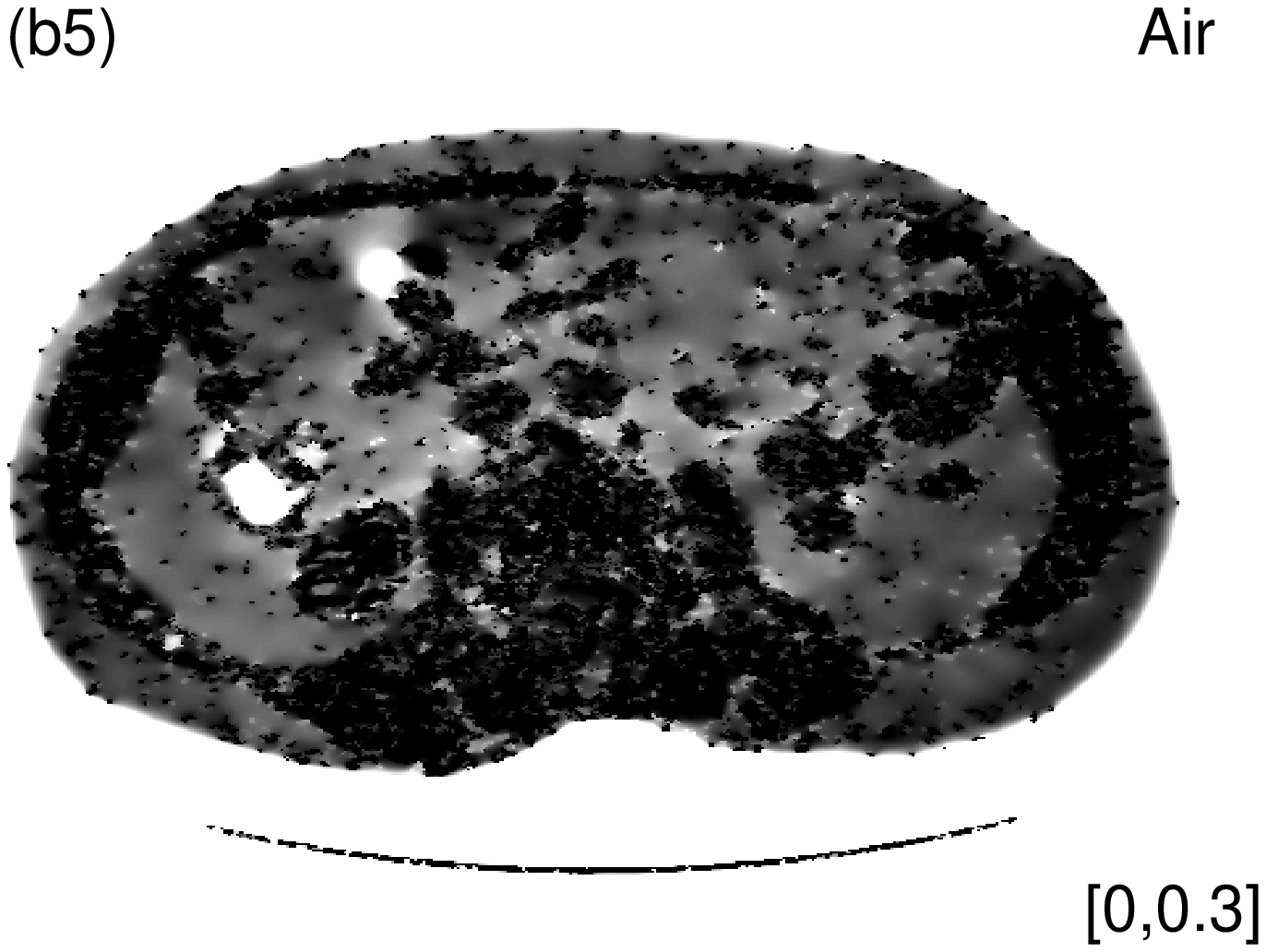}\\
\includegraphics[width=.2\linewidth,height=.2\linewidth]{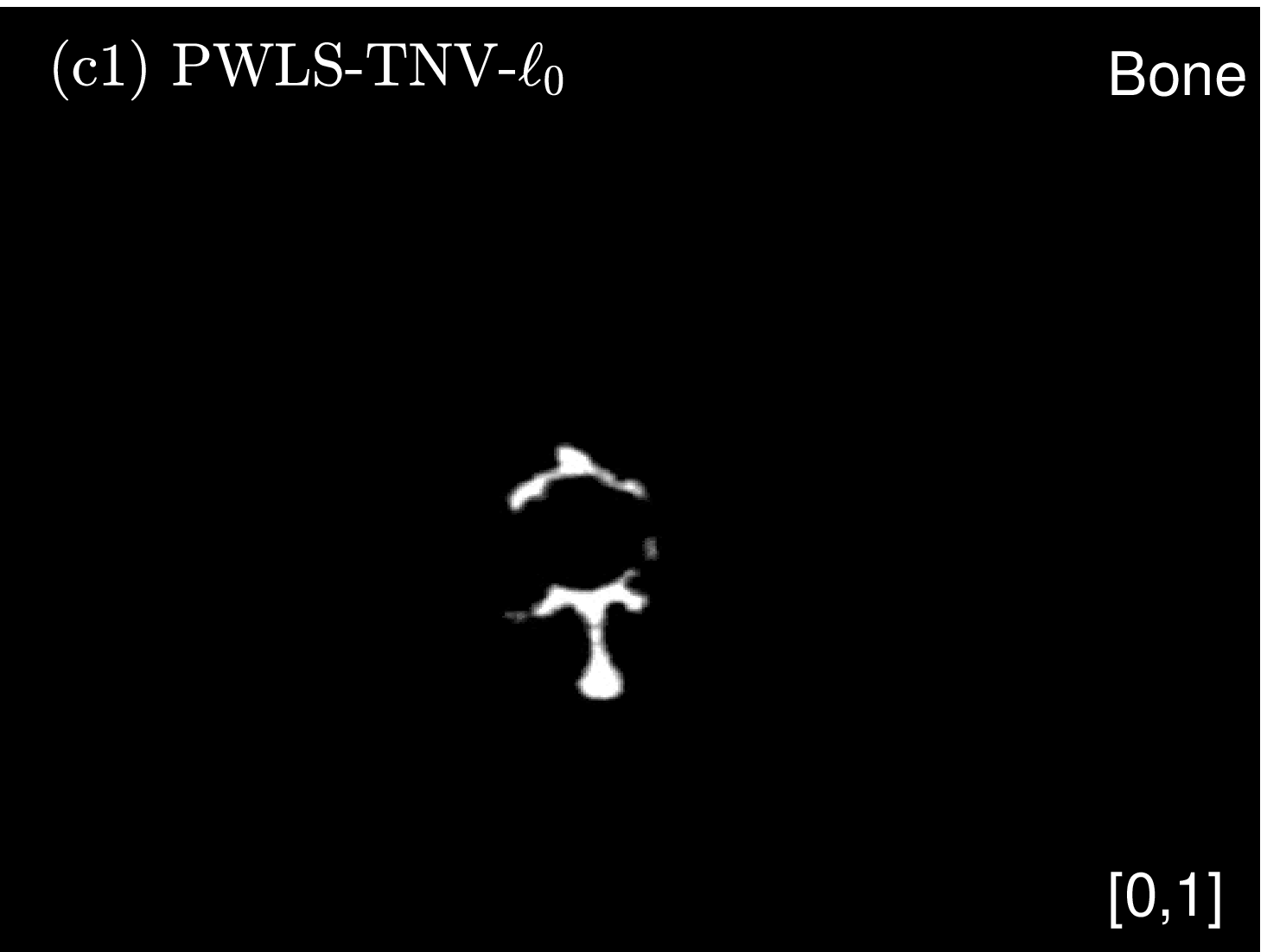}&
\includegraphics[width=.2\linewidth,height=.2\linewidth]{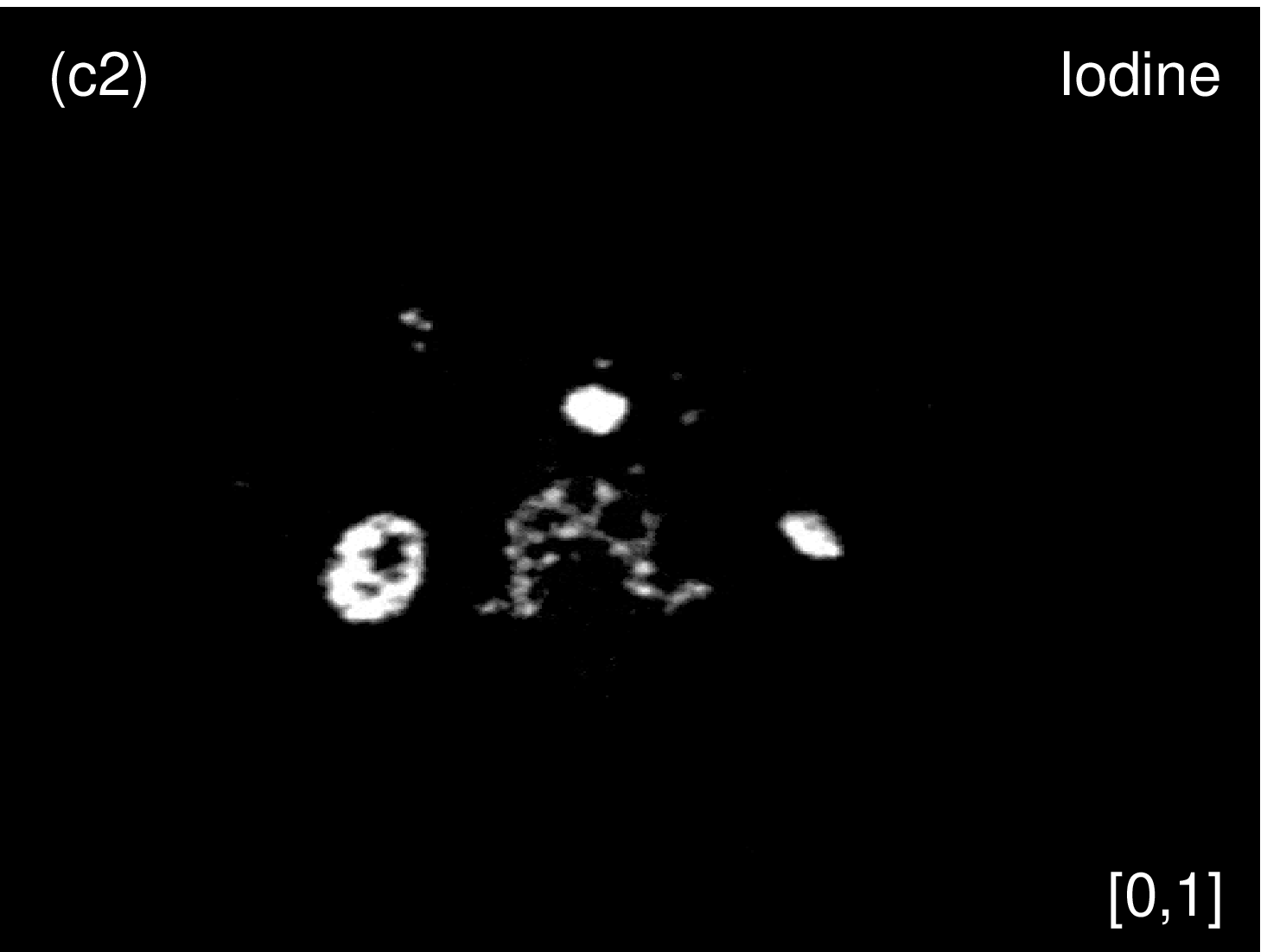}&
\includegraphics[width=.2\linewidth,height=.2\linewidth]{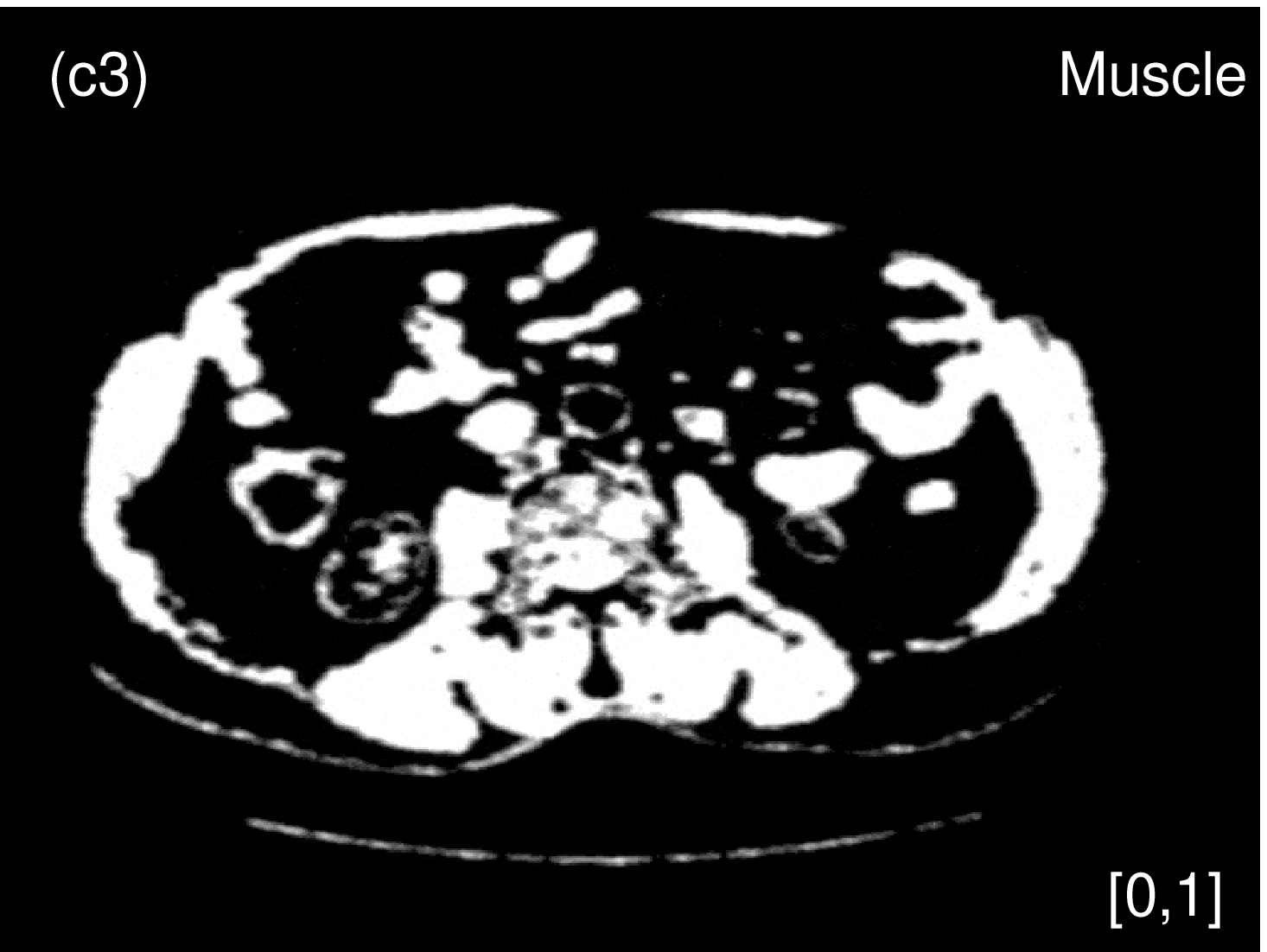}&
\includegraphics[width=.2\linewidth,height=.2\linewidth]{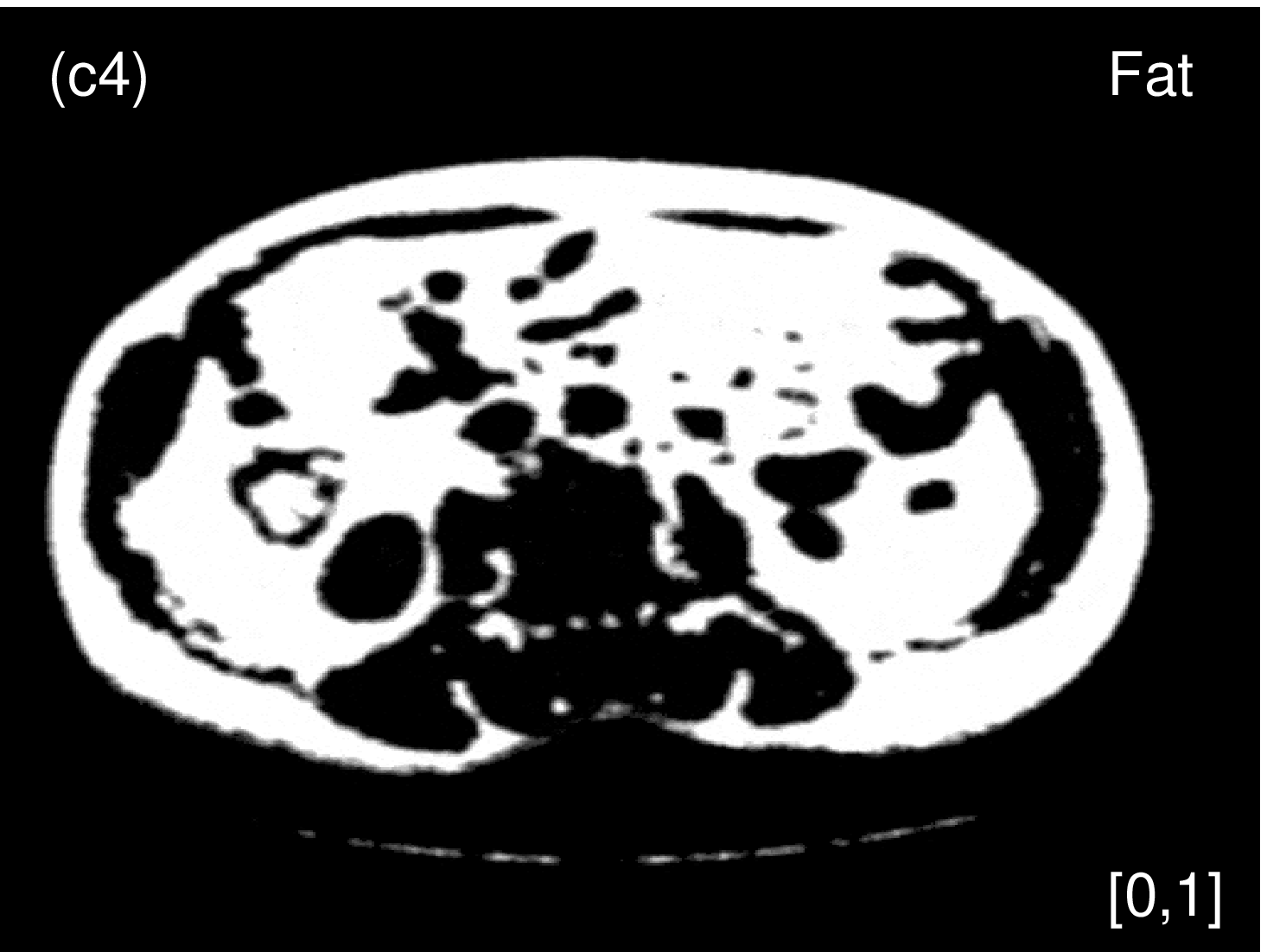}&
\includegraphics[width=.2\linewidth,height=.2\linewidth]{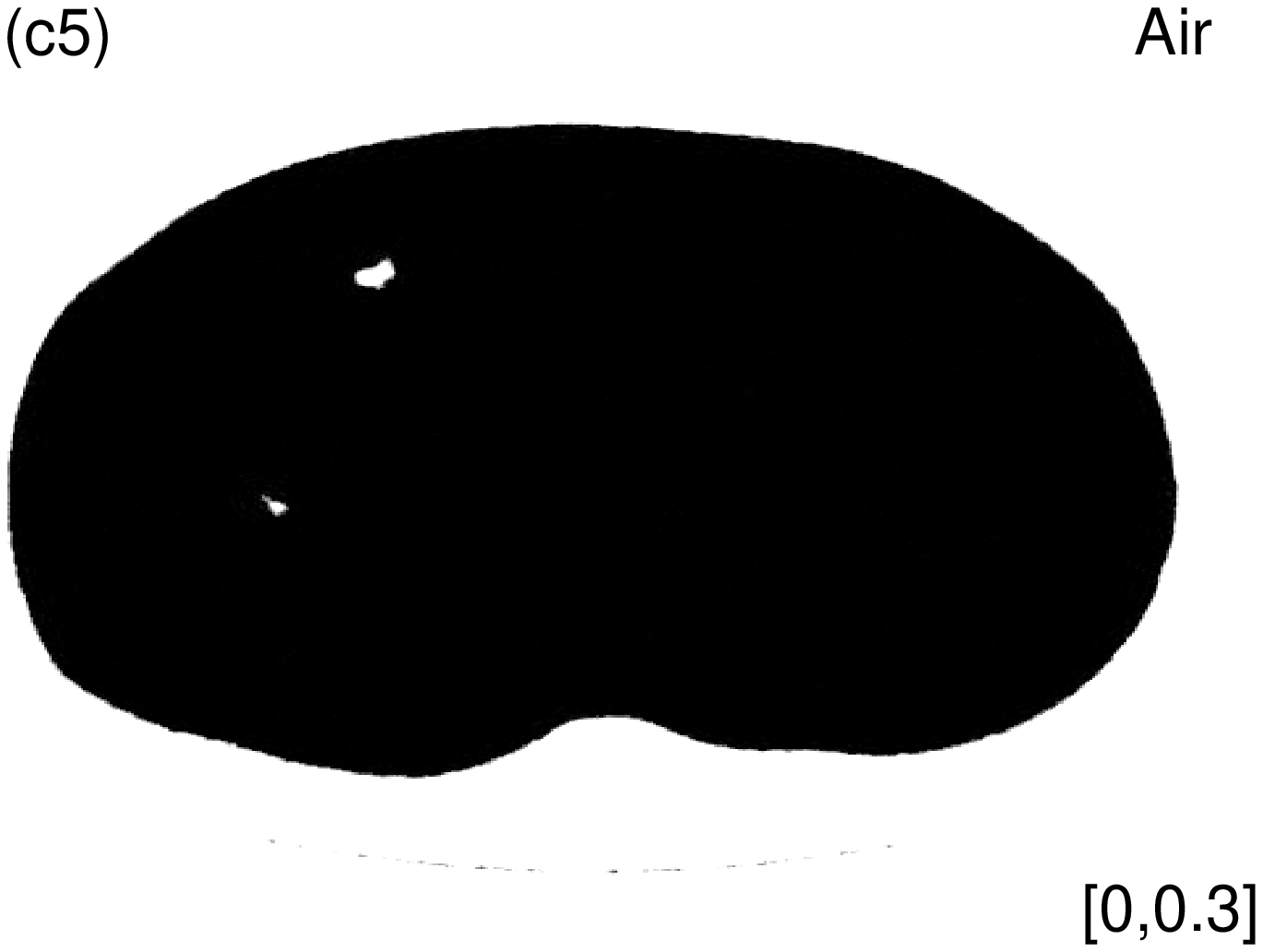}
\end{tabular}
\caption {  Material images of  Direct Inversion (the $1^{st}$ row), PWLS-EP-LOOP (the $2^{rd}$ row) and PWLS-TNV-$\ell_0$ (the $3^{nd}$ row).
The decomposed bone (the $1^{st}$  column), iodine (the $2^{ed}$  column), muscle (the $3^{nd}$ column), fat (the $4^{th}$ column) and air (the $5^{th}$  column) images.
 The display windows are shown in the bottom-right corners.}
\label{PelvisMM}
\end{center}
\end{figure*}

We also evaluated the proposed PWLS-TNV-$\ell_0$ method
using clinical pelvis data.
The patient's pelvis data was acquired
by Siemens SOMATOM Definition flash CT scanner
using DECT imaging protocol.
Table~\ref{DataAcquire} lists acquisition parameters
in the pelvis data scan.
Fig.~\ref{fig:pelvis} shows the high- and low-energy CT images
of the pelvis data.
Fig.\ref{fig:pelvis}~(b) shows selected basis materials,
bone, iodine, muscle, fat and air, and their assosicated ROIs
highlightened in white dashed
line circles.
We implemented the Direct Inversion
method in \cite{mendonca2014a} and used its results
as the initialization for the PWLS-EP-LOOP \cite{xue2017statistical} and the
proposed PWLS-TNV-$\ell_0$ method.
Fig.~\ref{PelvisMM} shows the decomposed material images by the
Direct Inversion, the PWLS-EP-LOOP and the PWLS-TNV-$\ell_0$ method.
Table~\ref{PelvisSTD} summarizes
the means and noise STDs of the decomposed material images by the above three
methods.
The volume fraction (VF) accuracies are $80.48\%$, $86.50\%$, and $99.96\%$
for the Direct Inversion method, the PWLS-EP-LOOP method and the
proposed PWLS-TNV-$\ell_0$, respectively.
Compared with the Direct Inversion and PWLS-EP-LOOP method,
the proposed method improves the VF accuracy by $19.48\%$ and $13.46\%$
respectively.
The proposed PWLS-TNV-$\ell_0$ method decomposes basis material images more
accurately, suppresses noise and decreases crosstalk, while
retaining spatial resolution of the decomposed images
 compared to
the other two methods.

\begin{table*}[htbp!]
\caption{\leftline{The means and STDs of decomposed images within each ROI of pelvis data.}}
\scalebox{1.0}{
\begin{tabular}{p{2.9cm}p{2.9cm}p{2.9cm}p{2.9cm}p{2.9cm}p{2.4cm}}
\hline
\hline
\multirow{2}{*}{Methods} & ROI1&  ROI2&     ROI3&    ROI4&  ROI5\\
                         & Bone&  Iodine&   Muscle&   Fat&  Air\\
\hline
Direct Inversion &       $1.000\pm0.000$&  $0.6380\pm0.2692$&     $0.6623\pm0.2603$&    $0.7237\pm0.2711$&  $1.0000\pm0.0000$\\
PWLS-EP-LOOP&            $0.8868\pm0.0055$&$0.7844\pm0.1595$&     $0.7914\pm0.0194$&    $0.8623\pm0.0208$&  $1.0000\pm0.0000$\\
PWLS-TNV-$\ell_0$&       $1.0012\pm0.0129$&$0.9998\pm0.0121$&     $1.0003\pm0.0084$&    $1.0002\pm0.0099$&  $1.0000\pm0.0003$\\
\hline
\hline
\end{tabular}
}
\label{PelvisSTD}
\end{table*}
\section{Discussion }
\label{discussion}

We proposed a statistical image-domain MMD method for DECT,
named PWLS-TNV-$\ell_0$.
Its cost function is in the form of PWLS estimation with a negative
log-likelihood term and three regularization terms.
The first TNV regularization term considers structural correlation
among basis material images, \emph{i.e.},
different material images share common or complementary edges and
material images are piecewise constant.
The second regularization term encourages
sparsity of material types in each pixel,
which is different from previous work
\cite{mendonca2014a,long2014multi-material}
that imposes a constraint that
each pixel contains at most three materials.
Considering volume and mass conservation,
the third  regularization term
includes sum-to-one and box constraint
which are imposed in the optimization process in
previous work
\cite{mendonca2014a,long2014multi-material,xue2017statistical}.
We applied the popular algorithm, ADMM,
to optimizate the proposed PWLS-TNV-$\ell_0$ problem.
Initialization is important for the PWLS-TNV-$\ell_0$ method
since its cost function is non-convex.
We set results of the Direct Inversion method
\cite{mendonca2014a}
as initialization for the proposed PWLS-TNV-$\ell_0$ method
to help with converging to a decent local minimum.

The PWLS-TNV-$\ell_0$ method requires
to tune two regularization parameters
and several other parameters when
optimizing its cost function using ADMM.
The choice of parameters significantly influences
the decomposed material images.
We need to determine appropriate combination of parameters
for each DECT dataset.
With the appropriate combination of parameters,
the propose PWLS-TNV-$\ell_0$ method decreases
noise while maintaining resolution of decomposed material images.
How to choose the parameters is still a challenge problem
and future work will investigate how to chose these parameters.
The most time consuming operation in the proposed method is
solving problem \eqref{a}
which requires SVD operation for
every pixel in each iteration.
We will investigate acceleration methods to
speed up the SVD operation in future work.
Similar to our previous work \cite{xue2017statistical,niu2014iterative},
the statistical weight of the proposed
PWLS-TNV-$\ell_0$ method was estimated
by the calculated numerical variance of two manually selected
homogeneous regions with a single material in both the high- and
low-energy CT image.
This variance estimation method assumes that the noise
in the high- and low-energy CT images are uncorrelated,
noise in pixels are uncorrelated and every pixel has the same
noise variance.
More accurate pixel-wise noise variance
can be estimated on a serial of DECT images
acquired from repeated scans on the same object.
This method is not practical
to implement on clinical patients due to accumulated high radiation dose.
Zhang-O'Connor and Fessler proposed a fast method to predict variance images
of PWLS or PL reconstructions with quadratic regularization from
sinograms or pre-log data \cite{zhang2007fast}.
Li \emph{et al.} proposed a computationally efficient technique for local noise
estimation directly from CT images \cite{li2014adaptive}.
We will investigate noise covariance estimation methods and apply them to the
PWLS-TNV-$\ell_0$ method in future work.

\section{Conclusion}
\label{conclusion}
We proposed an image-domain MMD method using DECT measurements and named it the
PWLS-TNV-$\ell_0$ method.
We imposed low rank property of material image gradients,
sparsity of material composition and mass and volume conservation
to help the proposed PWLS-TNV-$\ell_0$ method with estimating
multiple material images from DECT measurements.
To minimize the proposed cost function,
we introduced auxiliary variables so that the
original optimization problem can be divided into
solvable subproblems by the ADMM
method.
Testing on simulated digital phantom, Catphan\copyright600 phantom and
 clinical data, we concluded that the proposed PWLS-TNV-$\ell_0$ method
 suppresses noise and crosstalk, increases decomposition accuracy and
 maintains image resolution in the decomposed material images,
 compared to existing image-domain MMD methods using DECT measurements,
 the Direct Inversion and the PWLS-EP-LOOP method.

\begin{acknowledgments}
Xiaoqun Zhang and Qiaoqiao Ding are supported in part by Chinese 973 Program (Grant No. 2015CB856000) and National Youth Top-notch Talent program in China.
Tianye Niu is supported in part by  Zhejiang Provincial Natural Science Foundation of China (Grant No. LR16F010001), National High-tech R\&D Program for Young Scientists by the Ministry of Science and Technology of China (Grant No. 2015AA020917).
Yong Long is supported in part by NSFC (Grant No. 61501292) and the Interdisciplinary Program of Shanghai Jiao Tong University (Grant No. YG2015QN05).
\end{acknowledgments}

\section*{Appendix}
\label{Appendix}
The operators $\mathcal{D}$,  $\mathcal{H}$, $\mathcal{P}$ corresponding with  the subproblem of auxiliary variables, $u$,$v$ and $w$, are  (\ref{Subsola}),  (\ref{Subsolb}), (\ref{Subsolc}).
We will give the calculative methods in details.
\begin{itemize}
  \item The singular value thresholding operator, $\mathcal{D}_\cdot(\cdot)$, is the proximal operator  associated with the nuclear norm \cite{cai2010singular}. For  $\tau\geq0$ and $\bm{Y}\in \mathbb{R}^{n_1\times n_2}$, the  singular value shrinkage operator obeys
\begin{align}
\mathcal{D}_\tau(\bm{Y})=&\mathrm{prox}_{\lambda\|\cdot\|_\ast}(Y)\nonumber\\
=&\arg\min_{\bm{X}}\tau\|\bm{X}\|_\ast+\frac{1}{2}\|\bm{X}-\bm{Y}\|_F^2.
\end{align}
The singular value decomposition (SVD) of  $\bm{Y}$ is
\begin{align}
\bm{Y}=\bm{U}\bm{\Sigma}\bm{V}^\ast,
\end{align}
where $\bm{U}\in \mathbb{R}^{n_1\times r}$, $\bm{V}\in \mathbb{R}^{n_2\times r}$ with orthonormal
columns, and $\bm{\Sigma}=\rm{diag}(\{ \sigma_i\}_{1\leq i\leq r})$. We obtain
\begin{equation}
\mathcal{D}_\tau(\bm{Y}):=\bm{U}\mathcal{D}_\tau(\bm{\Sigma})\bm{V}^\ast,
\end{equation}
where $\mathcal{D}_\tau(\bm{\Sigma})=\rm{diag}(\{ \sigma_i-\tau\}_+)$, $\{ t \}_+= \max(0, t)$.

For each pixel $j$, we have
 \begin{align}
\bm{u}^{n+1}(:,:,j)&=\mathcal{D}_{\frac{\beta_1}{\gamma_1}}([\bm{D}\bm{x}^{n+1}+\frac{\bm{p}^n_1}{\gamma_1}](:,:,j)),\nonumber\\
j=&1,\cdots, N_p.
 \end{align}
\item
 For nonnegative $\lambda$ and vector $x$, the hard thresholding operator \cite{blumensath2008iterative,Trzasko2007Sparse} is defined as
    \begin{align}
     \mathcal{H}_{\lambda}(x)=\mathrm{prox}_{\lambda\|\cdot\|_0}(x)=\arg\min_y \lambda\|y\|_0+\frac{1}{2}\|y-x\|_2^2,
    \end{align}
with
\begin{equation}
 (\mathcal{H}_{\lambda}(x))_i=\left \{
 \begin{array}{lll}
   x_i      & \textrm{if  $|x_i|>\sqrt{2\lambda} $},\\
 \{0,x_i\}  &\textrm{if  $|x_i|=\sqrt{2\lambda} $},\\
 0         &\textrm{if  $|x_i|<\sqrt{2\lambda} $}.
 \end{array}
 \right .
  \end{equation}

  The closed-form solution  for \eqref{bb} is obtained by
    \begin{equation}
        \bm{v}^{n+1}=\mathcal{H}_{\frac{\beta_2}{\gamma_2}}(\nabla\bm{x}^{n+1}+\frac{\bm{p}^n_2}{\gamma_2}).\label{solb}
   \end{equation}
\item  For nonnegative $\lambda$ and vector $x$, we define
\begin{align}
\mathcal{P}_{\lambda^+}(x)=\mathrm{prox}_{\chi_{S}(\cdot)}(x)=\arg\min_y \chi_{S}(y)+\frac{1}{2}\|y-x\|_2^2,
\end{align}
where $S=\{x :\sum_i x_{i}=\lambda, x_{i} \geq 0\}$. Specifically,
\begin{align}
  (\mathcal{P}_{\lambda^+}(x))_i=\{x_i-\hat{t}\}_{+}
\end{align}
where \ $\hat{t}:=\frac{1}{n-k}(\sum_{j=k+1}^n x_{(j)} -\lambda)$ \ with\  $k:=\max\{p:x_{(p+1)}\geq\frac{1}{n-p}(\sum_{j=p+1}^n x_{(j)} -\lambda)\}$ and $x_{(1)}\leq \cdots \leq x_{(n)}$  is the permutation of  $x $ in ascending order \cite{kyrillidis2012sparse,chen2011projection}.

For each pixel $j$, subproblem \eqref{d} is the projection on to a simplex,
\begin{align}
(\bm{w}^{n+1})_j&=\mathcal{P}_{1^+}((\bm{x}^{n+1}+\frac{\bm{p}^n_3}{\gamma_3})_j ),\nonumber\\
 j&=1,\cdots N_p.
\end{align}
\end{itemize}
 \bibliographystyle{Plain}

\end{document}